\documentclass[11pt,a4paper]{article}

\usepackage[english]{babel}
\usepackage[utf8]{inputenc}
\usepackage[T1]{fontenc}
\usepackage{geometry}
\usepackage{setspace}
\usepackage{amsmath,amssymb,amsfonts}
\numberwithin{equation}{subsection}
\usepackage{graphicx}
\usepackage[hidelinks]{hyperref}
\hypersetup{pdfauthor={Lucas Ramírez Zalduendo},pdftitle={Worm Algorithms for the Schwinger Model},pdfsubject={Bachelor's Thesis in Physics}}
\usepackage{tikz}
\usepackage{wrapfig}
\usepackage{caption}
\usepackage{capt-of}
\usepackage{pgfplots}
\usepgfplotslibrary{groupplots}
\usetikzlibrary{calc}
\pgfplotsset{compat=1.18}
\usepackage{xcolor}
\definecolor{upzero}{RGB}{0,70,200}       
\definecolor{uppi}{RGB}{0,150,255}        
\definecolor{rhozero}{RGB}{200,0,0}       
\definecolor{rhopi}{RGB}{230,120,0}       

\pgfplotsset{compat=1.18}
\usepackage{subcaption}
\usepackage{csquotes}
\usetikzlibrary{arrows.meta,decorations.markings,calc,babel}
\usepackage{eso-pic}
\usepackage{transparent}
\usepackage{titlesec}

\titleformat{\section}[display]
  {\normalfont\bfseries}
  {\Large Chapter \thesection}
  {0.1cm}
  {\huge}

\titlespacing*{\section}
  {0pt}
  {0.2cm}
  {0.4cm}

\usepackage{float}
\usepackage[ruled,vlined,noend]{algorithm2e}

\SetKwInput{KwData}{Input}
\SetKwInput{KwResult}{Output}
\SetKw{KwReject}{reject}
\SetKw{KwAccept}{accept}
\SetKw{TRUE}{true}

\IncMargin{-0.4em}
\SetInd{0.4em}{0.7em}

\SetAlgoInsideSkip{smallskip}

\usepackage[
  backend=biber,
  style=numeric,
  sorting=none,
  url=true,
  doi=true,
  isbn=true,
  giveninits=true
]{biblatex}
\DefineBibliographyStrings{english}{
  references = {References},
}
\AddToHook{cmd/section/before}{\clearpage}

\title{Worm Algorithms for the Schwinger Model}\author{Author \\Lucas Ramírez Zalduendo\\Supervisor\\Alejandro Vaquero Avilés-Casco\\BSc in Physics\\Faculty of Science, University of Zaragoza}\date{Academic year 2025--2026}

\begin{document}

\pagenumbering{gobble}

\begin{titlepage}
\AddToShipoutPictureBG*{%
  \put(10,0){%
    \includegraphics[width=1\paperwidth]{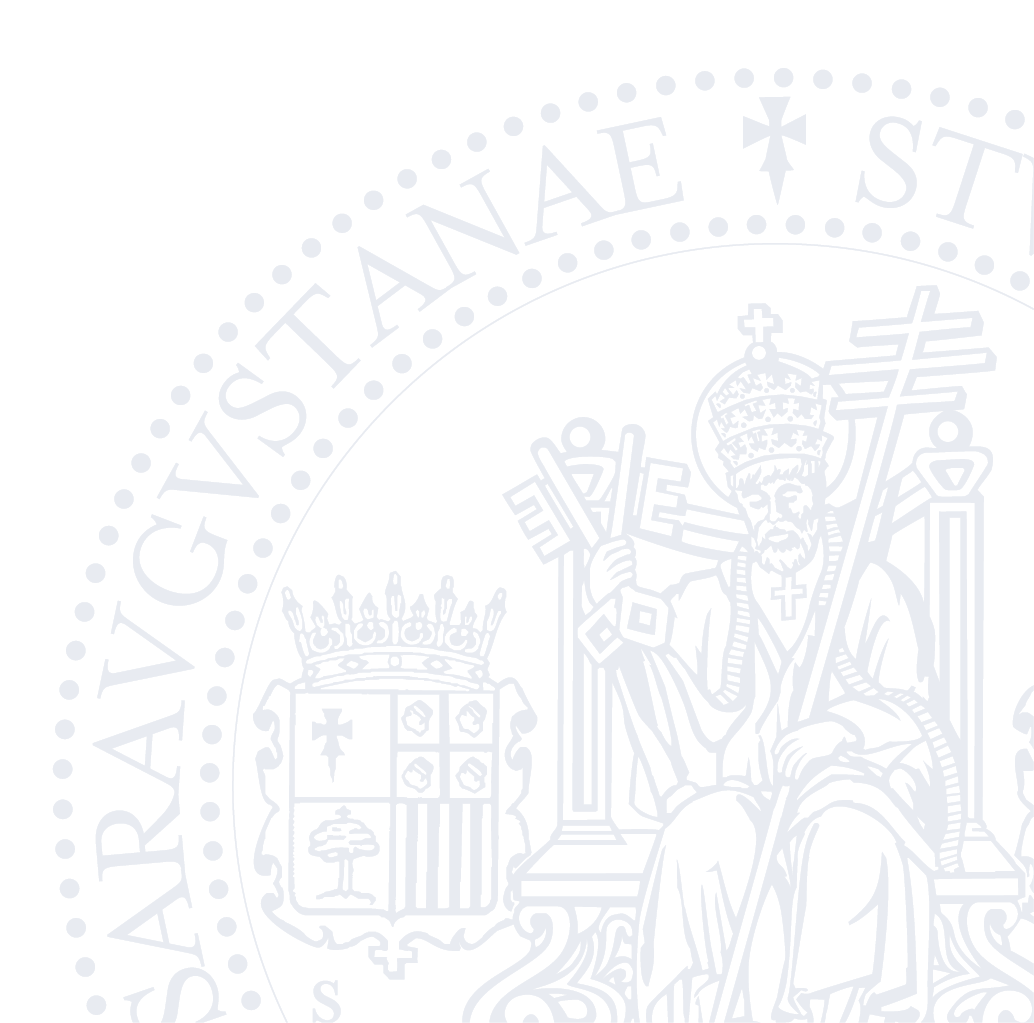}%
  }%
}

\begin{figure}[h]
    \centering
    \includegraphics[width=0.4\linewidth]{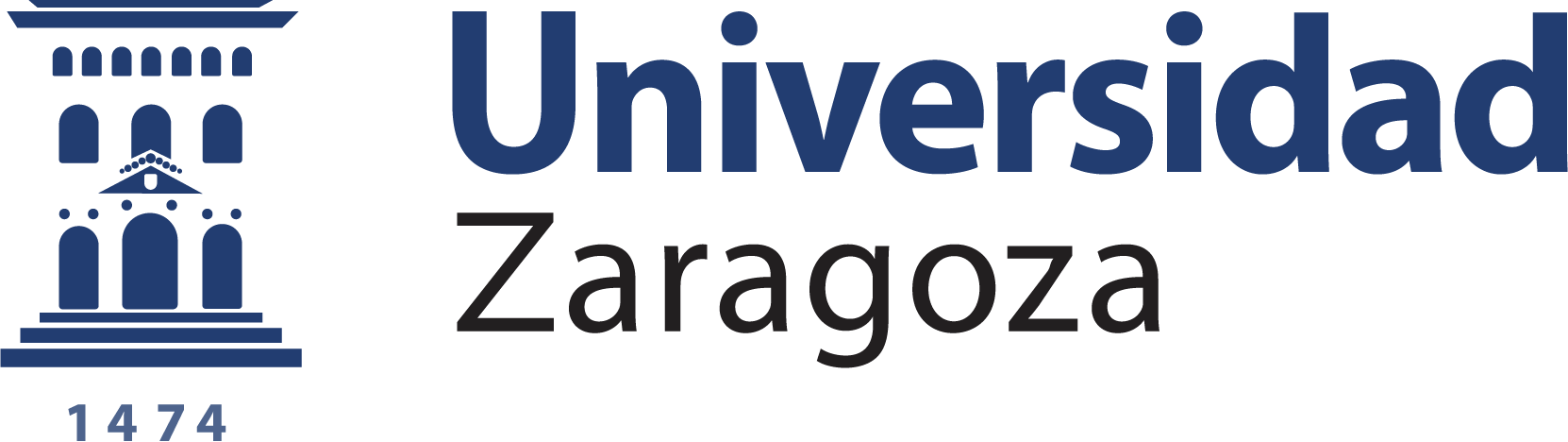}
\end{figure}
\vspace{1.2 cm}

\centering\Huge{\textbf{Bachelor's Thesis}}\vspace{0.2 cm}

\Large{\textbf{BSc in Physics}}
\vspace{1 cm}

\huge{\textbf{Worm Algorithms for the Schwinger Model}}

\vspace{3 cm}
\Large\textbf{{Author \\ Lucas Ramírez Zalduendo}} \\
\vspace{1 cm}
\Large\textbf{{Supervisor \\ Alejandro Vaquero Avilés-Casco}}

\vfill
\begin{figure}[h]
    \centering
    \includegraphics[width=0.4\linewidth]{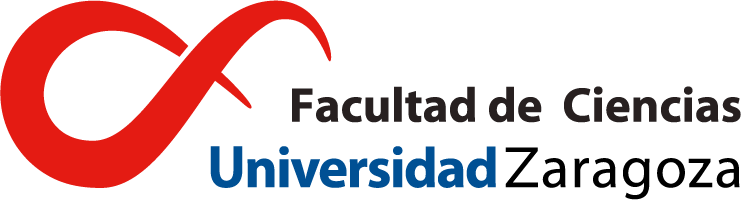}
\end{figure}
\Large{\textbf{Academic year 2025--2026}}

\end{titlepage}

\clearpage

\pagenumbering{roman}
\setcounter{page}{1}

\section*{Abstract}

This thesis studies the Schwinger model in $(1+1)$ dimensions using a lattice formulation and its dual representation. The model provides a simple yet non-trivial setting for investigating characteristic phenomena of gauge theories, including screening, confinement-related behaviour, the Schwinger effect and dependence on the topological angle ($\theta$). After introducing the Euclidean formulation with \textit{staggered} fermions, we describe the dual transformation in which the original degrees of freedom are replaced by configurations of plaquettes, dimers and, in the massive case, monomers.

\vspace{0.3 cm}

The main objective is to investigate the numerical behaviour of the model in the presence of a fermion mass and a topological angle. We implement Monte Carlo algorithms adapted to the dual representation, including local and non-local updates that allow admissible configurations to be sampled. For the massive model, we also analyse the emergence of the sign problem and its effect on the estimation of observables through reweighting.

\vspace{0.3 cm}

From the simulations we compute several quantities, including the average plaquette, the dimer density, the chiral condensate and the average sign. These observables characterise the dependence on the mass, coupling and topological angle ($\theta$), and identify the regions in which the sign problem becomes most significant. Finally, we discuss the numerical results and the usefulness of the dual formulation for investigating gauge theories in non-perturbative regimes.

\newpage

{\small\tableofcontents}

\newpage

\listoffigures

\newpage
\pagenumbering{arabic}
\setcounter{page}{1}

\section{Introduction}

Gauge theories play a central role in the modern description of fundamental interactions. However, some of their characteristic phenomena, including confinement, dynamical mass generation and the non-trivial structure of the vacuum, are difficult to investigate analytically. It is therefore useful to consider simpler models that retain part of this physics and allow it to be studied in a more controlled setting.

\vspace{0.3 cm}

One of the most important examples is the Schwinger model, corresponding to quantum electrodynamics in $1+1$ dimensions. The model exhibits relevant non-perturbative properties, including screening, confinement-related behaviour, dynamical mass generation and dependence on a topological angle $\theta$. It also provides a natural connection with the Schwinger effect, associated with particle--antiparticle pair creation in the presence of a strong electric field. Thus, although it is a simplified version of a four-dimensional gauge theory, it serves as a theoretical laboratory for controlled investigations of these phenomena.

\subsection{The Schwinger model and the Schwinger effect}

The Schwinger model is an Abelian $U(1)$ gauge theory with fermionic fields defined in one spatial and one temporal dimension. Its Lagrangian is
\begin{equation}
\label{modelo_con_masa}
\mathcal{L}
=
-\frac{1}{4}F_{\mu\nu}F^{\mu\nu}
+
\bar{\psi}
\left(i\gamma^\mu D_\mu - m\right)
\psi ,
\end{equation}

where $\psi$ is the fermionic field, $m$ the fermion mass and $A_\mu$ the gauge field. The covariant derivative and electromagnetic field-strength tensor are defined as
\begin{equation}
\label{deriv_cov}
D_\mu = \partial_\mu + i g A_\mu,
\qquad
F_{\mu\nu}=\partial_\mu A_\nu-\partial_\nu A_\mu .
\end{equation}

\vspace{-0.2 cm}

In QED, the coupling constant $g$ is the fermion charge. In $1+1$ dimensions the tensor $F_{\mu\nu}$ has only one independent component, associated with the electric field, and we adopt the convention $E\equiv F_{01}$.

\vspace{0.3 cm}

One of the most interesting properties of the model is dynamical mass generation. The photon acquires an effective mass without the need to introduce an explicit mass term for the gauge field. For the continuum theory with one Dirac fermion and vanishing fermion mass, $m=0$, this mass follows analytically from bosonisation (see Appendix~\ref{app:bosonizacion}) and is given by
\begin{equation}
\label{masa_boson_schwinger}
m_\gamma = \frac{g}{\sqrt{\pi}} .
\end{equation}

This one-flavour result illustrates dynamical mass generation and electric-field screening. It should not be identified quantitatively with the continuum spectrum of the unrooted staggered lattice formulation used below, which has two tastes in two dimensions (Section~\ref{Ferrmion_reticulo}). Another characteristic feature of two-dimensional gauge dynamics is the competition between confinement and screening.

Because there are no transverse directions into which the electric field can spread, the electric flux between two external charges is restricted to the segment joining them in the pure-gauge approximation. Dynamical fermions, however, may screen the charges and break the resulting electric string; in particular, the massless Schwinger model exhibits screening rather than an indefinitely rising asymptotic potential. If the charges are separated by a distance $L$ and the field between them is approximately constant, the stored energy is
\begin{equation}
U_{\mathrm{field}}
=
\frac{1}{2}\int dx\,E^2(x)
\simeq
\frac{1}{2}E^2L .
\end{equation}

This classical field configuration has an energy growing linearly with separation, as in a confining string. It is not, by itself, a statement about the asymptotic static potential of the full theory with dynamical fermions, where pair production can cause screening or string breaking. 

\vspace{0.3 cm}

This picture helps motivate the Schwinger effect: an external electric field can separate a virtual pair and supply the energy required for it to become a real particle--antiparticle pair. Although the phenomenon is not exclusive to $1+1$ dimensions, it arises naturally in the Schwinger model because of the linear relation between energy and charge separation. As discussed in \cite{Dunne2004HeisenbergEuler}, for a constant electric field the pair-production rate contains the exponential factor

\vspace{-0.2cm}

\begin{equation}
\Gamma \propto
\exp\left(
-\frac{\pi m^2}{gE}
\right).
\end{equation}

Consequently, pair production is suppressed at weak electric fields and increases with field strength. One may also add to the Lagrangian (\ref{modelo_con_masa}) a topological term associated with the angle $\theta$,
\begin{equation}
\label{eq:termino_theta}
\mathcal{L}_\theta
=
\frac{g\theta}{2\pi}F_{01}
=
\frac{g\theta}{2\pi}E ,
\end{equation}

which favours configurations with a background electric field. Varying $\theta$ does not directly fix $E$; rather, it changes the relative weights of configurations with different electric fields, connecting the vacuum structure with pair production and screening. This dependence is especially relevant near $\theta=\pi$, where an abrupt rearrangement of the vacuum structure may occur. This is discussed further in Appendix~\ref{app:termino_masa}.

\subsection{Motivation and objectives}

As discussed above, the Schwinger model captures several characteristic phenomena of gauge theories within a framework simple enough for controlled numerical investigation. It is therefore suitable for examining how the vacuum structure depends on gauge dynamics, the presence of fermions and the topological angle $\theta$.

\vspace{0.3 cm}

The aim of this thesis is to study the dependence of the Schwinger model on the fermion mass, gauge coupling and topological angle $\theta$, focusing on gauge-field observables and their relation to the background electric field and the Schwinger effect. Particular attention is given to the massive case, where numerical methods are especially valuable. To this end, we employ a lattice formulation and its dual representation in terms of monomers, dimers and loops, together with Monte Carlo simulations.

\section{Lattice Quantum Field Theory}
Having established the objectives of this thesis, this chapter introduces the framework needed to study the Schwinger model numerically, starting with its lattice formulation.
\subsection{From the continuum to the lattice}
In quantum field theory, a common approach to computing physical observables is perturbation theory. In this approach, an observable $\mathcal{O}(g)$ is expanded in powers of the coupling parameter,
\begin{equation}
\mathcal{O}(g)
\sim
\sum_{n=0}^{\infty} c_n g^n,
\qquad g\to 0.
\end{equation}

For weak coupling, the first few terms can provide a good approximation. However, the approach becomes unreliable in strongly coupled regimes. Moreover, perturbative series often fail to converge in a controlled way, so including terms of increasingly high order does not necessarily improve the result. This motivates the use of non-perturbative methods.

\vspace{0.3 cm}

Lattice field theory provides a particularly useful formulation. The central idea is to replace continuous spacetime with a discrete grid of points separated by a distance $a$, called the lattice spacing. In $d$ dimensions, the discretisation is defined by
\begin{equation}
x \in \mathbb{R}^d
\quad \longrightarrow \quad
x^\mu = a n^\mu,
\qquad n^\mu\in\mathbb{Z} ,
\end{equation}

where the index $\mu\in\{0,\ldots,d-1\}$ labels spacetime directions. Since the Schwinger model has $d=2$, we work with $\mu\in\{0,1\}$, representing the temporal and spatial directions, respectively. The fields are therefore defined only at integer multiples of the lattice spacing, called lattice sites. For convenience, we write
\begin{equation}
\psi(x)
\quad \longrightarrow \quad
\psi(an)\equiv \psi(n).
\end{equation}

On a finite lattice, the theory has finitely many degrees of freedom, enabling a numerical investigation. We therefore introduce the path integral.

\subsection{Path integral}

The path-integral formulation expresses expectation values as sums over all possible field configurations \cite{Rothe2012LatticeGaugeTheories}. Schematically, the expectation value of an observable $\mathcal{O}$ can be obtained in a manner analogous to the standard formulation of classical statistical mechanics:
\begin{equation}
\langle \mathcal{O} \rangle =
\frac{1}{Z}
\int \mathcal{D}\psi \,
\mathcal{O}[\psi] e^{iS[\psi]},
\qquad
Z =
\int \mathcal{D}\psi \, e^{iS[\psi]} .
\end{equation}

Here $Z$ is the partition function of the theory. Importantly, the factor $e^{iS}$ is complex and cannot directly be interpreted as a statistical weight.

To obtain a formulation suitable for numerical methods, we perform a Wick rotation to Euclidean time, starting from the metric $g_{\mu\nu}=\mathrm{diag}(1,-1)$, which becomes $\delta_{\mu\nu}$.
\begin{equation}
\label{wick_rot}
t\to -i\tau,
\qquad
iS_M\to -S_E,
\qquad
e^{iS_M}\to e^{-S_E}.
\end{equation}
Then,
\begin{equation}
\langle \mathcal{O} \rangle =
\frac{1}{Z}
\int \mathcal{D}\psi \,
\mathcal{O}[\psi] e^{-S_E[\psi]},
\qquad
Z =
\int \mathcal{D}\psi \, e^{-S_E[\psi]} .
\end{equation}

When the weights $e^{-S_E}$ are real and positive, they can be interpreted as a probability distribution and sampled using Monte Carlo methods. On the lattice, the functional integral is replaced by a multiple integral over the degrees of freedom defined at lattice sites,
\begin{equation}
\int \mathcal{D}\psi
\quad \longrightarrow \quad
\int \prod_n d\psi(n).
\end{equation}

\vspace{-0.2 cm}

If the weights are not real and positive, the so-called sign problem arises: they cannot be interpreted directly as probabilities, and methods such as reweighting are required. Furthermore, discretisation requires careful treatment of local gauge invariance and fermionic degrees of freedom, motivating the introduction of link variables and \textit{staggered} fermions in the following sections.

\subsection{Gauge fields on the lattice}
\label{Gauge_reticulo}

When moving to the lattice, a first attempt to discretise the covariant derivative (\ref{deriv_cov}) would be to replace the ordinary derivative by a finite difference and evaluate the gauge field at a single point. However, this compares fields at different sites directly and does not transform covariantly under local gauge transformations. A discretisation that connects the two sites in a gauge-covariant way is therefore needed.

\vspace{0.2cm}

\setlength{\columnsep}{0.55cm}
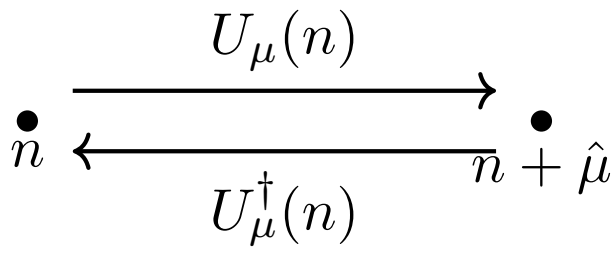
\begin{wrapfigure}[7]{r}{0.43\textwidth}
\vspace{-0.25cm}
\centering
\begin{tikzpicture}[scale=1.0]
    \fill (0,0) circle (2pt);
    \fill (3.4,0) circle (2pt);

    \node[below] at (0,0) {\(n\)};
    \node[below] at (3.4,0) {\(n+\hat{\mu}\)};

    \draw[->, thick] (0.3,0.2) -- (3.1,0.2);
    \node[above] at (1.7,0.2) {\(U_\mu(n)\)};

    \draw[->, thick] (3.1,-0.2) -- (0.3,-0.2);
    \node[below] at (1.7,-0.2) {\(U_\mu^\dagger(n)\)};
\end{tikzpicture}

{
\captionsetup{font=small}
\caption[Link-variable orientation]{Orientation of $U_\mu(n)$ and $U_\mu^\dagger(n)$ between two neighbouring lattice sites.}
}

\label{fig:enlace_conjugado}
\vspace{-0.15cm}
\end{wrapfigure}

\noindent
For this purpose, we introduce the link variable $U_\mu(n)$, which represents gauge parallel transport between neighbouring sites. It is associated with the oriented link from site $n$ to $n+\hat{\mu}$, where $\mu$ labels the link direction and $\hat{\mu}$ is the corresponding unit vector. As noted earlier, $\mu\in\{0,1\}$ in the Schwinger model, so there are two possible directions.

\vspace{0.4 cm}

The conjugate field $U_\mu^\dagger(n)=U_\mu(n)^{-1}$ represents the same link traversed in the opposite direction, with its endpoints interchanged. In this model the gauge group is $U(1)$, so

\vspace{-0.1 cm}

\begin{equation}
\label{enlace_Schw}
U_\mu(n)\in U(1),
\qquad
U_\mu(n)=e^{i\varphi_\mu(n)}.
\end{equation}

Schematically, these variables are related to the continuum gauge field through
\begin{equation}
\label{fase_en_continuo}
U_\mu(n)\sim e^{iagA_\mu(n)}.
\end{equation}

The discrete covariant derivative is then defined using these variables as
\begin{equation}
\label{deriv_cov_discret}
D_\mu\psi(n)
=
\frac{
U_\mu(n)\psi(n+\hat{\mu})
-
U_\mu^\dagger(n-\hat{\mu})\psi(n-\hat{\mu})
}{2a}.
\end{equation}

The link factors parallel-transport the neighbouring fields to site $n$, so that all terms can be compared at the same point. Under a local gauge transformation $\Omega(n)$, the link variables transform as
\begin{equation}
\label{trans_enlace}
U_\mu(n)\to U'_\mu(n)
=
\Omega(n)U_\mu(n)\Omega^\dagger(n+\hat{\mu})
\quad \Longrightarrow \quad
D_\mu\psi(n)\to \Omega(n)D_\mu\psi(n).
\end{equation}

Hence, the discrete covariant derivative transforms in the same way as the fermionic field, preserving the gauge structure of the theory on the lattice.

\subsection{Staggered fermions}
\label{Ferrmion_reticulo}
Discretising fermions introduces an additional difficulty: the \textit{fermion-doubling problem}. This arises because the discretised fermion operator may vanish at several momentum values, not only at the physical mode $p_\mu=0$. These additional zeros behave as local energy minima and lead to artificial copies of the original fermion.

\vspace{0.3 cm}

One way to mitigate this problem is to use \textit{staggered} fermions. In this formulation, the full set of spinor components is not introduced at each lattice site. Instead, one works with one-component Grassmann fields,
\begin{equation}
\label{grassman_unacomp}
\psi(n),\bar{\psi}(n)
\longrightarrow
\chi(n),\bar{\chi}(n).
\end{equation}

\vspace{-0.2 cm}

The spinorial structure of the original field is not removed but distributed across neighbouring lattice sites. Thus, the degrees of freedom represented by spinor components in the continuum formulation are rearranged on the lattice. As a consequence, the $\gamma^\mu$ matrices no longer appear explicitly in the action: their effect is encoded in site-dependent phase factors, called \textit{staggered phases},
\begin{equation}
\label{staggered_phase}
\eta^\mu(n)=(-1)^{n_0+\cdots+n_{\mu-1}}.
\end{equation}

\vspace{-0.2 cm}

These phases accompany hopping terms between neighbouring sites and reproduce the sign structure associated with the $\gamma^\mu$ matrices. In the presence of gauge fields, the hopping terms must include link variables to preserve local gauge invariance. The fermionic action therefore contains terms of the form
\begin{equation}
\bar{\chi}(n)\eta^\mu(n)U_\mu(n)\chi(n+\hat{\mu})
-
\bar{\chi}(n+\hat{\mu})\eta^\mu(n)U_\mu^\dagger(n)\chi(n).
\end{equation}

\vspace{-0.2 cm}

The link $U_\mu(n)$ enters the forward hop, whereas $U_\mu^\dagger(n)$ enters the reverse hop. The staggered formulation thus combines the rearrangement of fermionic degrees of freedom with the gauge structure introduced by the link variables.

\vspace{0.3 cm}

This formulation reduces, but does not eliminate, fermion doubling: in even dimension $d$, the number of species decreases from $2^d$ to $2^{d/2}$ \cite{Golterman2024StaggeredFermions}. Thus, for $d=2$, an unrooted staggered field describes two continuum tastes. The term ``one flavour'' in the dual lattice model refers to a single staggered field, not necessarily a single continuum flavour. Accordingly, the one-flavour bosonisation in Appendix~\ref{app:bosonizacion} serves as qualitative motivation rather than an exact description of the lattice theory.

\vspace{0.1 cm}

We have now introduced the ingredients required to turn a continuum fermionic theory into a lattice theory. The next step is to apply these transformations to the Schwinger model in order to use the lattice path integral.

\section{The Dual Schwinger Model}
\label{sec:schwingermono}
Starting from the discretisation rules introduced in the previous chapter, we construct a formulation of the Schwinger model suitable for Monte Carlo simulations.

\subsection{The Schwinger model on the lattice}

The starting point is the Lagrangian (\ref{modelo_con_masa}) with the topological term (\ref{eq:termino_theta}) added.
\begin{equation}
\label{modelo_completo_minkowski}
\mathcal{L}
=
-\frac{1}{4}F_{\mu\nu}F^{\mu\nu}
+
\bar{\psi}
\left(
i\gamma^\mu D_\mu - m
\right)
\psi
+
\frac{g\theta}{2\pi}F_{01}.
\end{equation}

To use the lattice path-integral formulation, we perform a Wick rotation as in (\ref{wick_rot}) and work in Euclidean spacetime. The corresponding Euclidean Lagrangian is
\begin{equation}
\label{modelo_completo_euclideo}
\mathcal{L}_E
=
\frac{1}{4}F_{\mu\nu}F^{\mu\nu}
+
\bar{\psi}
\left(
\gamma^\mu D_\mu + m
\right)
\psi
+
i\frac{g\theta}{2\pi}F_{01}.
\end{equation}

We construct the lattice action from this expression using the link variables and \textit{staggered} fermions introduced in the previous chapter. First, we separate the contributions to the Euclidean Lagrangian,
\begin{equation}
\mathcal{L}_E
=
\mathcal{L}_G
+
\mathcal{L}_F
+
\mathcal{L}_\theta,
\end{equation}
where
\begin{equation}
\label{terminos_Schwinger}
\mathcal{L}_G
=
\frac{1}{4}F_{\mu\nu}F^{\mu\nu},
\qquad
\mathcal{L}_F
=
\bar{\psi}
\left(
\gamma^\mu D_\mu + m
\right)
\psi,
\qquad
\mathcal{L}_\theta
=
i\frac{g\theta}{2\pi}F_{01}.
\end{equation}

The continuum Euclidean action is obtained by integrating over Euclidean spacetime,
\begin{equation}
S_E
=
\int d^2x\,
\mathcal{L}_E
=
S_G+S_F+S_\theta .
\end{equation}
\vspace{-0.6 cm}

On the lattice, the integral is replaced by a sum over sites $n=(n_0,n_1)$, following the direction convention established in the preceding chapter.
\begin{equation}
\label{int_to_sum}
\int d^2x \;\longrightarrow\; a^2 \sum_n ,
\end{equation}

\vspace{-0.2 cm}

Each term of the continuum action must therefore be replaced by its lattice counterpart. The details of this discretisation are given in Appendix~\ref{app:discretizado}, following the rules introduced above.

\vspace{0.3 cm}

First, with \textit{staggered} fermions, the fermionic contribution is written as
\begin{equation}
\label{S_F}
S_F
=
a^2\sum_n m\,\bar{\chi}(n)\chi(n)
+
\frac{a}{2}
\sum_{n,\mu}
\eta^\mu(n)
\left[
\bar{\chi}(n)U_\mu(n)\chi(n+\hat{\mu})
-
\bar{\chi}(n+\hat{\mu})U_\mu^\dagger(n)\chi(n)
\right].
\end{equation}

The gauge contribution is constructed from link variables. Under a local gauge transformation, a single link is not invariant but transforms according to (\ref{trans_enlace}). Thus, an individual link cannot by itself provide a gauge-invariant contribution to the action, because it carries transformation factors associated with both endpoints.

The simplest way to cancel these factors is to multiply links along a closed path. The gauge factors at intermediate sites then cancel. The smallest closed lattice path is a plaquette, corresponding in $1+1$ dimensions to a square of side $a$, defined by
\begin{equation}
\label{plaqueta}
U_{p}(n)
=
U_0(n)
U_1(n+\hat{0})
U_0^\dagger(n+\hat{1})
U_1^\dagger(n).
\end{equation}

\vspace{-0.1 cm}

This variable represents parallel transport around a lattice cell and is the discrete analogue of the field-strength tensor $F_{01}$. The lattice gauge action is not unique, but it must be local, gauge invariant and reproduce the continuum gauge contribution in the limit $a\to0$. A standard choice satisfying these requirements (see Appendix~\ref{T_gauge_anexo}) is
\begin{equation}
\label{S_G}
S_G
=
\beta
\sum_n
\left[
1-\operatorname{Re} U_p(n)
\right],
\qquad
\beta=\frac{1}{g^2a^2},
\end{equation}

\vspace{-0.2 cm}

where $\beta$ is the lattice gauge coupling. Finally, the topological term is discretised using the phase associated with the plaquette. Since the link variables belong to the group $U(1)$, which is closed under multiplication and conjugation, the plaquette also belongs to $U(1)$. Hence, we may write
\begin{equation}
\label{plaqueta_U1}
U_p(n)=e^{i\varphi_p(n)},
\end{equation}

where $\varphi_p(n)$ is the plaquette phase. In the continuum limit ($a\to0$), this phase is related to the field-strength tensor by
\begin{equation}
\label{fase_to_cont}
\varphi_p(n) \simeq a^2 g F_{01}(n).
\end{equation}
At the level of smooth continuum fields, the topological contribution is approximated by
\begin{equation}
S_\theta \simeq i\frac{\theta}{2\pi}\sum_n\varphi_p(n),
\qquad
Q_{\mathrm{phase}}=\frac{1}{2\pi}\sum_n\varphi_p(n).
\end{equation}
The phase expression depends on the branch used for the plaquette angles. Instead of identifying it with the charge in the simulated action at finite lattice spacing, we adopt the \emph{field-theoretical} lattice definition used in Ref.~\cite{Gattringer2015SchwingerDual}:
\begin{equation}
\label{carga_topo}
Q_{\mathrm{lat}}[U]
=\frac{1}{4\pi i}\sum_n\bigl[U_p(n)-U_p^{-1}(n)\bigr]
=\frac{1}{2\pi}\sum_n\sin\varphi_p(n).
\end{equation}
The lattice topological term used throughout the dual formulation is therefore
\begin{equation}
\label{S_theta}
S_\theta=i\theta Q_{\mathrm{lat}}[U]
=\frac{\theta}{4\pi}\sum_n\bigl[U_p(n)-U_p^{-1}(n)\bigr].
\end{equation}
The expressions $Q_{\mathrm{phase}}$ and $Q_{\mathrm{lat}}$ agree to leading order for smooth fields as $a\to0$, but are \emph{not identical} at finite lattice spacing. In particular, $Q_{\mathrm{lat}}$ is generally not integer-valued and exact $2\pi$ periodicity in $\theta$ is not guaranteed at finite $a$. A suitable continuum limit is needed to recover the continuum interpretation of $\theta$ as an angle \cite{Gattringer2015SchwingerDual,Kloiber2014ScalarQEDTheta}.
The displayed term is purely imaginary for $U_p\in U(1)$ and produces complex weights in the original representation.

We have now obtained the Euclidean lattice action studied in this thesis. However, the topological term makes the action complex, so the weight $e^{-S_E}$ cannot be interpreted directly as a probability. We therefore introduce a dual representation in terms of discrete geometric variables, inspired by monomer--dimer--loop formulations such as that of Karsch and Mütter for strong-coupling QCD \cite{Karsch1988StrongCouplingQCD}. The reformulation removes the sign problem in the massless case and provides the configuration space for the update algorithms. In the massive case, however, negative weights reappear and reweighting is required.

\subsection{Dual representation of the massless model}

We first consider the massless theory, $m=0$. The fermionic action then contains only hopping terms between neighbouring sites. The lattice partition function is
\begin{equation}
Z
=
\int \mathcal{D}U\,
\mathcal{D}\bar{\chi}\mathcal{D}\chi\,
e^{-S_G-S_\theta-S_F}.
\end{equation}

Although spacetime has been discretised, the degrees of freedom defined on the lattice must still be integrated over. The measures above are understood as
\begin{equation}
\int \mathcal{D}U
\equiv
\int\prod_{n,\mu}dU_\mu(n),
\qquad
\int \mathcal{D}\bar{\chi}\mathcal{D}\chi
\equiv
\int \prod_n d\bar{\chi}(n)d\chi(n).
\end{equation}

We now aim to rewrite the partition function in a dual representation that admits a probabilistic interpretation. We follow Gattringer, Kloiber and Sazonov \cite{Gattringer2015SchwingerDual}, who express the massless Schwinger model through discrete geometric variables. First, the staggered signs are absorbed into the link variables by the change of variables
\begin{equation}
\widetilde{U}_\mu(n)=\eta^\mu(n)U_\mu(n).
\end{equation}

Since $\eta^\mu(n)=\pm1$, $\widetilde{U}_\mu(n)$ still belongs to $U(1)$, leaving the gauge measure unchanged. However, in two dimensions the product of staggered signs around a plaquette is $-1$. Indeed, from the definition (\ref{staggered_phase}), we obtain
\begin{equation}
\eta^0(n)\eta^1(n+\hat{0})
\eta^0(n+\hat{1})^{-1}\eta^1(n)^{-1}
=
1\cdot (-1)^{n_0+1}\cdot 1\cdot (-1)^{-n_0}
=
-1.
\end{equation}
Therefore,
\begin{equation}
\widetilde{U}_p(n)=-U_p(n).
\end{equation}

From this point onwards, we omit the tildes to simplify the notation and rename $\widetilde{U}_\mu(n)\equiv U_\mu(n)$. With this convention, the staggered signs disappear from the fermionic action, which becomes
\begin{equation}
S_F
=
\frac{a}{2}
\sum_{n,\mu}
\left[
\bar{\chi}(n)U_\mu(n)\chi(n+\hat{\mu})
-
\bar{\chi}(n+\hat{\mu})U_\mu(n)^{-1}\chi(n)
\right].
\end{equation}

Because Grassmann fields are used, the integral at a lattice site is non-zero only when exactly one variable $\chi(n)$ and one variable $\bar\chi(n)$ occur.

With the usual normalisation,
\begin{equation}
\int d\bar{\chi}(n)d\chi(n)\,
\chi(n)\bar{\chi}(n)
=
1,
\end{equation}

Any term missing either variable, or containing one of them more than once, vanishes. Consequently, when $e^{-S_F}$ is expanded, only products of hopping terms that saturate every lattice site exactly once contribute. In the massless model no local term can saturate an isolated site, so saturation must arise entirely from hops between neighbouring sites.

\vspace{0.3 cm}

This gives rise to two kinds of fermionic objects. The first is a dimer, corresponding to occupying the same link in both directions. In that case the gauge factors cancel,
\begin{equation}
U_\mu(n)U_\mu(n)^{-1}=1,
\end{equation}

so the dimer carries no gauge flux. Its contribution is a constant factor,
\begin{equation}
W_D
=
\left(\frac{a}{2}\right)^2.
\end{equation}

In the dual representation, dimers act as neutral fermionic occupations: they saturate two neighbouring sites without modifying the gauge-flux constraint.

\vspace{0.3 cm}

The second object is a loop, formed by a closed sequence of oriented hops. For a contractible loop $l$, the fermionic contribution can be written as
\begin{equation}
\label{W_F}
W_F(l)
=
-
\left(\frac{a}{2}\right)^{L(l)}
(-1)^{L(l)/2}
\prod_{(n,\mu)\in l}
U_\mu(n)^{s_{n,\mu}},
\end{equation}
where $L(l)$ is the loop length and $s_{n,\mu}=\pm1$ specifies the orientation in which each link is traversed. The initial minus sign is the fermionic sign from loop closure, while $(-1)^{L(l)/2}$ counts the signs of oriented hops for contractible loops. Unlike a dimer, a loop leaves link variables and therefore carries gauge flux. For $L=2$, Eq.~\eqref{W_F} reduces to the dimer contribution.

For non-contractible loops, the hop-counting sign must also account for windings around the torus. On the even-sided $N_x\times N_t$ lattices used here, let $w_x(l)$ and $w_t(l)$ be the signed spatial and temporal winding numbers. The number of forward hops is then
\begin{equation}
N_+(l)=\frac{L(l)+N_xw_x(l)+N_tw_t(l)}{2}.
\end{equation}
Consequently, the expression in Eq.~\eqref{W_F} acquires the additional winding factor
\begin{equation}
\label{eq:winding_factor}
(-1)^{\frac{N_x}{2}w_x(l)+\frac{N_t}{2}w_t(l)+w_t(l)},
\end{equation}
where the final term comes from antiperiodic temporal fermion boundary conditions. This distinction matters for individual winding loops when a lattice extent is not a multiple of four, as noted in Ref.~\cite{Gattringer2015SchwingerDual}. In a fully gauge-admissible configuration on a periodic lattice, plaquette flux cannot compensate a net oriented winding: $\sum_l w_x(l)=\sum_l w_t(l)=0$. The product of the extra factors \eqref{eq:winding_factor} over all loops is therefore unity. 

Thus the compact overall sign used below remains valid, including for even-sided lattices such as $10\times10$, although the simple per-loop formula alone does not include winding signs.

\vspace{0.3 cm}

Integrating over the gauge variables now imposes a local constraint on every link:
\begin{equation}
\label{cond_gauge}
\int dU_\mu(n)\,U_\mu(n)^{j_\mu(n)}
=
\delta_{j_\mu(n),0}.
\end{equation}

A configuration contributes only when the total power of each link variable vanishes. Dimers satisfy this condition automatically because their gauge factors cancel. Fermionic loops, on the other hand, leave non-zero powers of $U_\mu(n)$. For the gauge integral to be non-zero, this flux must be compensated by additional link factors arising from the expansion of the gauge and topological terms. We start from
\begin{equation}
S_G
=
\beta
\sum_n
\left[
1-\operatorname{Re}U_p(n)
\right]
=
\beta N_{\text{plaq}}
-
\frac{\beta}{2}
\sum_n
\left[
U_p(n)+U_p(n)^{-1}
\right],
\end{equation}

where $N_{\text{plaq}}$ is the total number of lattice plaquettes. Its contribution to the statistical weight is
\begin{equation}
e^{-S_G}
=
e^{-\beta N_{\text{plaq}}}
\exp
\left\{
\frac{\beta}{2}
\sum_n
\left[
U_p(n)+U_p(n)^{-1}
\right]
\right\}.
\end{equation}

The first factor is constant and does not affect sampling, so it is omitted from now on. The topological term (\ref{S_theta}) contributes
\begin{equation}
\exp\left\{
-\frac{\theta}{4\pi}
\sum_n
\left[
U_p(n)-U_p(n)^{-1}
\right]
\right\}.
\end{equation}

Together, before accounting for the plaquette sign change, the gauge--topological weight contains
\begin{equation}
\exp\left\{
\sum_n
\left[
\left(\frac{\beta}{2}-\frac{\theta}{4\pi}\right)U_p(n)
+
\left(\frac{\beta}{2}+\frac{\theta}{4\pi}\right)U_p(n)^{-1}
\right]
\right\}.
\end{equation}
This motivates the definitions
\begin{equation}
\lambda
=
\frac{\beta}{2}
-
\frac{\theta}{4\pi},
\qquad
\bar{\lambda}
=
\frac{\beta}{2}
+
\frac{\theta}{4\pi}.
\end{equation}
For the real, non-negative Bessel weights used below, we restrict attention to $\lambda>0$ and $\bar\lambda>0$, i.e. $\beta>|\theta|/(2\pi)$. The numerical parameter ranges considered here satisfy this condition. This does not remove the separate fermionic sign problem in the massive case.
Recalling that the redefinition of links implies $\widetilde{U}_p(n)=-U_p(n)$, we obtain
\begin{equation}
e^{-S_G-S_\theta}
=
\prod_n
e^{-\lambda U_p(n)}
e^{-\bar{\lambda}U_p(n)^{-1}}.
\end{equation}\vspace{-0.1 cm}
We now expand each plaquette factor in modified Bessel functions. Using the identity derived in \cite{Gattringer2015SchwingerDual}, we have
\begin{equation}
e^{-\lambda U_p(n)}
e^{-\bar{\lambda}U_p(n)^{-1}}
=
\sum_{p(n)\in\mathbb{Z}}
(-1)^{p(n)}
I_{|p(n)|}
\left(
2\sqrt{\lambda\bar{\lambda}}
\right)
\left(
\sqrt{\frac{\lambda}{\bar{\lambda}}}
\right)^{p(n)}
U_p(n)^{p(n)}.
\end{equation}

The expansion introduces an integer variable $p(n)$, called the plaquette occupation number. It measures the power of $U_p(n)$: for $p(n)>0$ the factor $U_p(n)$ appears, for $p(n)<0$ its inverse appears, and for $p(n)=0$ the plaquette contributes no link factors. Thus $p(n)$ represents the discrete gauge flux through the plaquette, including the orientation and multiplicity of its boundary-link contributions. Its role is to compensate the flux generated by fermionic loops so that the total power of each link variable vanishes upon gauge integration.

\vspace{0.3 cm}

The factor $(-1)^{p(n)}$ arises from the plaquette sign change. Multiplying all plaquettes gives
\begin{equation}
\prod_n (-1)^{p(n)}
=
(-1)^{\sum_n p(n)}
\equiv
(-1)^{N_P},
\qquad
N_P=\sum_n p(n).
\end{equation}

With these definitions, the admissible configuration space $\mathcal{A}$ comprises arrangements of loops, dimers and plaquette occupations subject to two conditions: Grassmann variables must saturate every site, and the net gauge flux must cancel on every link.

The partition function therefore becomes
\begin{equation}
\label{Z_dual_sinmasa}
Z
=
\sum_{\{l,d,p\}\in\mathcal{A}}
\left(\frac{a}{2}\right)^{\sum_l L(l)+2N_D}
(-1)^{N_L+N_P+\frac{1}{2}\sum_l L(l)}
\prod_n
I_{|p(n)|}
\left(
2\sqrt{\lambda\bar{\lambda}}
\right)
\left(
\sqrt{\frac{\lambda}{\bar{\lambda}}}
\right)^{p(n)},
\end{equation}

where $N_L$ is the total number of loops and $N_D$ the number of dimers. In the massless model, Grassmann saturation implies $\sum_lL(l)+2N_D=V$, so the factor $(a/2)^V$ is global and will henceforth be omitted.

\vspace{0.3 cm}

At first sight, the weights entering $Z$ may still have an indefinite sign. However, following the geometrical argument of Gattringer, Kloiber and Sazonov \cite{Gattringer2015SchwingerDual}, the exponent of the sign in (\ref{Z_dual_sinmasa}) is always even for admissible configurations. Hence the dual partition function can finally be written as
\begin{equation}
Z
=
\sum_{\{l,d,p\}\in\mathcal{A}}
\prod_n
I_{|p(n)|}
\left(
2\sqrt{\lambda\bar{\lambda}}
\right)
\left(
\sqrt{\frac{\lambda}{\bar{\lambda}}}
\right)^{p(n)}.
\end{equation}

We will use diagrams to illustrate various concepts. Throughout these figures, loops are shown as thick lines with arrows, dimers as pairs of parallel lines on a link, and plaquette occupations as the integer $p(n)$ displayed inside each plaquette.

\vspace{0.3 cm}

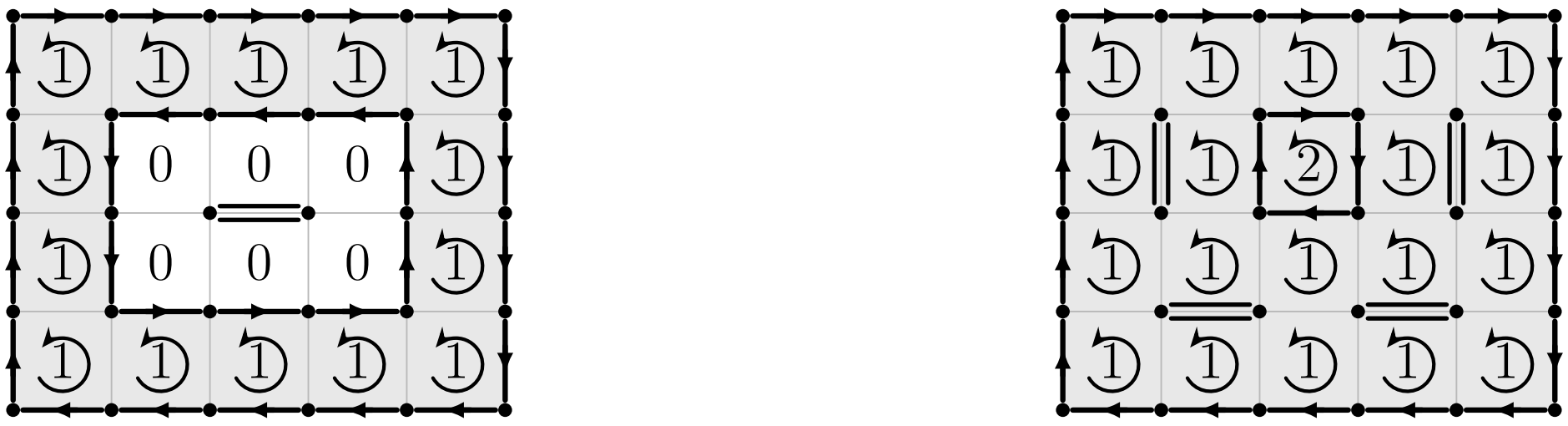
\begin{figure}[htbp]
\centering

\tikzset{
  dualsite/.style={circle, fill=black, inner sep=1.1pt},
  duallink/.style={gray!55, line width=0.35pt},
  dualloopline/.style={very thick, line cap=round},
  duallooparrow/.style={-{Latex[length=1.8mm,width=1.3mm]}, very thick, line cap=round},
  dualplaq/.style={fill=gray!18, draw=none},
  dualplaqarrow/.style={
    -{Stealth[length=1.6mm,width=1.3mm]},
    line width=0.8pt,
    line cap=round,
    line join=round
  },
  dualplaqnum/.style={font=\large},
  dualdimer/.style={line width=1.0pt, line cap=round}
}

\def\dualplaqccw#1#2#3{%
  \fill[dualplaq] (#1,#2) rectangle (#1+1,#2+1);
  \node[dualplaqnum] at (#1+0.5,#2+0.5) {$#3$};
  \draw[dualplaqarrow]
    ($(#1+0.5,#2+0.46)+(210:0.27)$)
    arc[start angle=210,end angle=500,radius=0.27];
}

\def\dualplaqzero#1#2{%
  \node[dualplaqnum] at (#1+0.5,#2+0.5) {$0$};
}

\def\dualhdimer#1#2{%
  \draw[dualdimer] (#1+0.10,#2+0.07) -- (#1+0.90,#2+0.07);
  \draw[dualdimer] (#1+0.10,#2-0.07) -- (#1+0.90,#2-0.07);
}

\def\dualvdimer#1#2{%
  \draw[dualdimer] (#1-0.07,#2+0.10) -- (#1-0.07,#2+0.90);
  \draw[dualdimer] (#1+0.07,#2+0.10) -- (#1+0.07,#2+0.90);
}

\begin{minipage}{0.48\textwidth}
\centering
\begin{tikzpicture}[x=0.78cm,y=0.78cm]

\foreach \x in {0,...,4}{
  \dualplaqccw{\x}{0}{1}
  \dualplaqccw{\x}{3}{1}
}
\foreach \y in {1,2}{
  \dualplaqccw{0}{\y}{1}
  \dualplaqccw{4}{\y}{1}
}
\foreach \x in {1,...,3}{
  \foreach \y in {1,2}{
    \dualplaqzero{\x}{\y}
  }
}

\foreach \x in {0,...,4}{
  \foreach \y in {0,...,4}{
    \draw[duallink] (\x,\y) -- (\x+1,\y);
  }
}
\foreach \x in {0,...,5}{
  \foreach \y in {0,...,3}{
    \draw[duallink] (\x,\y) -- (\x,\y+1);
  }
}

\foreach \x in {0,...,4}{
  \draw[dualloopline] (\x+0.10,0) -- (\x+0.90,0);
  \draw[dualloopline] (\x+0.90,4) -- (\x+0.10,4);
}
\foreach \y in {0,...,3}{
  \draw[dualloopline] (5,\y+0.10) -- (5,\y+0.90);
  \draw[dualloopline] (0,\y+0.90) -- (0,\y+0.10);
}
\foreach \x in {0,...,4}{
  \draw[duallooparrow] (\x+0.65,0) -- (\x+0.35,0);
  \draw[duallooparrow] (\x+0.35,4) -- (\x+0.65,4);
}
\foreach \y in {0,...,3}{
  \draw[duallooparrow] (5,\y+0.65) -- (5,\y+0.35);
  \draw[duallooparrow] (0,\y+0.35) -- (0,\y+0.65);
}

\foreach \x in {1,...,3}{
  \draw[dualloopline] (\x+0.10,1) -- (\x+0.90,1);
  \draw[dualloopline] (\x+0.10,3) -- (\x+0.90,3);
}
\foreach \y in {1,2}{
  \draw[dualloopline] (1,\y+0.10) -- (1,\y+0.90);
  \draw[dualloopline] (4,\y+0.10) -- (4,\y+0.90);
}
\foreach \x in {1,...,3}{
  \draw[duallooparrow] (\x+0.35,1) -- (\x+0.65,1);
  \draw[duallooparrow] (\x+0.65,3) -- (\x+0.35,3);
}
\foreach \y in {1,2}{
  \draw[duallooparrow] (1,\y+0.65) -- (1,\y+0.35);
  \draw[duallooparrow] (4,\y+0.35) -- (4,\y+0.65);
}

\dualhdimer{2}{2}

\foreach \x in {0,...,5}{
  \foreach \y in {0,...,4}{
    \node[dualsite] at (\x,\y) {};
  }
}

\end{tikzpicture}
\end{minipage}
\hfill
\begin{minipage}{0.48\textwidth}
\centering
\begin{tikzpicture}[x=0.78cm,y=0.78cm]

\foreach \x in {0,...,4}{
  \foreach \y in {0,...,3}{
    \dualplaqccw{\x}{\y}{1}
  }
}
\dualplaqccw{2}{2}{2}

\foreach \x in {0,...,4}{
  \foreach \y in {0,...,4}{
    \draw[duallink] (\x,\y) -- (\x+1,\y);
  }
}
\foreach \x in {0,...,5}{
  \foreach \y in {0,...,3}{
    \draw[duallink] (\x,\y) -- (\x,\y+1);
  }
}

\foreach \x in {0,...,4}{
  \draw[dualloopline] (\x+0.10,0) -- (\x+0.90,0);
  \draw[dualloopline] (\x+0.90,4) -- (\x+0.10,4);
}
\foreach \y in {0,...,3}{
  \draw[dualloopline] (5,\y+0.10) -- (5,\y+0.90);
  \draw[dualloopline] (0,\y+0.90) -- (0,\y+0.10);
}
\foreach \x in {0,...,4}{
  \draw[duallooparrow] (\x+0.65,0) -- (\x+0.35,0);
  \draw[duallooparrow] (\x+0.35,4) -- (\x+0.65,4);
}
\foreach \y in {0,...,3}{
  \draw[duallooparrow] (5,\y+0.65) -- (5,\y+0.35);
  \draw[duallooparrow] (0,\y+0.35) -- (0,\y+0.65);
}

\draw[dualloopline] (2.10,2) -- (2.90,2);
\draw[dualloopline] (3,2.10) -- (3,2.90);
\draw[dualloopline] (2.90,3) -- (2.10,3);
\draw[dualloopline] (2,2.90) -- (2,2.10);

\draw[duallooparrow] (2.65,2) -- (2.35,2);
\draw[duallooparrow] (3,2.65) -- (3,2.35);
\draw[duallooparrow] (2.35,3) -- (2.65,3);
\draw[duallooparrow] (2,2.35) -- (2,2.65);

\dualhdimer{1}{1}
\dualhdimer{3}{1}
\dualvdimer{1}{2}
\dualvdimer{4}{2}

\foreach \x in {0,...,5}{
  \foreach \y in {0,...,4}{
    \node[dualsite] at (\x,\y) {};
  }
}

\end{tikzpicture}
\end{minipage}

\caption[Examples of dual configurations]{Examples of dual configurations on a $6\times5$ lattice. The left panel shows a ring with $p=1$ and an interior region with $p=0$; the right panel shows a central plaquette with $p=2$.}
\label{fig:config_dual_p0_p2}
\end{figure}

The sign cancellation has a simple geometric interpretation: a plaquette occupation may be regarded as a counter-oriented flux that compensates the flux of a fermionic loop, satisfying the condition (\ref{cond_gauge}). Figure~\ref{fig:config_dual_p0_p2} gives two schematic examples, including plaquette occupation numbers larger than one.

\vspace{0.3 cm}

The main observables are now defined as derivatives of $\log Z$ with respect to the action parameters. We consider a rectangular lattice with $N_x$ sites in the spatial direction and $N_t$ in the temporal direction, so the total lattice volume is $V=N_xN_t$. The average real plaquette and its susceptibility follow from derivatives with respect to $\beta$. It is important to distinguish the full partition function $Z_{\mathrm{full}}$ from $\widetilde Z=e^{\beta V}Z_{\mathrm{full}}$, since the constant factor $e^{-\beta V}$ in the Wilson gauge action was dropped when deriving the dual weights:
\begin{equation}
\label{Up_obs}
\langle\operatorname{Re}U_p\rangle
=1+\frac{1}{V}\frac{\partial}{\partial\beta}\log Z_{\mathrm{full}}
=\frac{1}{V}\frac{\partial}{\partial\beta}\log\widetilde Z,
\qquad
\chi_p=\frac{1}{V}\frac{\partial^2}{\partial\beta^2}\log Z_{\mathrm{full}}
=\frac{1}{V}\frac{\partial^2}{\partial\beta^2}\log\widetilde Z .
\end{equation}
The plotted plaquette observable is $\langle\operatorname{Re}U_p\rangle$ (denoted $\langle U_p\rangle$ in figures for brevity). Its coupling dependence measures the gauge response, while $\chi_p$ is a connected fluctuation observable. Similarly, the topological charge and its susceptibility are obtained by differentiating with respect to the topological angle $\theta$,
\begin{equation}
\label{Obs_topo}
\langle Q_{\mathrm{lat}}\rangle
= i\,\frac{\partial\log Z}{\partial\theta},
\qquad
\langle q\rangle\equiv-\frac{1}{V}\frac{\partial\log Z}{\partial\theta}
=\frac{i}{V}\langle Q_{\mathrm{lat}}\rangle,
\qquad
\chi_{\theta}\equiv\frac{1}{V}\frac{\partial^2\log Z}{\partial\theta^2}
=-\frac{\partial\langle q\rangle}{\partial\theta}.
\end{equation}

The real quantity plotted as $\langle q\rangle$ in this work follows the dual-lattice response convention of Ref.~\cite{Goeschl2017SchwingerStrategies}. It is not the Euclidean expectation value $\langle Q_{\mathrm{lat}}\rangle/V$, which differs by a factor of $i$. With the definition above, $\chi_{\theta}$ is a curvature of the free energy, not automatically the variance of a Monte Carlo estimator. These responses help describe the change in the vacuum as the fermion mass is varied, without by themselves establishing a phase transition. For clarity, if $w(C;\theta)$ is the signed dual weight and $q_C=-V^{-1}\partial_\theta\log|w(C;\theta)|$ is the configuration-level response estimator where the sign is locally constant, then
\begin{equation}
\chi_\theta=V\bigl(\langle q_C^2\rangle-\langle q_C\rangle^2\bigr)-\left\langle\partial_\theta q_C\right\rangle.
\end{equation}
The averages in this identity use the full signed ensemble (implemented through reweighting). The explicit derivative term is required when the dual weight is not exponential-linear in $\theta$, as is the case for the modified-Bessel weights used here. A variance-only estimator cannot be identified with $\chi_\theta$ without checking this contribution.

\vspace{0.2 cm}

\subsection{Extension to the massive case and the sign problem}

In the massive case, the fermionic action includes the local term
\begin{equation}
a^2m\sum_n \bar{\chi}(n)\chi(n).
\end{equation}

Expanding the exponential of this term in the Grassmann integral introduces a new dual object: the monomer. The mass term already contains both Grassmann variables at the same site, $\chi(n)$ and $\bar\chi(n)$, and can therefore locally saturate the Grassmann integral.

\vspace{0.3 cm}

With monomers included, each lattice site can be saturated in three ways: by belonging to a loop, by being the endpoint of a dimer, or by hosting a monomer. If $N_M$ denotes the number of monomers in a configuration, each contributes a factor proportional to $a^2m$. 

Rearranging the weights and omitting global factors, the partition function becomes
\begin{equation}
\label{Z_dual_masiva}
Z
=
\sum_{\{l,d,p,s\}\in\mathcal{A}_m}
(2am)^{N_M(s)}
(-1)^{N_L+N_P+\frac{1}{2}\sum_l L(l)}
\prod_n
I_{|p(n)|}
\left(
2\sqrt{\lambda\bar{\lambda}}
\right)
\left(
\sqrt{\frac{\lambda}{\bar{\lambda}}}
\right)^{p(n)},
\end{equation}

where $s$ denotes the monomer distribution and $\mathcal{A}_m$ is the space of admissible configurations containing loops, dimers, plaquettes and monomers.

\vspace{0.3 cm}

Monomers disappear in the limit $m\to0$ because their weight is proportional to the mass. The massless representation is then recovered, in which sites must be saturated solely by loops or dimers. As $m$ increases, configurations with monomers acquire greater relative weight and modify the statistics of loops and dimers. The fermion mass thus not only changes the weight of each configuration but can also change which configurations dominate the ensemble.

\vspace{0.3 cm}

\begin{figure}[htbp]
\centering
\label{config_mono}

\tikzset{
  masssite/.style={circle, fill=black, inner sep=1.0pt},
  masslink/.style={gray!55, line width=0.35pt},
  massloopline/.style={very thick, line cap=round},
  masslooparrow/.style={-{Latex[length=1.7mm,width=1.25mm]}, very thick, line cap=round},
  massplaq/.style={fill=gray!18, draw=none},
  massplaqarrow/.style={
    -{Stealth[length=1.5mm,width=1.2mm]},
    line width=0.8pt,
    line cap=round,
    line join=round
  },
  massplaqnum/.style={font=\normalsize},
  massdimer/.style={line width=1.0pt, line cap=round}
}

\def\massplaqcw#1#2#3{%
  \fill[massplaq] (#1,#2) rectangle (#1+1,#2+1);
  \node[font=\scriptsize] at (#1+0.5,#2+0.5) {$#3$};
  \draw[massplaqarrow]
    ($(#1+0.5,#2+0.46)+(-30:0.30)$)
    arc[start angle=-30,end angle=-320,radius=0.30];
}

\def\massplaqzero#1#2{%
  \node[massplaqnum] at (#1+0.5,#2+0.5) {$0$};
}

\def\masshdimer#1#2{%
  \draw[massdimer] (#1+0.10,#2+0.07) -- (#1+0.90,#2+0.07);
  \draw[massdimer] (#1+0.10,#2-0.07) -- (#1+0.90,#2-0.07);
}

\def\massvdimer#1#2{%
  \draw[massdimer] (#1-0.07,#2+0.10) -- (#1-0.07,#2+0.90);
  \draw[massdimer] (#1+0.07,#2+0.10) -- (#1+0.07,#2+0.90);
}

\def\massmonomer#1#2{%
  \node[font=\Large] at (#1,#2) {$\times$};
}

\begin{minipage}{0.48\textwidth}
\centering
\begin{tikzpicture}[x=0.72cm,y=0.72cm]


\massplaqzero{0}{0}
\massplaqzero{1}{0}
\massplaqzero{2}{0}
\massplaqzero{3}{0}
\massplaqzero{4}{0}

\massplaqzero{0}{1}
\massplaqcw{1}{1}{-1}
\massplaqcw{2}{1}{-1}
\massplaqzero{3}{1}
\massplaqzero{4}{1}

\massplaqzero{0}{2}
\massplaqcw{1}{2}{-1}
\massplaqcw{2}{2}{-1}
\massplaqzero{3}{2}
\massplaqzero{4}{2}

\massplaqzero{0}{3}
\massplaqzero{1}{3}
\massplaqzero{2}{3}
\massplaqzero{3}{3}
\massplaqzero{4}{3}

\foreach \x in {0,...,4}{
  \foreach \y in {0,...,4}{
    \draw[masslink] (\x,\y) -- (\x+1,\y);
  }
}

\foreach \x in {0,...,5}{
  \foreach \y in {0,...,3}{
    \draw[masslink] (\x,\y) -- (\x,\y+1);
  }
}

\draw[massloopline] (1.10,1) -- (1.90,1);
\draw[massloopline] (2.10,1) -- (2.90,1);

\draw[massloopline] (3,1.10) -- (3,1.90);
\draw[massloopline] (3,2.10) -- (3,2.90);

\draw[massloopline] (2.90,3) -- (2.10,3);
\draw[massloopline] (1.90,3) -- (1.10,3);

\draw[massloopline] (1,2.90) -- (1,2.10);
\draw[massloopline] (1,1.90) -- (1,1.10);

\draw[masslooparrow] (1.35,1) -- (1.65,1);
\draw[masslooparrow] (2.35,1) -- (2.65,1);

\draw[masslooparrow] (3,1.35) -- (3,1.65);
\draw[masslooparrow] (3,2.35) -- (3,2.65);

\draw[masslooparrow] (2.65,3) -- (2.35,3);
\draw[masslooparrow] (1.65,3) -- (1.35,3);

\draw[masslooparrow] (1,2.65) -- (1,2.35);
\draw[masslooparrow] (1,1.65) -- (1,1.35);


\masshdimer{0}{0}
\masshdimer{2}{0}
\masshdimer{4}{0}

\masshdimer{0}{4}
\masshdimer{2}{4}
\masshdimer{4}{4}

\massvdimer{0}{1}
\massvdimer{4}{1}
\massvdimer{5}{1}

\foreach \x in {0,...,5}{
  \foreach \y in {0,...,4}{
    \node[masssite] at (\x,\y) {};
  }
}

\massmonomer{0}{3}
\massmonomer{2}{2}
\massmonomer{4}{3}
\massmonomer{5}{3}

\end{tikzpicture}

\end{minipage}
\hfill
\begin{minipage}{0.48\textwidth}
\centering
\begin{tikzpicture}[x=0.72cm,y=0.72cm]


\massplaqzero{0}{0}
\massplaqzero{1}{0}
\massplaqzero{2}{0}
\massplaqzero{3}{0}
\massplaqzero{4}{0}

\massplaqzero{0}{1}
\massplaqzero{1}{1}
\massplaqzero{2}{1}
\massplaqcw{3}{1}{-1}
\massplaqzero{4}{1}

\massplaqzero{0}{2}
\massplaqzero{1}{2}
\massplaqzero{2}{2}
\massplaqzero{3}{2}
\massplaqzero{4}{2}

\massplaqzero{0}{3}
\massplaqzero{1}{3}
\massplaqzero{2}{3}
\massplaqzero{3}{3}
\massplaqzero{4}{3}

\foreach \x in {0,...,4}{
  \foreach \y in {0,...,4}{
    \draw[masslink] (\x,\y) -- (\x+1,\y);
  }
}

\foreach \x in {0,...,5}{
  \foreach \y in {0,...,3}{
    \draw[masslink] (\x,\y) -- (\x,\y+1);
  }
}

\draw[massloopline] (3.10,1) -- (3.90,1);
\draw[massloopline] (4,1.10) -- (4,1.90);
\draw[massloopline] (3.90,2) -- (3.10,2);
\draw[massloopline] (3,1.90) -- (3,1.10);

\draw[masslooparrow] (3.35,1) -- (3.65,1);
\draw[masslooparrow] (4,1.35) -- (4,1.65);
\draw[masslooparrow] (3.65,2) -- (3.35,2);
\draw[masslooparrow] (3,1.65) -- (3,1.35);


\masshdimer{0}{0}
\masshdimer{2}{0}
\masshdimer{4}{0}

\masshdimer{0}{4}
\masshdimer{2}{4}
\masshdimer{4}{4}

\masshdimer{0}{1}
\masshdimer{0}{2}
\masshdimer{0}{3}
\masshdimer{2}{3}
\masshdimer{4}{3}

\foreach \x in {0,...,5}{
  \foreach \y in {0,...,4}{
    \node[masssite] at (\x,\y) {};
  }
}

\massmonomer{2}{1}
\massmonomer{5}{1}
\massmonomer{2}{2}
\massmonomer{5}{2}

\end{tikzpicture}

\end{minipage}

\caption[Examples of massive-model configurations]{Examples of admissible configurations in the massive case. Monomers are marked with crosses at the sites they occupy. The left configuration contains a $2\times2$ loop that could not occur in the massless case.}
\label{fig:config_masiva_monomeros}
\end{figure}
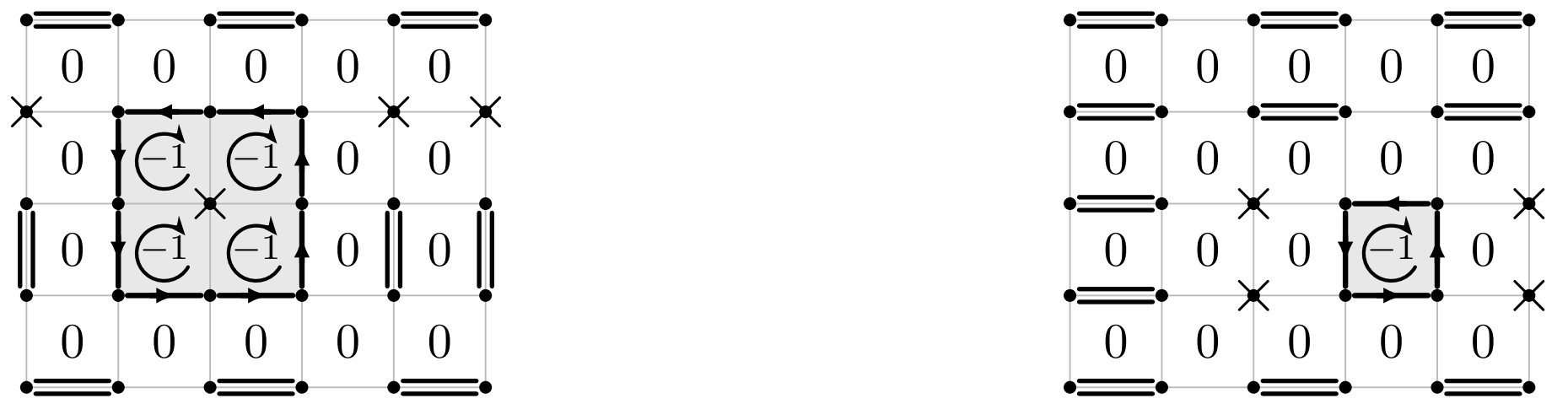

Monomers also have an important consequence: the weights are no longer necessarily positive. Consider, for example, the left-hand configuration in Figure~\ref{fig:config_masiva_monomeros}. It contains one loop, so $N_L=1$. The loop has length $L=8$ and encloses four plaquettes with occupation $p=-1$, giving $N_P=-4$.

The exponent of the configuration sign is therefore
\begin{equation}
N_L+N_P+\frac{1}{2}\sum_l L(l)
=
1-4+4
=
1.
\end{equation}

\vspace{-0.15 cm}

The weight in (\ref{Z_dual_masiva}) is consequently negative, giving rise to the sign problem. To handle it, we separate the sign of the weight from its absolute value,
\begin{equation}
w(C)=\mathrm{sign}(C)\,|w(C)|.
\end{equation}

\vspace{-0.15 cm}

Sampling is then performed with the positive weight $|w(C)|$, defining the \textit{sign-quenched} ensemble \cite{Giordano2020SignReweighting}. Physical expectation values are recovered in this ensemble by reweighting:
\begin{equation}
\label{reweighting_signo}
\langle \mathcal{O}\rangle
=
\frac{
\langle \mathcal{O}\,\mathrm{sign}\rangle_{|w|}
}{
\langle \mathrm{sign}\rangle_{|w|}
}.
\end{equation}

\vspace{-0.05 cm}

Here $\langle\cdot\rangle_{|w|}$ denotes expectations taken with the absolute weight. The denominator $\langle\mathrm{sign}\rangle_{|w|}$ measures the severity of the sign problem: the smaller it becomes, the stronger the cancellations between configurations of opposite signs and the more difficult it is to obtain accurate estimates.

\vspace{0.3 cm}

The introduction of monomers allows us to define the chiral condensate. In the continuum, this observable is associated with $\langle\bar\psi\psi\rangle$, whereas in the staggered formulation it is related to $\langle\bar\chi\chi\rangle$. Physically, the chiral condensate measures the response of the vacuum to a fermion mass.

\vspace{0.3 cm}

In the dual representation, this relation has a simple interpretation. Each monomer originates from the mass term and contributes a factor proportional to $m$, so differentiating the partition function with respect to mass amounts to counting monomers. The chiral condensate and its susceptibility can therefore be defined as
\begin{equation}
\langle \bar{\chi}\chi\rangle
=
\frac{1}{V}
\frac{\partial}{\partial m}
\log Z,
\qquad
\chi_{\bar{\chi}\chi}
=
\frac{1}{V}
\frac{\partial^2}{\partial m^2}
\log Z .
\end{equation}

At non-zero mass, the derivative of the dual partition function gives $\partial_m\log Z=\langle N_M\rangle/m$, evaluated in the reweighted ensemble. In our convention the condensate is therefore $\langle N_M\rangle/(mV)$ (up to the conventional sign chosen for the fermionic bilinear). The second derivative includes a contact term,
\[
\partial_m^2\log Z=\frac{\mathrm{Var}(N_M)-\langle N_M\rangle}{m^2},
\]
where expectations refer to the full signed ensemble. Thus the chiral susceptibility is not simply the variance of the monomer density. Here $m$ denotes the mass parameter differentiated in the dual partition function; when expressing the physical condensate in continuum units, the lattice normalisation (including powers of $a$) and the Euclidean sign convention must also be restored. The connected moments in this identity are evaluated with the full signed measure and need not behave as non-negative variances. The physical expectations require the reweighting in Eq.~\eqref{reweighting_signo}.

\vspace{0.3 cm}


\section{Simulation Methodology}

Having formulated the lattice model and obtained its dual representation, the next step is to generate configurations representative of the partition function. As discussed above, we employ Monte Carlo algorithms, which sample configuration space with probabilities proportional to the statistical weights. This chapter discusses the technical aspects of the simulations and their physical significance.

\subsection{Physical quantities and boundary conditions}
On the lattice, the spacing $a$ sets the scale of all measured quantities. We therefore work in lattice units, setting $a=1$. The parameters appearing in the program are dimensionless: lengths are measured in lattice sites and masses in units of $1/a$. If $m$ denotes the dimensionful mass in the Lagrangian, then the mass used in the simulations is $m_{\mathrm{lattice}}=am$, written simply as $m_{\mathrm{lattice}}$ in the code. Recalling the definition of $\beta$ in (\ref{S_G}), we have
\begin{equation}
ag=\frac{1}{\sqrt{\beta}}.
\end{equation}

\vspace{-0.15 cm}

Varying $\beta$ is thus equivalent to changing the lattice spacing measured in units of $1/g$. Increasing $\beta$ decreases $a$ and moves the simulation towards the continuum limit. The relevant dimensionless combinations are
\begin{equation}
\label{magni_fisi}
\frac{m}{g}
=
\frac{m_{\mathrm{lattice}}}{ag}
=
m_{\mathrm{lattice}}\sqrt{\beta},
\qquad
Lg
=
N_xag
=
\frac{N_x}{\sqrt{\beta}},
\qquad
Tg
=
N_tag
=
\frac{N_t}{\sqrt{\beta}}.
\end{equation}

Here $L=N_xa$ and $T=N_ta$ are the spatial and temporal lattice sizes. These combinations allow simulations with different values of $\beta$ to be compared and relate lattice volume to finite physical volume. We must also specify the lattice boundaries. Periodic boundary conditions are imposed on the gauge variables,
\begin{equation}
n+N_\mu\hat{\mu}\equiv n.
\end{equation}

This avoids physical boundaries, gives each site the same number of neighbours and allows loops in the dual representation to close across lattice edges. Fermionic fields obey periodic boundary conditions in the spatial direction and antiperiodic conditions in the temporal direction,
\[
\chi(n_0+N_t,n_1)=-\chi(n_0,n_1).
\]

Antiperiodicity initially contributes a factor $(-1)^{w_t}$, where $w_t$ is the total temporal winding number of fermion loops. Closure of an individual loop does not make its winding number even. However, the \emph{gauge-link constraints} in the dual ensemble imply that the total oriented winding flux vanishes: loops with net positive and negative temporal winding must compensate each other, because periodic plaquette occupations cannot provide an uncompensated flux around the time direction. Thus the total winding number has even parity and the antiperiodic factor equals unity for admissible configurations. This is the cancellation shown for the corresponding dual formulation in Ref.~\cite{Gattringer2015SchwingerDual}. It is a consequence of gauge-flux neutrality, not merely of loop closure; no separate temporal-boundary sign is therefore needed in the dual weights written below.

We now describe how the admissible configuration space $\mathcal{A}_m$ is explored. A Monte Carlo simulation does not sum explicitly over all possible configurations, but constructs a Markov chain through successive updates. We therefore introduce algorithms that connect admissible configurations while respecting the constraints of the dual representation.

\subsection{Update algorithms}

The implemented simulation samples admissible configurations according to the non-negative \emph{sign-quenched} weight $\pi(C)\propto |W(C)|$. The general detailed-balance condition can be expressed using a proposal probability $T(C\to C')$ and the Metropolis--Hastings acceptance rule
\begin{equation}
\label{metropolis_hastings}
P_{\mathrm{acc}}(C\to C')
=\min\!\left(1,\frac{|W(C')|\,T(C'\to C)}{|W(C)|\,T(C\to C')}\right).
\end{equation}
For symmetric proposals, $T(C\to C')=T(C'\to C)$, this reduces to the ratio of absolute weights used in the usual Metropolis rule. The sign of the original statistical weight enters the expectation values through reweighting, not through a negative acceptance probability. For compound or path-dependent proposals, detailed balance concerns the complete proposal and its reverse, rather than only the ratio of configuration weights. Appendix~\ref{pseudo} summarises the move sets; the numerical implementation and the validation tests are described separately.

\vspace{0.3cm}

The updates preserve the constraints of $\mathcal{A}_m$. Plaquette and worm moves are combined to improve sampling and access configurations that may be inaccessible through local updates alone. Although the numerical tests in Section~\ref{sec:validation} support the implementation, they do not establish ergodicity. The update rules are summarised in Appendix~\ref{pseudo}.

\vspace{0.3 cm}

The plaquette update modifies the links around an elementary plaquette. Starting from the massless updates proposed by G{\"o}schl \cite{Goeschl2017SchwingerStrategies}, we extend them to include monomers in the massive case, allowing local changes in gauge flux and fermionic configurations.

\begin{figure}[htbp]
\centering

\tikzset{
  mdpsite/.style={circle, fill=black, inner sep=1.0pt},
  mdplink/.style={gray!55, line width=0.35pt},
  mdpdimer/.style={line width=1.0pt, line cap=round},
  mdploop/.style={very thick, line cap=round},
  mdparrow/.style={-latex, very thick, line cap=round},
  mdplabl/.style={font=\small},
  mdpmono/.style={font=\Large},
  mdpconn/.style={blue, -latex, line width=0.75pt}
}

\newcommand{\MDPPlaquetteBase}{
  \fill[white] (-0.10,-0.10) rectangle (1.10,1.10);
  \draw[mdplink] (0,0) rectangle (1,1);
  \node[mdpsite] at (0,0) {};
  \node[mdpsite] at (1,0) {};
  \node[mdpsite] at (0,1) {};
  \node[mdpsite] at (1,1) {};
}

\newcommand{\MDPPlacePlaquette}[2]{
  \begin{scope}[shift={($(#1)+(-0.5,-0.5)$)}]
    \MDPPlaquetteBase
    #2
  \end{scope}
}

\newcommand{\MDPConn}[2]{
  \draw[mdpconn] ($(#1)!0.30!(#2)$) -- ($(#1)!0.70!(#2)$);
  \draw[mdpconn] ($(#2)!0.30!(#1)$) -- ($(#2)!0.70!(#1)$);
}

\newcommand{\MDPConnB}[2]{
  \draw[mdpconn] ($(#1)!0.26!(#2)$) -- ($(#1)!0.74!(#2)$);
  \draw[mdpconn] ($(#2)!0.26!(#1)$) -- ($(#2)!0.74!(#1)$);
}

\newcommand{\MDPConnSmall}[2]{
  \draw[mdpconn] ($(#1)!0.24!(#2)$) -- ($(#1)!0.76!(#2)$);
  \draw[mdpconn] ($(#2)!0.24!(#1)$) -- ($(#2)!0.76!(#1)$);
}

\newcommand{\MDPMonomersFour}{
  \node[mdpmono] at (0,0) {$\times$};
  \node[mdpmono] at (1,0) {$\times$};
  \node[mdpmono] at (0,1) {$\times$};
  \node[mdpmono] at (1,1) {$\times$};
}

\newcommand{\MDPDimerH}{
  \draw[mdpdimer] (0.12,0.06) -- (0.88,0.06);
  \draw[mdpdimer] (0.12,-0.06) -- (0.88,-0.06);
  \draw[mdpdimer] (0.12,1.06) -- (0.88,1.06);
  \draw[mdpdimer] (0.12,0.94) -- (0.88,0.94);
}

\newcommand{\MDPDimerV}{
  \draw[mdpdimer] (0.06,0.12) -- (0.06,0.88);
  \draw[mdpdimer] (-0.06,0.12) -- (-0.06,0.88);
  \draw[mdpdimer] (1.06,0.12) -- (1.06,0.88);
  \draw[mdpdimer] (0.94,0.12) -- (0.94,0.88);
}

\newcommand{\MDPLoopCW}{
  \draw[mdploop] (0.10,0) -- (0.90,0);
  \draw[mdploop] (1,0.10) -- (1,0.90);
  \draw[mdploop] (0.90,1) -- (0.10,1);
  \draw[mdploop] (0,0.90) -- (0,0.10);
  \draw[mdparrow] (0.30,0) -- (0.60,0);
  \draw[mdparrow] (1,0.30) -- (1,0.60);
  \draw[mdparrow] (0.70,1) -- (0.40,1);
  \draw[mdparrow] (0,0.70) -- (0,0.40);
}

\newcommand{\MDPLoopCCW}{
  \draw[mdploop] (0.10,0) -- (0.90,0);
  \draw[mdploop] (1,0.10) -- (1,0.90);
  \draw[mdploop] (0.90,1) -- (0.10,1);
  \draw[mdploop] (0,0.90) -- (0,0.10);
  \draw[mdparrow] (0.60,0) -- (0.30,0);
  \draw[mdparrow] (1,0.60) -- (1,0.30);
  \draw[mdparrow] (0.40,1) -- (0.70,1);
  \draw[mdparrow] (0,0.40) -- (0,0.70);
}

\newcommand{\MDPMplusL}{
  \node[mdpmono] at (0,0) {$\times$};
  \node[mdpmono] at (1,0) {$\times$};
  \draw[mdploop] (0.10,1) -- (0.90,1);
}

\newcommand{\MDPDplusL}{
  \draw[mdpdimer] (0.12,0.06) -- (0.88,0.06);
  \draw[mdpdimer] (0.12,-0.06) -- (0.88,-0.06);
  \draw[mdploop] (0.10,1) -- (0.90,1);
}

\newcommand{\MDPCshape}{
  \draw[mdploop] (0.10,0) -- (0.90,0);
  \draw[mdploop] (0,0.10) -- (0,0.90);
  \draw[mdploop] (1,0.10) -- (1,0.90);
}

\newcommand{\MDPAntiH}{
  \draw[mdploop] (0.10,0) -- (0.90,0);
  \draw[mdparrow] (0.30,0) -- (0.60,0);
  \draw[mdploop] (0.90,1) -- (0.10,1);
  \draw[mdparrow] (0.70,1) -- (0.40,1);
}

\newcommand{\MDPAntiV}{
  \draw[mdploop] (0,0.10) -- (0,0.90);
  \draw[mdparrow] (0,0.30) -- (0,0.60);
  \draw[mdploop] (1,0.90) -- (1,0.10);
  \draw[mdparrow] (1,0.70) -- (1,0.40);
}

\newcommand{\MDPCornerBL}{
  \node[mdpmono] at (0,0) {$\times$};
  \draw[mdploop] (0.10,1) -- (0.90,1);
  \draw[mdploop] (1,0.10) -- (1,0.90);
}

\newcommand{\MDPCornerTR}{
  \node[mdpmono] at (1,1) {$\times$};
  \draw[mdploop] (0.10,0) -- (0.90,0);
  \draw[mdploop] (0,0.10) -- (0,0.90);
}

\makebox[\textwidth][l]{%
\hspace{-0.5cm}%
\begin{minipage}{0.47\textwidth}
\centering
\begin{tikzpicture}[x=0.74cm,y=0.74cm]

\def\r{2.2}

\coordinate (V1) at (90:\r);
\coordinate (V2) at (162:\r);
\coordinate (V3) at (234:\r);
\coordinate (V4) at (306:\r);
\coordinate (V5) at (18:\r);

\useasboundingbox (-3.35,-3.05) rectangle (3.35,3.45);

\node[mdplabl] at (0,3.15) {(a) States \(M,D,P\)};

\MDPConn{V1}{V2}
\MDPConn{V1}{V3}
\MDPConn{V1}{V4}
\MDPConn{V1}{V5}
\MDPConn{V2}{V3}
\MDPConn{V2}{V4}
\MDPConn{V2}{V5}
\MDPConn{V3}{V4}
\MDPConn{V3}{V5}
\MDPConn{V4}{V5}

\MDPPlacePlaquette{V1}{\MDPMonomersFour}
\MDPPlacePlaquette{V2}{\MDPDimerH}
\MDPPlacePlaquette{V3}{\MDPLoopCW}
\MDPPlacePlaquette{V4}{\MDPLoopCCW}
\MDPPlacePlaquette{V5}{\MDPDimerV}

\end{tikzpicture}
\end{minipage}%
\hspace{-0.05\textwidth}%
\begin{minipage}{0.25\textwidth}
\centering
\begin{tikzpicture}[x=0.74cm,y=0.74cm]

\def\r{1.85}

\coordinate (T1) at (90:\r);
\coordinate (T2) at (210:\r);
\coordinate (T3) at (330:\r);

\useasboundingbox (-2.70,-2.40) rectangle (2.70,3.45);

\node[mdplabl] at (0,3.15) {(b) Type \(C\)};

\MDPConnB{T1}{T2}
\MDPConnB{T2}{T3}
\MDPConnB{T3}{T1}

\MDPPlacePlaquette{T1}{\MDPMplusL}
\MDPPlacePlaquette{T2}{\MDPDplusL}
\MDPPlacePlaquette{T3}{\MDPCshape}

\end{tikzpicture}
\end{minipage}%
\hspace{0.02\textwidth}%
\begin{minipage}{0.24\textwidth}
\centering

\begin{tikzpicture}[x=0.74cm,y=0.74cm]

\useasboundingbox (-1.9,-0.55) rectangle (3.9,1.95);

\node[mdplabl] at (1.0,1.65) {(c) Antiparallel links};

\coordinate (A1) at (0,0.45);
\coordinate (A2) at (2.8,0.45);

\MDPConnSmall{A1}{A2}

\MDPPlacePlaquette{A1}{\MDPAntiH}
\MDPPlacePlaquette{A2}{\MDPAntiV}

\end{tikzpicture}

\vspace{0.35cm}

\begin{tikzpicture}[x=0.74cm,y=0.74cm]

\useasboundingbox (-1.9,-0.55) rectangle (3.9,1.95);

\node[mdplabl] at (1.0,1.65) {(d) Corner rotation};

\coordinate (G1) at (0,0.45);
\coordinate (G2) at (2.8,0.45);

\MDPConnSmall{G1}{G2}

\MDPPlacePlaquette{G1}{\MDPCornerBL}
\MDPPlacePlaquette{G2}{\MDPCornerTR}

\end{tikzpicture}

\end{minipage}%
}

\caption[Plaquette update schemes]{Illustration of local plaquette updates. $M$ denotes four monomers at the corners, $D$ two parallel dimers and $P$ an oriented $1\times1$ loop. Type-$C$ updates, antiparallel links and diagonal corner rotations are also shown. Blue directional arrows indicate the proposed transitions between local configurations; their acceptance is determined by the Metropolis criterion.}
\label{fig:update_plaqueta}

\end{figure}
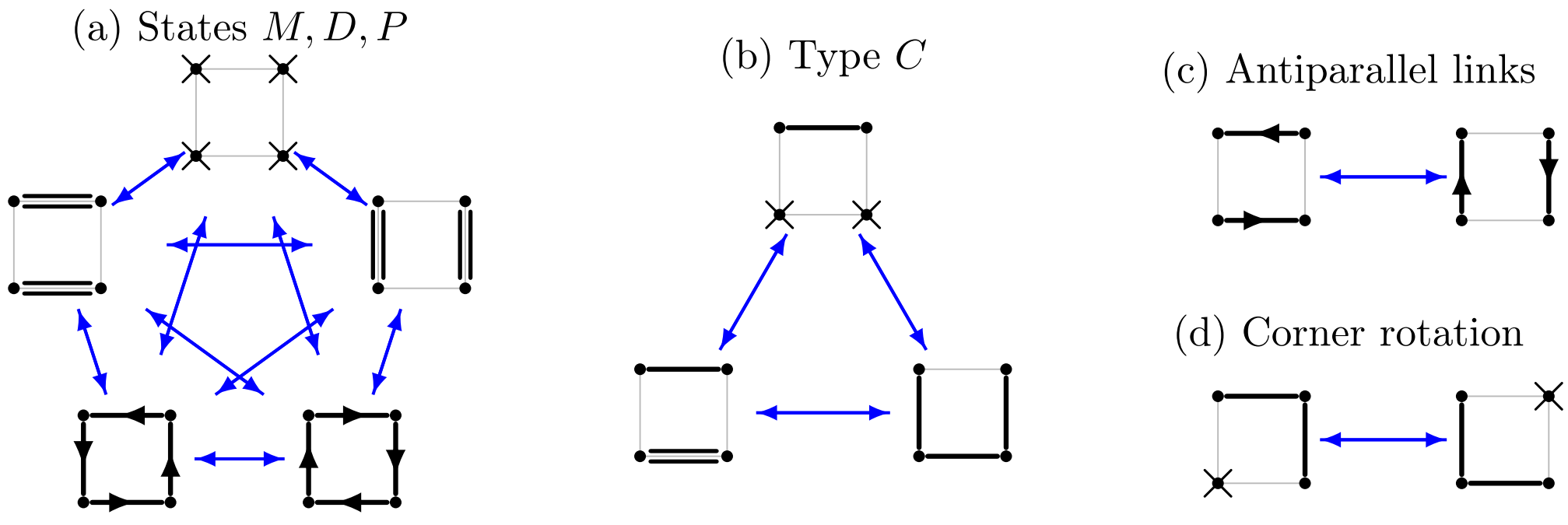

\newpage

Figure~\ref{fig:update_plaqueta} summarises the local plaquette updates. We consider four fully connected groups of states, so any local configuration in a group can be proposed as a destination from any other state in that group. In the limit $m\to0$, monomer configurations disappear, and moves involving them explicitly, such as the corner rotation in Figure~\ref{fig:update_plaqueta}(d), are no longer present. Plaquette updates are simple and computationally inexpensive, but their strictly local nature may make them inefficient at changing extended structures or loops winding around the periodic lattice.

\vspace{0.3 cm}

Worm updates are used to explore configurations not easily connected by local plaquette moves. The idea is to temporarily break the local Grassmann saturation at two lattice sites. These sites act as the worm endpoints and move through the lattice, modifying the traversed links until saturation is restored. The final result is a new admissible configuration. \vspace{0.3 cm}

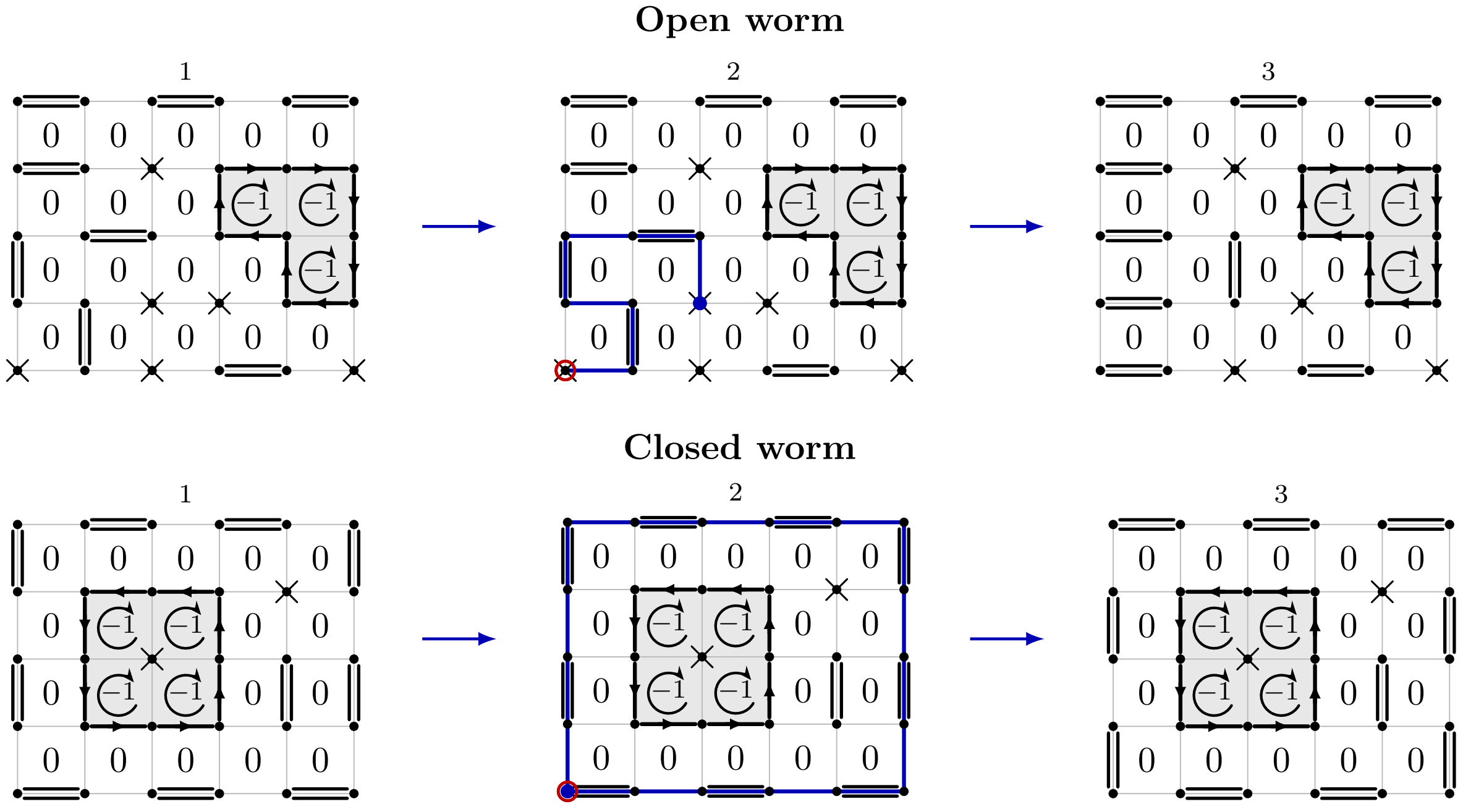
\begin{figure}[htbp]
\centering

\tikzset{
  masssite/.style={circle, fill=black, inner sep=1.0pt},
  masslink/.style={gray!55, line width=0.35pt},
  massloopline/.style={very thick, line cap=round},
  masslooparrow/.style={-{Latex[length=1.7mm,width=1.25mm]}, very thick, line cap=round},
  massplaq/.style={fill=gray!18, draw=none},
  massplaqarrow/.style={
    -{Stealth[length=1.5mm,width=1.2mm]},
    line width=0.8pt,
    line cap=round,
    line join=round
  },
  massplaqnum/.style={font=\normalsize},
  massdimer/.style={line width=1.0pt, line cap=round},
  wormpath/.style={blue!70!black, line width=1.2pt, line cap=round, line join=round},
  wormstep/.style={blue!70!black, -{Latex[length=2.0mm,width=1.5mm]}, line width=0.9pt},
  wormhead/.style={circle, fill=blue!70!black, draw=blue!70!black, inner sep=1.35pt},
  wormtail/.style={circle, draw=red!75!black, line width=0.9pt, inner sep=2.0pt}
}

\def\massplaqcw#1#2#3{%
  \fill[massplaq] (#1,#2) rectangle (#1+1,#2+1);
  \node[font=\scriptsize] at (#1+0.5,#2+0.5) {$#3$};
  \draw[massplaqarrow]
    ($(#1+0.5,#2+0.46)+(-30:0.30)$)
    arc[start angle=-30,end angle=-320,radius=0.30];
}

\def\massplaqzero#1#2{%
  \node[massplaqnum] at (#1+0.5,#2+0.5) {$0$};
}

\def\masshdimer#1#2{%
  \draw[massdimer] (#1+0.10,#2+0.07) -- (#1+0.90,#2+0.07);
  \draw[massdimer] (#1+0.10,#2-0.07) -- (#1+0.90,#2-0.07);
}

\def\massvdimer#1#2{%
  \draw[massdimer] (#1-0.07,#2+0.10) -- (#1-0.07,#2+0.90);
  \draw[massdimer] (#1+0.07,#2+0.10) -- (#1+0.07,#2+0.90);
}

\def\massmonomer#1#2{%
  \node[font=\Large] at (#1,#2) {$\times$};
}

\def\wormtailmark#1#2{%
  \node[wormtail] at (#1,#2) {};
}

\def\wormheadmark#1#2{%
  \node[wormhead] at (#1,#2) {};
}

\def\wormplaquetteszero{%
  \massplaqzero{0}{0}
  \massplaqzero{1}{0}
  \massplaqzero{2}{0}
  \massplaqzero{3}{0}
  \massplaqzero{4}{0}

  \massplaqzero{0}{1}
  \massplaqzero{1}{1}
  \massplaqzero{2}{1}
  \massplaqzero{3}{1}
  \massplaqzero{4}{1}

  \massplaqzero{0}{2}
  \massplaqzero{1}{2}
  \massplaqzero{2}{2}
  \massplaqzero{3}{2}
  \massplaqzero{4}{2}

  \massplaqzero{0}{3}
  \massplaqzero{1}{3}
  \massplaqzero{2}{3}
  \massplaqzero{3}{3}
  \massplaqzero{4}{3}
}

\def\wormlattice{%
  \foreach \x in {0,...,4}{
    \foreach \y in {0,...,4}{
      \draw[masslink] (\x,\y) -- (\x+1,\y);
    }
  }

  \foreach \x in {0,...,5}{
    \foreach \y in {0,...,3}{
      \draw[masslink] (\x,\y) -- (\x,\y+1);
    }
  }
}

\def\wormsites{%
  \foreach \x in {0,...,5}{
    \foreach \y in {0,...,4}{
      \node[masssite] at (\x,\y) {};
    }
  }
}

\def\openLplaquettes{%
  \massplaqcw{3}{2}{-1}
  \massplaqcw{4}{2}{-1}
  \massplaqcw{4}{1}{-1}
}

\def\openLloop{%
  \draw[massloopline] (3,2.10) -- (3,2.90);
  \draw[massloopline] (3.10,3) -- (3.90,3);
  \draw[massloopline] (4.10,3) -- (4.90,3);
  \draw[massloopline] (5,2.90) -- (5,2.10);
  \draw[massloopline] (5,1.90) -- (5,1.10);
  \draw[massloopline] (4.90,1) -- (4.10,1);
  \draw[massloopline] (4,1.10) -- (4,1.90);
  \draw[massloopline] (3.90,2) -- (3.10,2);

  \draw[masslooparrow] (3,2.35) -- (3,2.65);
  \draw[masslooparrow] (3.35,3) -- (3.65,3);
  \draw[masslooparrow] (4.35,3) -- (4.65,3);
  \draw[masslooparrow] (5,2.65) -- (5,2.35);
  \draw[masslooparrow] (5,1.65) -- (5,1.35);
  \draw[masslooparrow] (4.65,1) -- (4.35,1);
  \draw[masslooparrow] (4,1.35) -- (4,1.65);
  \draw[masslooparrow] (3.65,2) -- (3.35,2);
}

\def\closedLoopPlaquettes{%
  \massplaqcw{1}{1}{-1}
  \massplaqcw{2}{1}{-1}
  \massplaqcw{1}{2}{-1}
  \massplaqcw{2}{2}{-1}
}

\def\closedLoop{%
  \draw[massloopline] (1.10,1) -- (1.90,1);
  \draw[massloopline] (2.10,1) -- (2.90,1);

  \draw[massloopline] (3,1.10) -- (3,1.90);
  \draw[massloopline] (3,2.10) -- (3,2.90);

  \draw[massloopline] (2.90,3) -- (2.10,3);
  \draw[massloopline] (1.90,3) -- (1.10,3);

  \draw[massloopline] (1,2.90) -- (1,2.10);
  \draw[massloopline] (1,1.90) -- (1,1.10);

  \draw[masslooparrow] (1.35,1) -- (1.65,1);
  \draw[masslooparrow] (2.35,1) -- (2.65,1);

  \draw[masslooparrow] (3,1.35) -- (3,1.65);
  \draw[masslooparrow] (3,2.35) -- (3,2.65);

  \draw[masslooparrow] (2.65,3) -- (2.35,3);
  \draw[masslooparrow] (1.65,3) -- (1.35,3);

  \draw[masslooparrow] (1,2.65) -- (1,2.35);
  \draw[masslooparrow] (1,1.65) -- (1,1.35);
}

\def\openInit{%
  \openLloop

  \massmonomer{0}{0}
  \massmonomer{2}{1}

  \massvdimer{1}{0}
  \massvdimer{0}{1}
  \masshdimer{1}{2}

  \massmonomer{2}{0}
  \massmonomer{3}{1}
  \massmonomer{5}{0}

  \masshdimer{3}{0}

  \masshdimer{0}{3}
  \massmonomer{2}{3}

  \masshdimer{0}{4}
  \masshdimer{2}{4}
  \masshdimer{4}{4}
}

\def\openPath{%
  \draw[wormpath]
    (0,0) -- (1,0) -- (1,1) -- (0,1) --
    (0,2) -- (1,2) -- (2,2) -- (2,1);

  \wormtailmark{0}{0}
  \wormheadmark{2}{1}
}

\def\openFinal{%
  \openLloop

  \masshdimer{0}{0}
  \masshdimer{0}{1}
  \masshdimer{0}{2}
  \massvdimer{2}{1}

  \massmonomer{2}{0}
  \massmonomer{3}{1}
  \massmonomer{5}{0}

  \masshdimer{3}{0}

  \masshdimer{0}{3}
  \massmonomer{2}{3}

  \masshdimer{0}{4}
  \masshdimer{2}{4}
  \masshdimer{4}{4}
}

\def\closedFixed{%
  \closedLoop

  \massmonomer{2}{2}

  \massvdimer{4}{1}
  \massmonomer{4}{3}
}

\def\closedInit{%
  \closedFixed


  \masshdimer{0}{0}
  \masshdimer{2}{0}
  \masshdimer{4}{0}

  \massvdimer{5}{1}
  \massvdimer{5}{3}

  \masshdimer{3}{4}
  \masshdimer{1}{4}

  \massvdimer{0}{3}
  \massvdimer{0}{1}
}

\def\closedPath{%
  \draw[wormpath]
    (0,0) -- (1,0) -- (2,0) -- (3,0) -- (4,0) -- (5,0) --
    (5,1) -- (5,2) -- (5,3) -- (5,4) --
    (4,4) -- (3,4) -- (2,4) -- (1,4) -- (0,4) --
    (0,3) -- (0,2) -- (0,1) -- (0,0);

  \wormtailmark{0}{0}
  \wormheadmark{0}{0}
}

\def\closedFinal{%
  \closedFixed


  \masshdimer{1}{0}
  \masshdimer{3}{0}

  \massvdimer{5}{0}
  \massvdimer{5}{2}

  \masshdimer{4}{4}
  \masshdimer{2}{4}
  \masshdimer{0}{4}

  \massvdimer{0}{2}
  \massvdimer{0}{0}
}

\noindent\makebox[\textwidth][c]{\textbf{Open worm}}

\vspace{0.1cm}

\makebox[\textwidth][c]{%
\begin{minipage}{0.27\textwidth}
\centering
\begin{tikzpicture}[x=0.72cm,y=0.72cm]
\wormplaquetteszero
\openLplaquettes
\wormlattice
\openInit
\wormsites
\node[font=\scriptsize] at (2.5,4.45) {1};
\end{tikzpicture}
\end{minipage}
\hspace{0.10cm}
\raisebox{0.0cm}{\tikz{\draw[wormstep] (0,0) -- (0.8,0);}}
\hspace{0.10cm}
\begin{minipage}{0.27\textwidth}
\centering
\begin{tikzpicture}[x=0.72cm,y=0.72cm]
\wormplaquetteszero
\openLplaquettes
\wormlattice
\openInit
\openPath
\wormsites
\wormtailmark{0}{0}
\wormheadmark{2}{1}
\node[font=\scriptsize] at (2.5,4.45) {2};
\end{tikzpicture}
\end{minipage}
\hspace{0.10cm}
\raisebox{0.0cm}{\tikz{\draw[wormstep] (0,0) -- (0.8,0);}}
\hspace{0.10cm}
\begin{minipage}{0.27\textwidth}
\centering
\begin{tikzpicture}[x=0.72cm,y=0.72cm]
\wormplaquetteszero
\openLplaquettes
\wormlattice
\openFinal
\wormsites
\node[font=\scriptsize] at (2.5,4.45) {3};
\end{tikzpicture}
\end{minipage}
}

\vspace{0.35cm}

\noindent\makebox[\textwidth][c]{\textbf{Closed worm}}

\vspace{0.1cm}

\makebox[\textwidth][c]{%
\begin{minipage}{0.27\textwidth}
\centering
\begin{tikzpicture}[x=0.72cm,y=0.72cm]
\wormplaquetteszero
\closedLoopPlaquettes
\wormlattice
\closedInit
\wormsites
\node[font=\scriptsize] at (2.5,4.45) {1};
\end{tikzpicture}
\end{minipage}
\hspace{0.10cm}
\raisebox{0.0cm}{\tikz{\draw[wormstep] (0,0) -- (0.8,0);}}
\hspace{0.10cm}
\begin{minipage}{0.27\textwidth}
\centering
\begin{tikzpicture}[x=0.72cm,y=0.72cm]
\wormplaquetteszero
\closedLoopPlaquettes
\wormlattice
\closedInit
\closedPath
\wormsites
\wormtailmark{0}{0}
\wormheadmark{0}{0}
\node[font=\scriptsize] at (2.5,4.45) {2};
\end{tikzpicture}
\end{minipage}
\hspace{0.10cm}
\raisebox{0.0cm}{\tikz{\draw[wormstep] (0,0) -- (0.8,0);}}
\hspace{0.10cm}
\begin{minipage}{0.27\textwidth}
\centering
\begin{tikzpicture}[x=0.72cm,y=0.72cm]
\wormplaquetteszero
\closedLoopPlaquettes
\wormlattice
\closedFinal
\wormsites
\node[font=\scriptsize] at (2.5,4.45) {3};
\end{tikzpicture}
\end{minipage}
}

\caption[Examples of open and closed worms]{Illustration of worm updates. The endpoints are marked with a red circle and a blue disk in each row. The upper row shows an open worm: the blue path propagates through the lower-left region and connects two monomers, producing a new admissible configuration. The lower row shows a closed worm: its path surrounds the central loop without crossing it and rearranges the dimers along the path.}
\label{fig:update_gusano}

\end{figure}

A closed or open worm is distinguished by the way the update terminates. In a closed worm, the two endpoints propagate until they meet again, closing a path. An open worm, by contrast, may close through a change in monomer occupation at its endpoints. In both cases, the worm moves by alternating between links occupied by dimers and unoccupied links, avoiding closed loops. When encountering a monomer, the alternating order is reversed to avoid introducing additional violations of saturation. Figure~\ref{fig:update_gusano} gives examples of both kinds of worm on a lattice of size $(N_x,N_t)=(6,5)$.


Combining plaquette and worm updates allows both local and non-local changes, facilitating exploration of admissible configurations. A Monte Carlo step is defined here as a sequential sweep over plaquettes, followed by $\max\{1,\mathrm{int}(V/10)\}$ worm attempts and a second plaquette sweep.

\subsection{Thermalisation and error estimation}

At the start of a Monte Carlo simulation, configurations depend on the initial state and do not yet represent equilibrium. For simplicity, we initialise each chain in a pure-dimer configuration and discard an initial thermalisation period. This is monitored through the evolution of two observables: the average plaquette $\langle U_p\rangle$, defined in (\ref{Up_obs}), and the dimer density $\langle N_d/V\rangle$. The first probes the gauge sector, while the second is sensitive to fermionic degrees of freedom. Based on simulations at different masses and $\theta$ values, some of which are shown in Appendix~\ref{Termalizado_anexo}, the thermalisation period is fixed at $10^5$ Monte Carlo steps.

\vspace{0.3 cm}

After the initial transient is discarded, successive measurements cannot be regarded as fully independent, because each configuration is obtained from the preceding one. To account for autocorrelations, measurements are grouped into blocks and statistical errors are estimated using the jackknife method. If $\mathcal{O}^{(i)}$ is the observable evaluated with block $i$ omitted and $N_b$ is the number of blocks, the error is
\begin{equation}
\sigma_{\mathcal{O}}
=
\sqrt{
\frac{N_b-1}{N_b}
\sum_{i=1}^{N_b}
\left(
\mathcal{O}^{(i)}-\overline{\mathcal{O}}
\right)^2
},
\qquad
\overline{\mathcal{O}}
=
\frac{1}{N_b}
\sum_{i=1}^{N_b}
\mathcal{O}^{(i)}.
\end{equation}

\begin{figure}[htbp]
\centering
\includegraphics[width=0.94\textwidth]{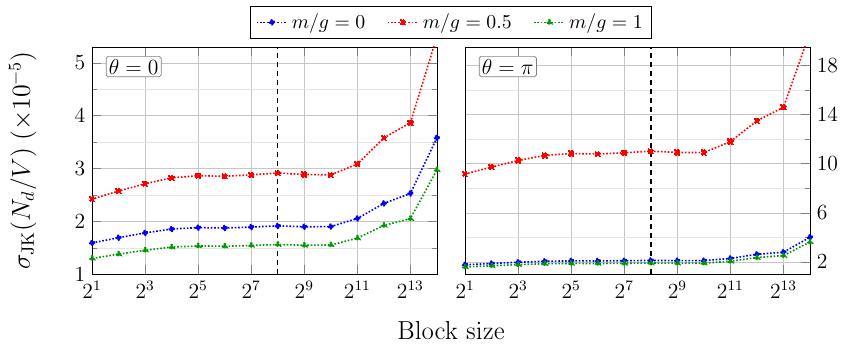}
\caption[Jackknife uncertainty versus block size]{Jackknife uncertainty of $\rho_d$ versus block size on a $V=16\times16$ lattice at $\beta=2$, comparing $\theta=0$ and $\theta=\pi$. Small blocks may underestimate errors due to autocorrelation, while very large blocks reduce the available statistics. The dashed black line indicates the selected block size.}
\label{fig:jackknife_rhod_bloques_masas}
\end{figure}

For the \textit{sign-quenched} ensemble defined in (\ref{reweighting_signo}), the jackknife method is applied directly to the ratio, recalculating numerator and denominator for every resample. This propagates uncertainties in the numerator and denominator together. To choose the block size, we examine the stability of the jackknife uncertainty of the dimer density $\rho_d$ as the block size is varied; Figure~\ref{fig:jackknife_rhod_bloques_masas} shows the corresponding numerical estimates for different masses and topological angles. The simulations use blocks of $2^8$ measurements, indicated by the dashed line. The plot provides an empirical check of error stability; a separate quantitative estimate of the integrated autocorrelation time would be needed to establish that this choice is conservative for every observable.

\section{Results}

With the simulation principles established, we now analyse the Schwinger model numerically. Before exploring the full parameter space, we first perform several checks of the implementation.

\subsection{Validation and consistency checks}\label{sec:validation}

The code can be tested in the massless limit using the reference results of \cite{Goeschl2017SchwingerStrategies}. We digitise the points shown in the figures of that reference and compare them with our simulations in the same regime, $m=0$. This is not an exact numerical validation because the reference data are extracted from published plots; nevertheless, it allows us to check whether the expected $\theta$ dependence of common observables is reproduced.


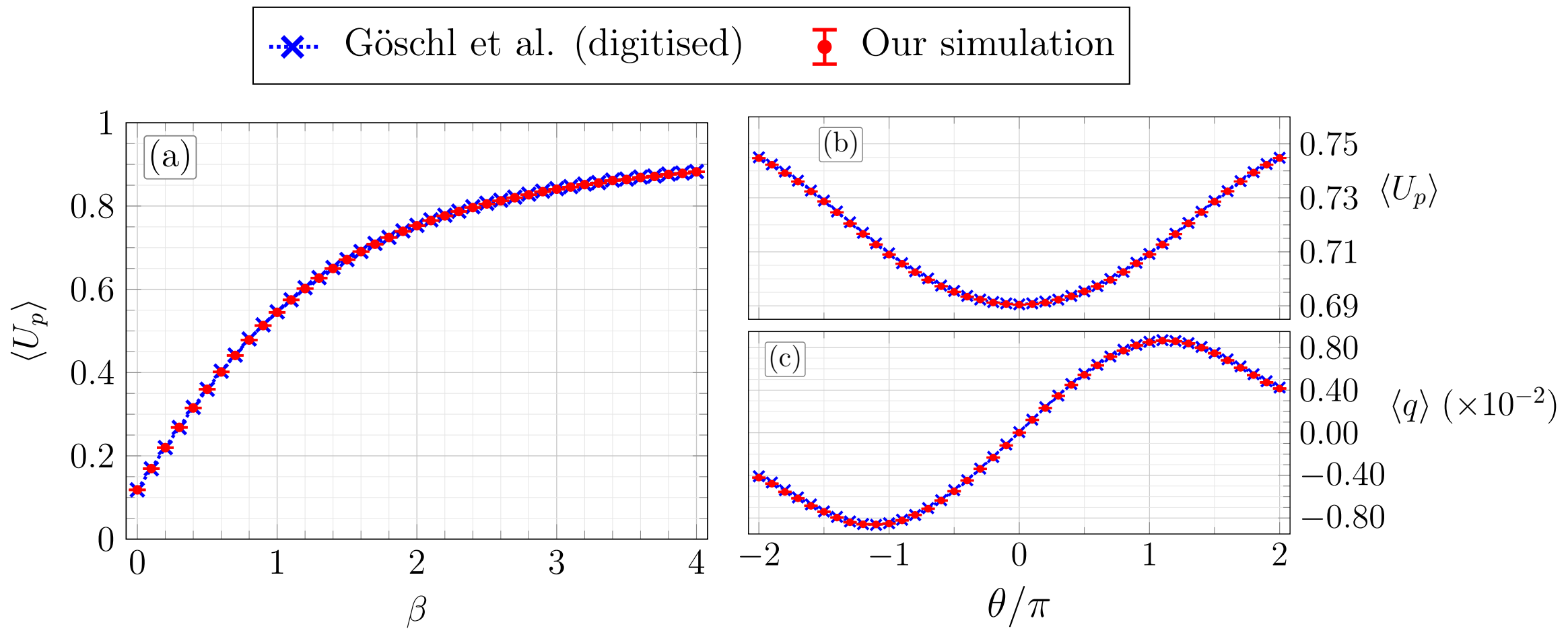
\begin{figure}[htbp]
\centering
\hspace*{0pt}
\resizebox{\textwidth}{!}{%
\begin{minipage}{\textwidth}
\centering

\makebox[\textwidth][c]{%
\hspace*{-0.12\textwidth}%
\begin{tikzpicture}
\node[
    draw=black,
    fill=white,
    inner sep=4pt,
    font=\normalsize
] {
\begin{tabular}{@{}c@{\hspace{0.22cm}}l@{\hspace{0.65cm}}c@{\hspace{0.22cm}}l@{}}

\begin{tikzpicture}[baseline=-0.5ex, x=1cm, y=1cm]
    \draw[blue, densely dotted, line width=1.1pt] (0,0) -- (0.50,0);
    \draw[blue, line width=1.4pt] (0.13,-0.10) -- (0.33,0.10);
    \draw[blue, line width=1.4pt] (0.13,0.10) -- (0.33,-0.10);
\end{tikzpicture}
&
Göschl et al. (digitised)
&
\begin{tikzpicture}[baseline=-0.5ex, x=1cm, y=1cm]
    \draw[red, line width=1.0pt] (0.23,-0.17) -- (0.23,0.17);
    \draw[red, line width=1.0pt] (0.11,-0.17) -- (0.35,-0.17);
    \draw[red, line width=1.0pt] (0.11,0.17) -- (0.35,0.17);
    \fill[red] (0.23,0) circle (2.0pt);
\end{tikzpicture}
&
Our simulation

\end{tabular}
};
\end{tikzpicture}%
}

\vspace{0.12cm}

\begin{minipage}[c]{0.45\textwidth}
\centering
\resizebox{\textwidth}{!}{%
\begin{tikzpicture}

\begin{axis}[
    width=0.95\textwidth,
    height=0.68\textwidth,
    scale only axis,
    axis lines=box,
    xmin=-0.08,
    xmax=4.08,
    ymin=0.0,
    ymax=1.0,
    xtick={0,1,2,3,4},
    ytick={0,0.2,0.4,0.6,0.8,1.0},
    tick align=inside,
    major tick length=5pt,
    minor x tick num=4,
    minor y tick num=3,
    grid=both,
    major grid style={draw=gray!45, line width=0.3pt},
    minor grid style={draw=gray!20, line width=0.2pt},
    label style={font=\large},
    ticklabel style={font=\large},
    xlabel={$\beta$},
    ylabel={$\langle U_p\rangle$},
    enlarge x limits=false,
    enlarge y limits=false
]

\addplot[
    blue,
    densely dotted,
    line width=1.1pt,
    mark=x,
    mark size=3.2pt,
    mark options={solid, draw=blue, line width=1.1pt},
    forget plot
]
table[
    col sep=space,
    comment chars={\#},
    x=beta,
    y=Up_digitized
]{comprobaciones/reference_goeschl_plaquette_beta_digitized.dat};

\addplot+[
    red,
    only marks,
    mark=*,
    mark size=1.4pt,
    mark options={fill=red, draw=none},
    error bars/.cd,
        y dir=both,
        y explicit,
        error mark=none,
        error bar style={red, line width=0.7pt},
    /pgfplots/forget plot
]
table[
    col sep=space,
    comment chars={\#},
    x=beta,
    y=Up,
    y error=err_Up
]{comprobaciones/simulation_plaquette_charge_beta_scan.dat};

\addplot[
    red,
    only marks,
    mark=|,
    mark size=2.8pt,
    mark options={red, rotate=90, line width=0.8pt},
    forget plot
]
table[
    col sep=space,
    comment chars={\#},
    x=beta,
    y expr=\thisrow{Up}+\thisrow{err_Up}
]{comprobaciones/simulation_plaquette_charge_beta_scan.dat};

\addplot[
    red,
    only marks,
    mark=|,
    mark size=2.8pt,
    mark options={red, rotate=90, line width=0.8pt},
    forget plot
]
table[
    col sep=space,
    comment chars={\#},
    x=beta,
    y expr=\thisrow{Up}-\thisrow{err_Up}
]{comprobaciones/simulation_plaquette_charge_beta_scan.dat};

\node[
    anchor=north west,
    font=\large,
    draw=black!50,
    fill=white,
    rounded corners=1pt,
    inner sep=1.2pt
] at (rel axis cs:0.03,0.97)
{(a)};

\end{axis}
\end{tikzpicture}
}
\end{minipage}
\hspace{-0.015\textwidth}
\begin{minipage}[c]{0.525\textwidth}
\centering
\vspace{0.13 cm}
\resizebox{\textwidth}{!}{%
\begin{tikzpicture}

\begin{axis}[
    name=panelB,
    width=0.95\textwidth,
    height=0.355\textwidth,
    scale only axis,
    axis lines=box,
    xmin=-2.08,
    xmax=2.08,
    ymin=0.685,
    ymax=0.760,
    xtick={-2,-1,0,1,2},
    xticklabels=\empty,
    ytick={0.69,0.71,0.73,0.75},
    yticklabel pos=right,
    ytick pos=right,
    tick align=inside,
    major tick length=5pt,
    minor x tick num=1,
    minor y tick num=3,
    grid=both,
    major grid style={draw=gray!45, line width=0.3pt},
    minor grid style={draw=gray!20, line width=0.2pt},
    label style={font=\Large},
    ticklabel style={font=\Large},
    enlarge x limits=false,
    enlarge y limits=false,
    ylabel={$\langle U_p\rangle$},
    ylabel style={
        font=\Large,
        rotate=-90,
        at={(axis description cs:1.22,0.5)},
        anchor=south
    }
]

\addplot[
    blue,
    densely dotted,
    line width=1.1pt,
    mark=x,
    mark size=3.2pt,
    mark options={solid, draw=blue, line width=1.1pt},
    forget plot
]
table[
    col sep=space,
    comment chars={\#},
    x=theta_pi,
    y=Up_digitized
]{comprobaciones/reference_goeschl_plaquette_charge_theta_digitized.dat};

\addplot+[
    red,
    only marks,
    mark=*,
    mark size=1.4pt,
    mark options={fill=red, draw=none},
    each nth point={2},
    error bars/.cd,
        y dir=both,
        y explicit,
        error mark=none,
        error bar style={red, line width=0.7pt},
    /pgfplots/forget plot
]
table[
    col sep=space,
    comment chars={\#},
    x=theta_pi,
    y=Up,
    y error=err_Up
]{comprobaciones/simulation_plaquette_charge_theta_scan.dat};

\addplot[
    red,
    only marks,
    mark=|,
    mark size=2.8pt,
    mark options={red, rotate=90, line width=0.8pt},
    each nth point={2},
    forget plot
]
table[
    col sep=space,
    comment chars={\#},
    x=theta_pi,
    y expr=\thisrow{Up}+\thisrow{err_Up}
]{comprobaciones/simulation_plaquette_charge_theta_scan.dat};

\addplot[
    red,
    only marks,
    mark=|,
    mark size=2.8pt,
    mark options={red, rotate=90, line width=0.8pt},
    each nth point={2},
    forget plot
]
table[
    col sep=space,
    comment chars={\#},
    x=theta_pi,
    y expr=\thisrow{Up}-\thisrow{err_Up}
]{comprobaciones/simulation_plaquette_charge_theta_scan.dat};

\node[
    anchor=north west,
    font=\large,
    draw=black!50,
    fill=white,
    rounded corners=1pt,
    inner sep=1.2pt
] at (rel axis cs:0.13,0.95)
{(b)};

\end{axis}

\begin{axis}[
    name=panelC,
    at={(panelB.south west)},
    anchor=north west,
    yshift=-0.18cm,
    width=0.95\textwidth,
    height=0.355\textwidth,
    scale only axis,
    axis lines=box,
    xmin=-2.08,
    xmax=2.08,
    ymin=-9.5,
    ymax=9.5,
    xtick={-2,-1,0,1,2},
    ytick={-8,-4,0,4,8},
    yticklabels={$-0.80$,$-0.40$,$0.00$,$0.40$,$0.80$},
    yticklabel pos=right,
    ytick pos=right,
    tick align=inside,
    major tick length=5pt,
    minor x tick num=1,
    minor y tick num=3,
    grid=both,
    major grid style={draw=gray!45, line width=0.3pt},
    minor grid style={draw=gray!20, line width=0.2pt},
    label style={font=\LARGE},
    ticklabel style={font=\Large},
    xticklabel style={font=\Large},
    enlarge x limits=false,
    enlarge y limits=false,
    xlabel={$\theta/\pi$},
    ylabel={$\langle q\rangle\;(\times 10^{-2})$},
    ylabel style={
        font=\Large,
        rotate=-90,
        at={(axis description cs:1.34,0.5)},
        anchor=south
    }
]

\addplot[
    blue,
    densely dotted,
    line width=1.1pt,
    mark=x,
    mark size=3.2pt,
    mark options={solid, draw=blue, line width=1.1pt},
    forget plot
]
table[
    col sep=space,
    comment chars={\#},
    x=theta_pi,
    y expr=1000*\thisrow{q_digitized}
]{comprobaciones/reference_goeschl_plaquette_charge_theta_digitized.dat};

\addplot+[
    red,
    only marks,
    mark=*,
    mark size=1.4pt,
    mark options={fill=red, draw=none},
    each nth point={2},
    error bars/.cd,
        y dir=both,
        y explicit,
        error mark=none,
        error bar style={red, line width=0.7pt},
    /pgfplots/forget plot
]
table[
    col sep=space,
    comment chars={\#},
    x=theta_pi,
    y expr=1000*\thisrow{q},
    y error expr=1000*\thisrow{err_q}
]{comprobaciones/simulation_plaquette_charge_theta_scan.dat};

\addplot[
    red,
    only marks,
    mark=|,
    mark size=2.8pt,
    mark options={red, rotate=90, line width=0.8pt},
    each nth point={2},
    forget plot
]
table[
    col sep=space,
    comment chars={\#},
    x=theta_pi,
    y expr=1000*(\thisrow{q}+\thisrow{err_q})
]{comprobaciones/simulation_plaquette_charge_theta_scan.dat};

\addplot[
    red,
    only marks,
    mark=|,
    mark size=2.8pt,
    mark options={red, rotate=90, line width=0.8pt},
    each nth point={2},
    forget plot
]
table[
    col sep=space,
    comment chars={\#},
    x=theta_pi,
    y expr=1000*(\thisrow{q}-\thisrow{err_q})
]{comprobaciones/simulation_plaquette_charge_theta_scan.dat};

\node[
    anchor=north west,
    font=\large,
    draw=black!50,
    fill=white,
    rounded corners=1pt,
    inner sep=1.2pt
] at (rel axis cs:0.03,0.95)
{(c)};

\end{axis}

\end{tikzpicture}
}
\end{minipage}

\end{minipage}
}

\caption[Comparison with reference results in the massless case]{Comparison with reference results in the massless case at fixed lattice volume $V=4\times4$. Panel (a) shows $\langle U_p\rangle$ as a function of $\beta$, comparing digitised data from \cite{Goeschl2017SchwingerStrategies} with our simulations on a $4\times4$ lattice, including jackknife error bars. Panels (b) and (c) show $\langle U_p\rangle$ and $\langle q\rangle$ (as defined in Eq.~\eqref{Obs_topo}), respectively, as functions of $\theta$ at $\beta=1.6$. Digitised reference points and our results are compared in both panels.}
\label{fig:comparacion_triple_goschl}

\end{figure}

The measurements reproduce the reference behaviour satisfactorily. In Figure~\ref{fig:comparacion_triple_goschl}, the average plaquette from our simulations follows the same dependence on $\beta$ as the digitised data of \cite{Goeschl2017SchwingerStrategies}. In particular, $\langle U_p\rangle$ increases monotonically with $\beta$, consistent with the gauge term favouring more ordered plaquette configurations.

\vspace{0.3 cm}

The expected dependence on the topological parameter is reproduced qualitatively as well. In the explored interval, the observable $\langle U_p\rangle$ displays an approximately symmetric dependence on $\theta$, while $\langle q\rangle$ changes sign under $\theta\to-\theta$ and vanishes near $\theta=0$. This comparison shows that the implementation captures the expected gauge behaviour and topological response in the massless case.

\newpage

The massive Schwinger model at non-zero $\theta$ has also been investigated with other numerical formulations, including the study of Azcoiti \emph{et al.}~\cite{Azcoiti2018MassiveTheta}. Our data do not provide a like-for-like quantitative benchmark against that work because the discretisation and parameter choices differ. We therefore restrict the direct comparison in Figure~\ref{fig:comparacion_triple_goschl} to the massless reference and perform internal consistency checks for the massive case. In the large-mass limit, the mass term dominates the statistical weight and favours monomer-saturated configurations, as observed in the simulations.

\vspace{0.3 cm}

In this regime, low acceptance probabilities can slow the Monte Carlo evolution and increase autocorrelations. Greater statistics are therefore needed to explore configuration space adequately. This loss of sampling efficiency is not, by itself, evidence of critical slowing down associated with a phase transition.

\subsection{Exploration of the \texorpdfstring{$(m/g,\theta)$}{(m/g,theta)} plane}
We study the massive model in the parameter plane defined by the dimensionless mass $m/g$ and the topological angle $\theta$. Simulations are performed on a grid of values for both parameters, and observables are displayed as heat maps. A cubic interpolation between grid points is used exclusively for visualisation; it does not increase the resolution or statistical content of the underlying Monte Carlo data. This makes it possible to examine their joint dependence on fermion mass and the topological term, rather than only one-dimensional slices.

\vspace{0.3 cm}

We first examine the average plaquette $\langle U_p\rangle$ and the chiral condensate $\langle\bar\chi\chi\rangle$. The former characterises the gauge sector, whereas the chiral condensate is related to monomer occupation and probes local fermionic saturation.

\begin{figure}[htbp]
    \centering

    \begin{subfigure}[t]{0.49\textwidth}
        \centering
        \includegraphics[width=\textwidth]{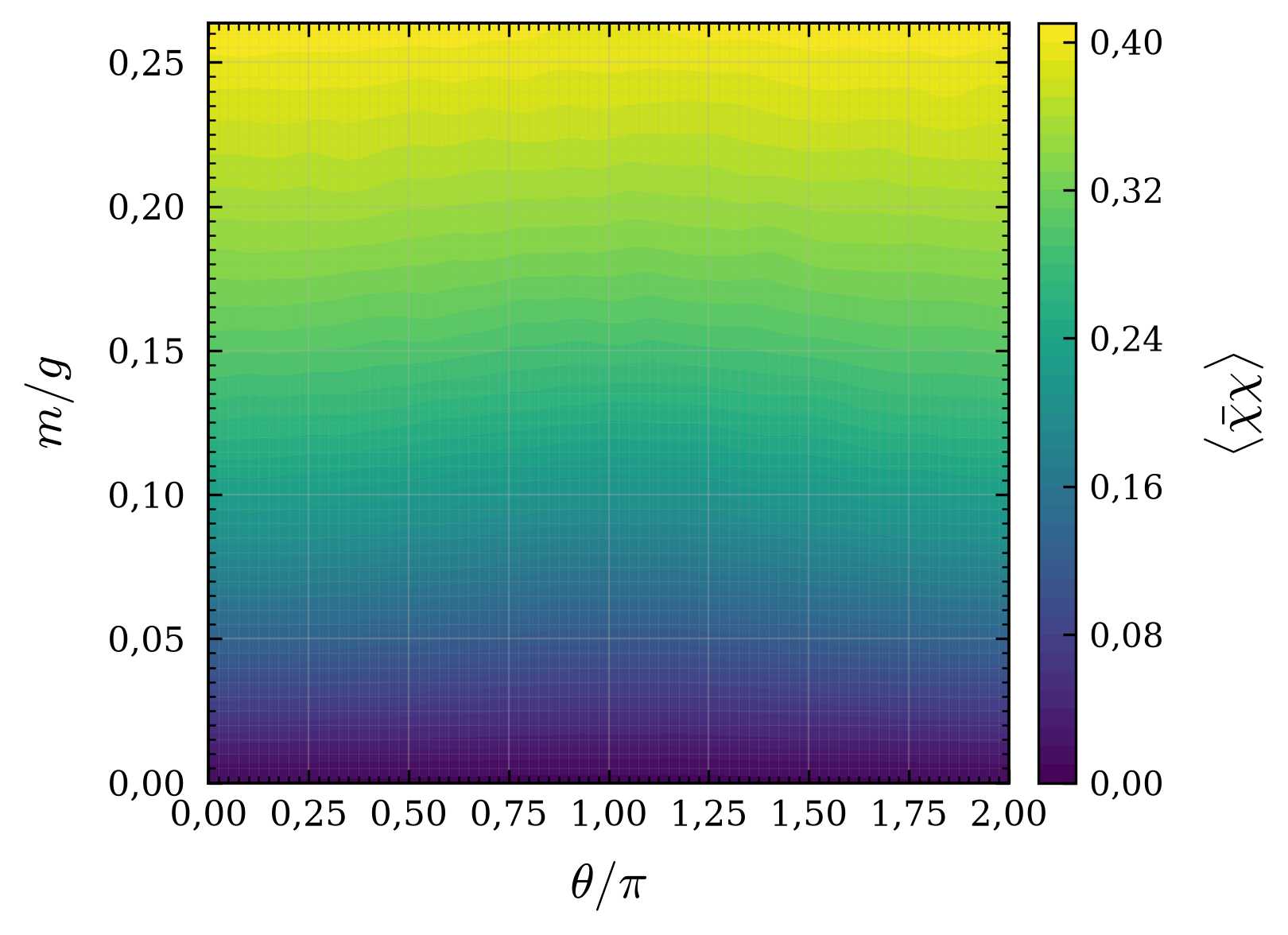}
    \end{subfigure}
    \hfill
    \begin{subfigure}[t]{0.49\textwidth}
        \centering
        \includegraphics[width=\textwidth]{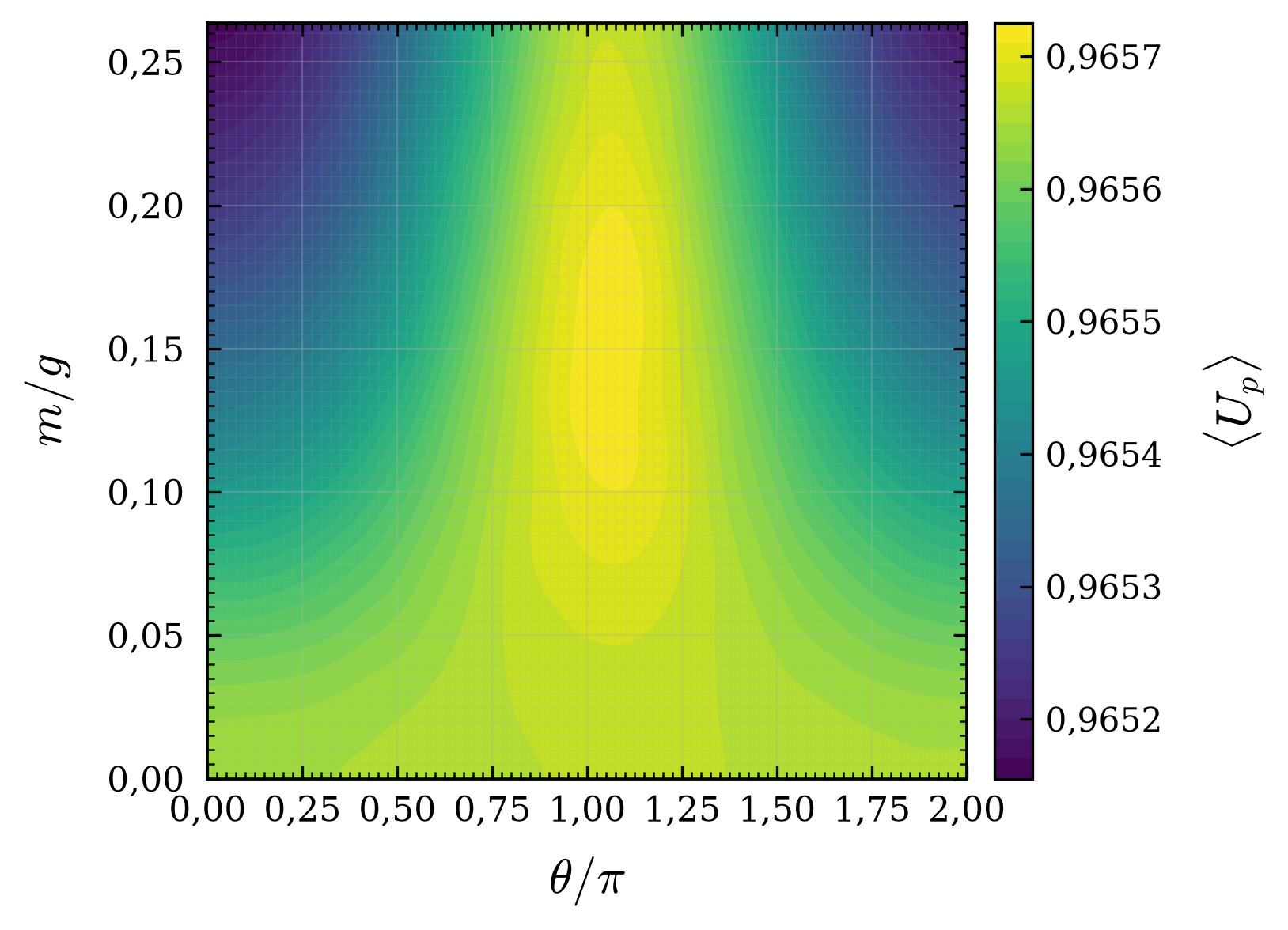}
    \end{subfigure}
    \vspace{-0.2 cm}
    \caption[Heat maps of the chiral condensate and average plaquette]{Heat maps of the chiral condensate $\langle\bar{\chi}\chi\rangle$ and the average plaquette $\langle U_p\rangle$ in the $(\theta/\pi,m/g)$ plane, for $V=12\times12$ and $\beta=14.4$. The condensate depends mainly on the mass, increasing with $m/g$, whereas the average plaquette has a more pronounced dependence on the topological angle, with higher values near $\theta/\pi\simeq1$.}
    \label{fig:heatmaps_theta_m2}
\end{figure}

To examine the heat maps in more detail, we consider selected slices, shown in Appendix~\ref{Cortes_plano_anexo}. Horizontal slices of $\langle U_p\rangle$ display a stronger response near $\theta=\pi$. The plotted curves use smoothing for visualisation of the underlying numerical observations; the smoothing is not an additional set of independent Monte Carlo measurements. Vertical slices of the chiral condensate $\langle\bar\chi\chi\rangle$ illustrate the fermionic response as $m/g$ varies. Its growth indicates that configurations associated with the fermionic degrees of freedom contribute more strongly to the observable at higher masses.

\begin{figure}[htbp]
    \centering

    \begin{subfigure}[t]{0.49\textwidth}
        \centering
        \includegraphics[width=\textwidth]{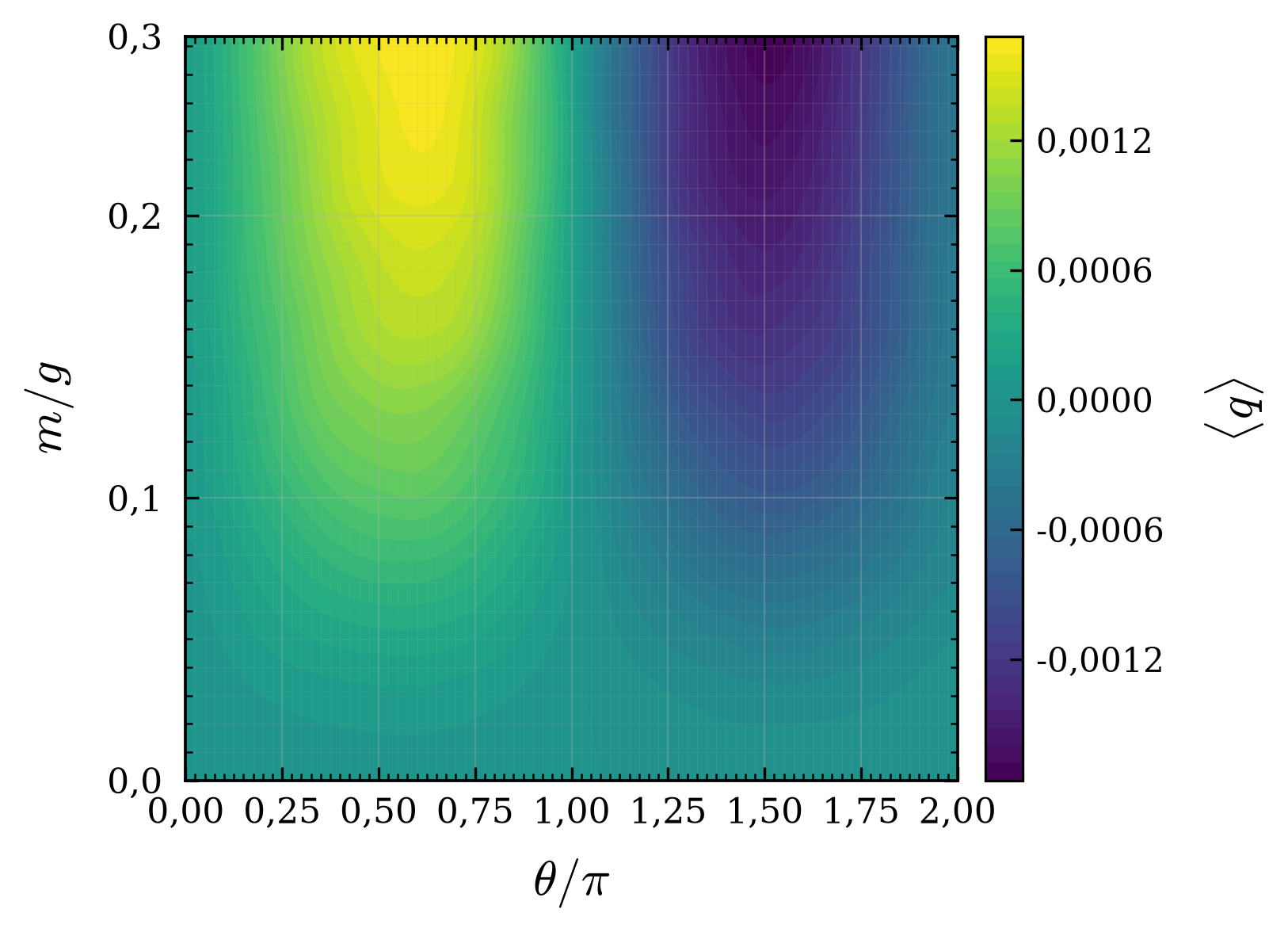}
    \end{subfigure}
    \hfill
    \begin{subfigure}[t]{0.49\textwidth}
        \centering
        \includegraphics[width=\textwidth]{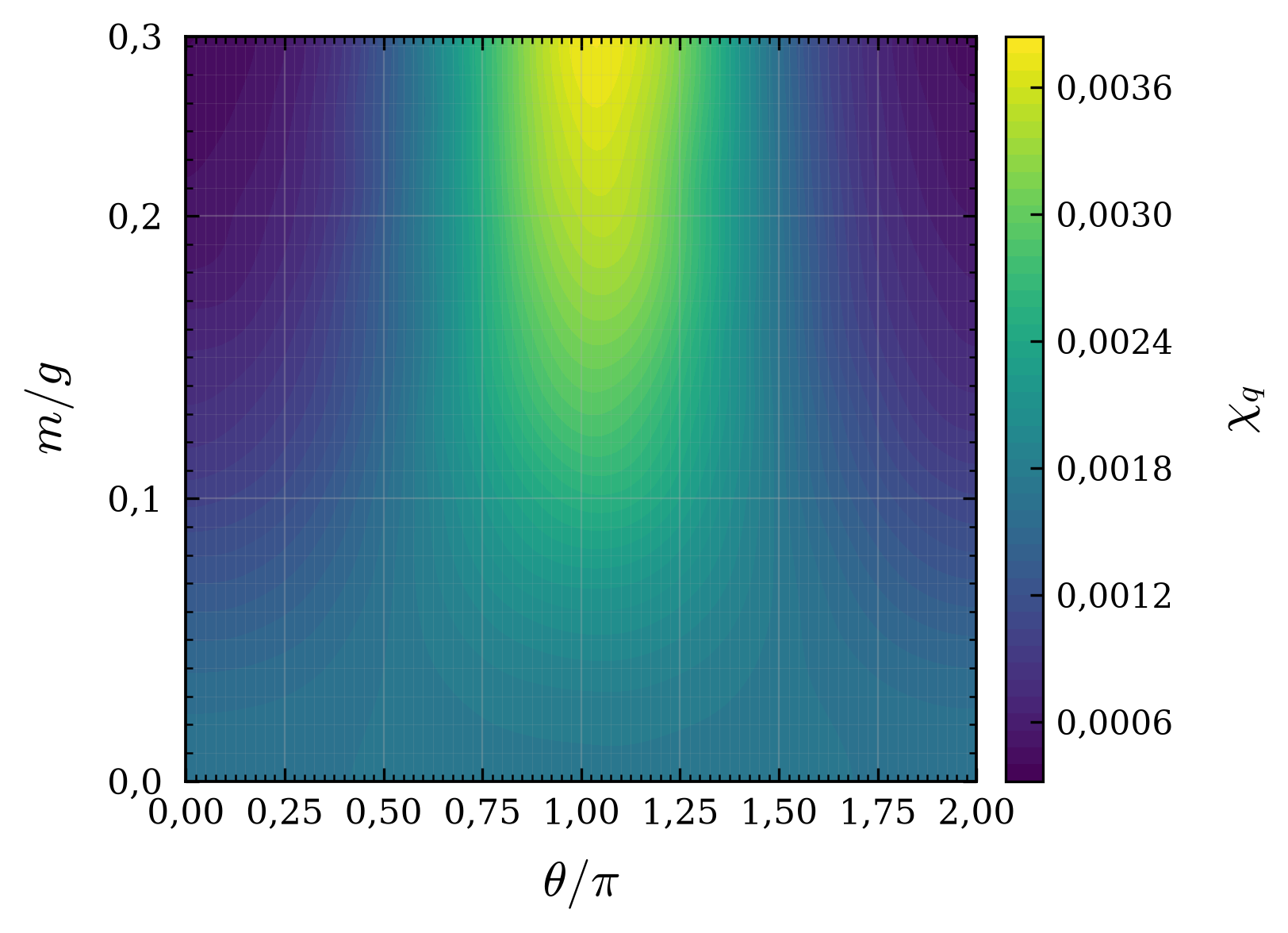}
    \end{subfigure}
    \vspace{-0.2 cm}
    \caption[Heat maps of topological charge density and susceptibility]{Heat maps of the topological charge density $\langle q\rangle$ and topological susceptibility $\chi_q$ in the $(\theta/\pi,m/g)$ plane, at $V=12\times12$ and $\beta=14.4$. The plotted interval for $\theta$ is $[0,2\pi]$.}
\label{fig:heatmaps_topologicos_theta_m}
\end{figure}

It is also useful to repeat the study using observables related, albeit indirectly, to the electric field. In particular, we consider the real topological response density $\langle q\rangle=-V^{-1}\partial_\theta\log Z$ and the corresponding response curvature $\chi_\theta$, defined in (\ref{Obs_topo}). The corresponding heat maps are shown in Figure~\ref{fig:heatmaps_topologicos_theta_m}. The historical symbol $\chi_q$ in the figures denotes the reported numerical topological-response diagnostic; identifying it quantitatively with the free-energy curvature $\chi_\theta$ requires checking the complete estimator, including its explicit derivative term. The supplied plotted tables do not contain the configuration-level derivative $\partial_\theta q_C$, so this identification cannot be independently reconstructed from the LaTeX package alone. We therefore use the sharpening of these plots only as a qualitative diagnostic.

The continuum theory is $2\pi$-periodic in the renormalised vacuum angle. The lattice charge used here does not guarantee this periodicity at finite lattice spacing. We display $[0,2\pi]$ as a physically motivated comparison window, not as an exact finite-spacing period. Two regions are visible in $\langle q\rangle$ around $\theta=0$ and $\theta=\pi$, while $\chi_q$ exhibits structures centred near these values, separated by transition regions. This pattern becomes more pronounced for larger masses.

\vspace{0.3 cm}

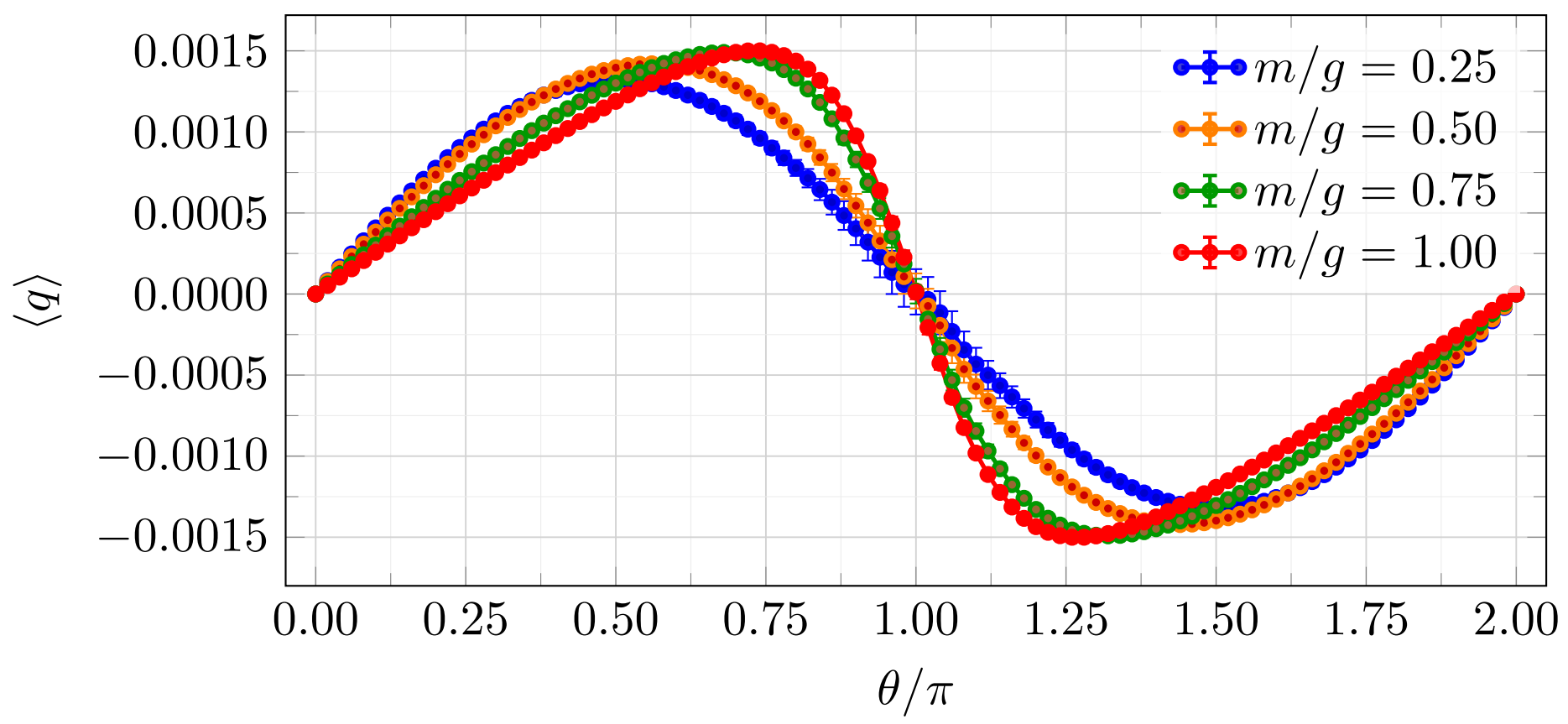
\begin{figure}[htbp]
\centering
\hspace*{-0.85cm}%
\begin{tikzpicture}
\begin{axis}[
    width=0.72\textwidth,
    height=0.38\textwidth,
    xlabel={$\theta/\pi$},
    ylabel={$\langle q\rangle$},
    xmin=-0.05,
    xmax=2.05,
    ymin=-0.00180,
    ymax=0.00172,
    grid=both,
    tick align=inside,
    tick pos=both,
    scaled y ticks=false,
    xtick={0,0.25,0.5,0.75,1,1.25,1.5,1.75,2},
    xticklabels={$0.00$,$0.25$,$0.50$,$0.75$,$1.00$,$1.25$,$1.50$,$1.75$,$2.00$},
    ytick={-0.0015,-0.0010,-0.0005,0,0.0005,0.0010,0.0015},
    yticklabels={$-0.0015$,$-0.0010$,$-0.0005$,$0.0000$,$0.0005$,$0.0010$,$0.0015$},
    minor tick num=1,
    major grid style={line width=0.35pt, draw=gray!35},
    minor grid style={line width=0.20pt, draw=gray!15},
    legend style={
        draw=none,
        fill=white,
        fill opacity=0.75,
        text opacity=1,
        at={(0.98,0.98)},
        anchor=north east
    },
    legend cell align=left,
    legend image code/.code={
        \draw[#1, line width=1.0pt] plot coordinates {(0cm,0cm) (0.45cm,0cm)};
        \draw[#1, line width=0.8pt] (0.225cm,-0.12cm) -- (0.225cm,0.12cm);
        \draw[#1, line width=0.8pt] (0.17cm,-0.12cm) -- (0.28cm,-0.12cm);
        \draw[#1, line width=0.8pt] (0.17cm,0.12cm) -- (0.28cm,0.12cm);
        \filldraw[#1] (0.225cm,0cm) circle (1.4pt);
    },
    error bars/y dir=both,
    error bars/y explicit,
]

\addplot+[
    blue,
    mark=*,
    mark size=1.4pt,
    line width=1.0pt,
    error bars/.cd,
    y dir=both,
    y explicit
]
table[
    x index=0,
    y index=1,
    y error index=2,
    col sep=space,
    comment chars={\#}
]
{cortes/topological_charge_theta_mass_scan.dat};
\addlegendentry{$m/g=0.25$}

\addplot+[
    orange,
    mark=*,
    mark size=1.4pt,
    line width=1.0pt,
    error bars/.cd,
    y dir=both,
    y explicit
]
table[
    x index=0,
    y index=3,
    y error index=4,
    col sep=space,
    comment chars={\#}
]
{cortes/topological_charge_theta_mass_scan.dat};
\addlegendentry{$m/g=0.50$}

\addplot+[
    green!60!black,
    mark=*,
    mark size=1.4pt,
    line width=1.0pt,
    error bars/.cd,
    y dir=both,
    y explicit
]
table[
    x index=0,
    y index=5,
    y error index=6,
    col sep=space,
    comment chars={\#}
]
{cortes/topological_charge_theta_mass_scan.dat};
\addlegendentry{$m/g=0.75$}

\addplot+[
    red,
    mark=*,
    mark size=1.4pt,
    line width=1.0pt,
    error bars/.cd,
    y dir=both,
    y explicit
]
table[
    x index=0,
    y index=7,
    y error index=8,
    col sep=space,
    comment chars={\#}
]
{cortes/topological_charge_theta_mass_scan.dat};
\addlegendentry{$m/g=1.00$}

\end{axis}
\end{tikzpicture}

\caption[Slices of the topological charge density]{Topological charge density $\langle q\rangle$ versus $\theta/\pi$ for different fermion masses, with $V=12\times12$ and $\beta=14.4$. As $m/g$ increases, the curves depart progressively from harmonic behaviour: regions away from $\theta/\pi\simeq1$ approach nearly linear branches, while the variation around $\theta/\pi=1$ becomes sharper.}
\label{fig:cortes_q_theta_m_altas}

\end{figure}

The distinction between the regions near $\theta=0$ and $\theta=\pi$, together with the increasingly sharp variation between them, is reminiscent of a possible phase change. We therefore examine the response around $\theta=\pi$ more closely for several masses, as shown in Figure~\ref{fig:cortes_q_theta_m_altas}.

The variation of $\langle q\rangle$ becomes sharper near $\theta\simeq\pi$, particularly as the mass increases. This suggests that the topological response becomes concentrated around this region, whereas away from $\theta=\pi$ the dependence is smoother. Such behaviour is compatible with an approach to non-analytic behaviour in the thermodynamic limit, potentially associated with a phase transition.

\vspace{0.3 cm}

The results are also consistent with the theoretical picture presented in Appendix~\ref{app:termino_masa}: increasing the fermion mass may lead to an abrupt rearrangement of the vacuum near $\theta=\pi$. This can be understood qualitatively in terms of the Schwinger effect. The topological angle acts as a background electric field, and near $\theta=\pi$ different electric-flux sectors compete. Fermion--antifermion pair production can partially screen that field and alter the preferred vacuum sector. For smaller masses, pair production is more favourable and the crossover between sectors is smoother; for larger masses, pair production is more costly and the change becomes increasingly localised around $\theta=\pi$. This provides a qualitative interpretation of the evolution of $\langle q\rangle$ from an approximately harmonic dependence to a much sharper variation around $\theta=\pi$.

\vspace{0.3 cm}

\begin{figure}[htbp]
\centering
\includegraphics[width=0.72\textwidth]{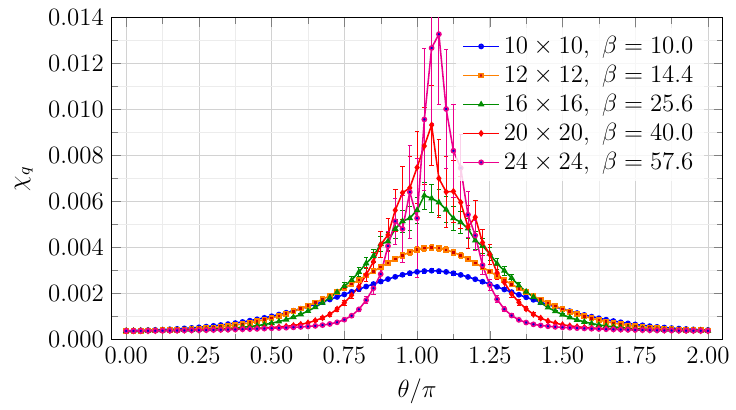}
\caption[Topological susceptibility in a scan towards the continuum]{Topological susceptibility $\chi_q$ versus $\theta/\pi$ for increasing lattice resolutions, with $\beta$ and lattice dimensions varied together to keep the physical extent approximately fixed. The peak sharpens with increasing resolution, while its uncertainty becomes large near the maximum. This sequence is not a thermodynamic finite-size-scaling test.}
\label{fig:chi_q_pico_volumen}
\end{figure}

To investigate this enhancement further, we perform a scan towards the continuum limit, increasing $\beta$ while adjusting the lattice dimensions so that the physical extent remains approximately fixed according to Eq.~(\ref{magni_fisi}). The plotted data file records the lattice dimensions, $\beta$ and the numerical uncertainty but does not document the fixed $m/g$ value or the configuration-level construction of the curvature estimator; those metadata should be supplied to support a quantitative interpretation. Figure~\ref{fig:chi_q_pico_volumen} shows a susceptibility peak that becomes higher and narrower as the lattice resolution increases. At the largest lattice sizes the uncertainty also grows strongly near the peak, preventing a controlled quantitative continuum extrapolation. This trajectory approaches the continuum at fixed physical volume and is \emph{not} a thermodynamic finite-size-scaling analysis; peak sharpening alone therefore does not establish a phase transition. The field-theoretical definition of lattice topological charge moreover requires care when comparing topological susceptibilities across lattice spacings \cite{Gattringer2015SchwingerDual}.

\subsection{Limitations due to the sign problem}

The continuum-oriented scan and the parameter maps show strong variations around $\theta=\pi$, accompanied by increasing uncertainty on larger lattices. A major obstacle to quantitative measurements in this region is the sign problem, which amplifies the uncertainties of reweighted observables.

\vspace{0.3 cm}

The difficulty arises because the region of strongest variation coincides with a pronounced reduction in the average sign. As the average sign becomes small, cancellations between opposite-sign configurations destabilise the reweighted estimates in (\ref{reweighting_signo}). The difficulty is therefore not uniform across the parameter plane, but is strongest precisely where a potential transition would be investigated.

\vspace{0.3 cm}

This is characteristic of a severe sign problem: if the average sign decreases exponentially with lattice volume at fixed physical parameters, the statistical cost of maintaining a given relative precision generally grows exponentially. Hence, increasing lattice size to approach the continuum or thermodynamic limit makes regions with small $\langle\mathrm{sign}\rangle$ increasingly difficult to explore.

\vspace{0.3 cm}


\begin{figure}[htbp]
\centering
\hspace*{-0.7cm}%
\resizebox{0.730\textwidth}{!}{%
\begin{tikzpicture}

\begin{axis}[
    width=0.95\textwidth,
    height=0.62\textwidth,
    scale only axis,
    axis lines=box,
    xmin=0,
    xmax=2400,
    ymin=0.11,
    ymax=1.05,
    ymode=log,
    xtick={0,500,1000,1500,2000},
    xticklabels={$0$,$500$,$1000$,$1500$,$2000$},
    ytick={0.1,0.2,0.3,0.4,0.6,1.0},
    yticklabels={$10^{-1}$,$2\times10^{-1}$,$3\times10^{-1}$,$4\times10^{-1}$,$6\times10^{-1}$,$10^{0}$},
    tick align=inside,
    major tick length=5pt,
    minor tick num=1,
    grid=both,
    major grid style={draw=gray!40, line width=0.3pt},
    minor grid style={draw=gray!20, line width=0.2pt},
    label style={font=\LARGE},
    ticklabel style={font=\LARGE},
    xlabel={$V=L^2$},
    ylabel={$\langle \mathrm{sign}\rangle$},
    enlarge x limits=false,
    enlarge y limits=false
]

\addplot+[
    blue,
    densely dashed,
    line width=1.0pt,
    mark=*,
    mark size=1.8pt,
    mark options={fill=blue, draw=blue},
    error bars/.cd,
        y dir=both,
        y explicit,
        error mark=none,
        error bar style={blue, line width=0.65pt},
    /pgfplots/forget plot
]
coordinates {
    (64,0.99)   +- (0,0.003)
    (144,0.97)  +- (0,0.003)
    (576,0.88)  +- (0,0.004)
    (2304,0.68) +- (0,0.005)
};

\addplot[
    blue,
    only marks,
    mark=|,
    mark size=2.2pt,
    mark options={blue, rotate=90, line width=0.65pt},
    forget plot
]
coordinates {
    (64,0.993)
    (144,0.973)
    (576,0.884)
    (2304,0.685)
};

\addplot[
    blue,
    only marks,
    mark=|,
    mark size=2.2pt,
    mark options={blue, rotate=90, line width=0.65pt},
    forget plot
]
coordinates {
    (64,0.987)
    (144,0.967)
    (576,0.876)
    (2304,0.675)
};

\addplot+[
    orange,
    densely dashed,
    line width=1.0pt,
    mark=*,
    mark size=1.8pt,
    mark options={fill=orange, draw=orange},
    error bars/.cd,
        y dir=both,
        y explicit,
        error mark=none,
        error bar style={orange, line width=0.65pt},
    /pgfplots/forget plot
]
coordinates {
    (64,0.98)   +- (0,0.003)
    (144,0.96)  +- (0,0.003)
    (576,0.84)  +- (0,0.004)
    (2304,0.58) +- (0,0.005)
};

\addplot[
    orange,
    only marks,
    mark=|,
    mark size=2.2pt,
    mark options={orange, rotate=90, line width=0.65pt},
    forget plot
]
coordinates {
    (64,0.983)
    (144,0.963)
    (576,0.844)
    (2304,0.585)
};

\addplot[
    orange,
    only marks,
    mark=|,
    mark size=2.2pt,
    mark options={orange, rotate=90, line width=0.65pt},
    forget plot
]
coordinates {
    (64,0.977)
    (144,0.957)
    (576,0.836)
    (2304,0.575)
};

\addplot+[
    green!60!black,
    densely dashed,
    line width=1.0pt,
    mark=*,
    mark size=1.8pt,
    mark options={fill=green!60!black, draw=green!60!black},
    error bars/.cd,
        y dir=both,
        y explicit,
        error mark=none,
        error bar style={green!60!black, line width=0.65pt},
    /pgfplots/forget plot
]
coordinates {
    (64,0.97)   +- (0,0.003)
    (144,0.93)  +- (0,0.003)
    (576,0.76)  +- (0,0.004)
    (2304,0.37) +- (0,0.004)
};

\addplot[
    green!60!black,
    only marks,
    mark=|,
    mark size=2.2pt,
    mark options={green!60!black, rotate=90, line width=0.65pt},
    forget plot
]
coordinates {
    (64,0.973)
    (144,0.933)
    (576,0.764)
    (2304,0.374)
};

\addplot[
    green!60!black,
    only marks,
    mark=|,
    mark size=2.2pt,
    mark options={green!60!black, rotate=90, line width=0.65pt},
    forget plot
]
coordinates {
    (64,0.967)
    (144,0.927)
    (576,0.756)
    (2304,0.366)
};

\addplot+[
    red,
    densely dashed,
    line width=1.0pt,
    mark=*,
    mark size=1.8pt,
    mark options={fill=red, draw=red},
    error bars/.cd,
        y dir=both,
        y explicit,
        error mark=none,
        error bar style={red, line width=0.65pt},
    /pgfplots/forget plot
]
coordinates {
    (64,0.95)   +- (0,0.003)
    (144,0.89)  +- (0,0.003)
    (576,0.59)  +- (0,0.003)
    (2304,0.13) +- (0,0.002)
};

\addplot[
    red,
    only marks,
    mark=|,
    mark size=2.2pt,
    mark options={red, rotate=90, line width=0.65pt},
    forget plot
]
coordinates {
    (64,0.953)
    (144,0.893)
    (576,0.593)
    (2304,0.132)
};

\addplot[
    red,
    only marks,
    mark=|,
    mark size=2.2pt,
    mark options={red, rotate=90, line width=0.65pt},
    forget plot
]
coordinates {
    (64,0.947)
    (144,0.887)
    (576,0.587)
    (2304,0.128)
};

\addplot+[
    violet,
    densely dashed,
    line width=1.0pt,
    mark=*,
    mark size=1.8pt,
    mark options={fill=violet, draw=violet},
    error bars/.cd,
        y dir=both,
        y explicit,
        error mark=none,
        error bar style={violet, line width=0.65pt},
    /pgfplots/forget plot
]
coordinates {
    (64,0.96)   +- (0,0.003)
    (144,0.92)  +- (0,0.003)
    (576,0.66)  +- (0,0.004)
    (2304,0.20) +- (0,0.003)
};

\addplot[
    violet,
    only marks,
    mark=|,
    mark size=2.2pt,
    mark options={violet, rotate=90, line width=0.65pt},
    forget plot
]
coordinates {
    (64,0.963)
    (144,0.923)
    (576,0.664)
    (2304,0.203)
};

\addplot[
    violet,
    only marks,
    mark=|,
    mark size=2.2pt,
    mark options={violet, rotate=90, line width=0.65pt},
    forget plot
]
coordinates {
    (64,0.957)
    (144,0.917)
    (576,0.656)
    (2304,0.197)
};

\node[
    draw=black!30,
    fill=white,
    anchor=south west,
    inner sep=5pt,
    font=\large
] at (axis description cs:0.04,0.05)
{
\begin{tabular}{@{}c@{}}
\multicolumn{1}{c}{$\theta/\pi$}\\[4pt]

\begin{tikzpicture}[baseline=-0.6ex, x=1cm, y=1cm]
    \draw[blue, densely dashed, line width=1.0pt] (0.00,0) -- (0.55,0);
    \fill[blue] (0.28,0) circle (1.2pt);
    \draw[blue, line width=0.65pt] (0.28,-0.10) -- (0.28,0.10);
    \draw[blue, line width=0.65pt] (0.20,-0.10) -- (0.36,-0.10);
    \draw[blue, line width=0.65pt] (0.20,0.10) -- (0.36,0.10);
\end{tikzpicture}\;\; $0.00$
\\[2pt]

\begin{tikzpicture}[baseline=-0.6ex, x=1cm, y=1cm]
    \draw[orange, densely dashed, line width=1.0pt] (0.00,0) -- (0.55,0);
    \fill[orange] (0.28,0) circle (1.2pt);
    \draw[orange, line width=0.65pt] (0.28,-0.10) -- (0.28,0.10);
    \draw[orange, line width=0.65pt] (0.20,-0.10) -- (0.36,-0.10);
    \draw[orange, line width=0.65pt] (0.20,0.10) -- (0.36,0.10);
\end{tikzpicture}\;\; $0.50$
\\[2pt]

\begin{tikzpicture}[baseline=-0.6ex, x=1cm, y=1cm]
    \draw[green!60!black, densely dashed, line width=1.0pt] (0.00,0) -- (0.55,0);
    \fill[green!60!black] (0.28,0) circle (1.2pt);
    \draw[green!60!black, line width=0.65pt] (0.28,-0.10) -- (0.28,0.10);
    \draw[green!60!black, line width=0.65pt] (0.20,-0.10) -- (0.36,-0.10);
    \draw[green!60!black, line width=0.65pt] (0.20,0.10) -- (0.36,0.10);
\end{tikzpicture}\;\; $1.00$
\\[2pt]

\begin{tikzpicture}[baseline=-0.6ex, x=1cm, y=1cm]
    \draw[red, densely dashed, line width=1.0pt] (0.00,0) -- (0.55,0);
    \fill[red] (0.28,0) circle (1.2pt);
    \draw[red, line width=0.65pt] (0.28,-0.10) -- (0.28,0.10);
    \draw[red, line width=0.65pt] (0.20,-0.10) -- (0.36,-0.10);
    \draw[red, line width=0.65pt] (0.20,0.10) -- (0.36,0.10);
\end{tikzpicture}\;\; $1.50$
\\[2pt]

\begin{tikzpicture}[baseline=-0.6ex, x=1cm, y=1cm]
    \draw[violet, densely dashed, line width=1.0pt] (0.00,0) -- (0.55,0);
    \fill[violet] (0.28,0) circle (1.2pt);
    \draw[violet, line width=0.65pt] (0.28,-0.10) -- (0.28,0.10);
    \draw[violet, line width=0.65pt] (0.20,-0.10) -- (0.36,-0.10);
    \draw[violet, line width=0.65pt] (0.20,0.10) -- (0.36,0.10);
\end{tikzpicture}\;\; $2.00$

\end{tabular}
};

\end{axis}
\end{tikzpicture}
}
\caption[Severity of the sign problem]{Average sign $\langle\mathrm{sign}\rangle$ as a function of square lattice volume $V=L^2$, for different $\theta/\pi$ values at fixed mass $m/g=0.9$ and $\beta=10$.}
\label{fig:signo_volumen_log}

\end{figure}
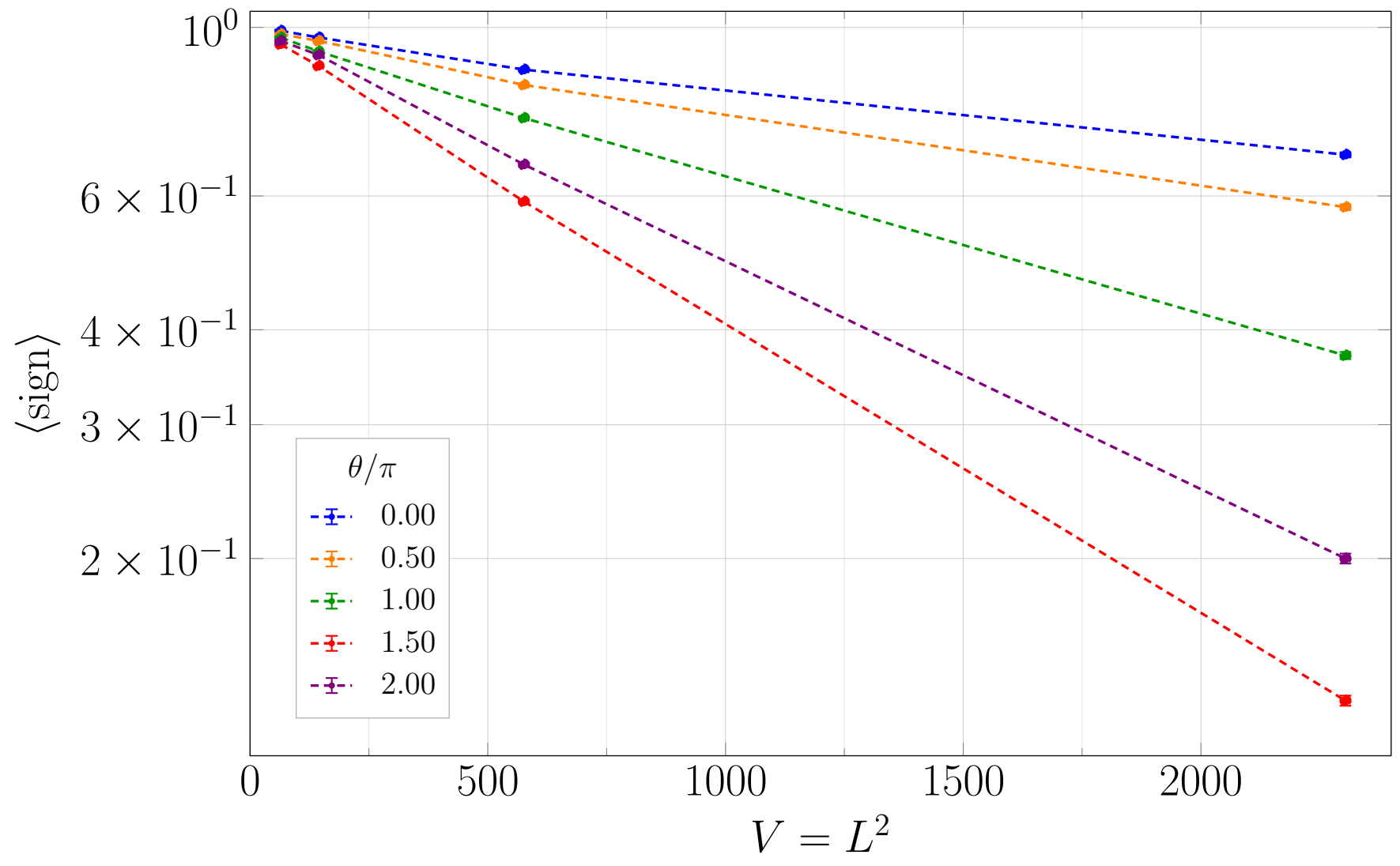

Figure~\ref{fig:signo_volumen_log} illustrates this mechanism. Where the average sign approaches zero, statistical fluctuations are amplified and physical observables become progressively harder to determine reliably. This explains why the continuum-oriented scan of the previous section does not yield a controlled quantitative extrapolation.

\vspace{0.3 cm}

Nevertheless, the average sign itself can serve as a diagnostic observable. Rather than examining only selected parameter slices, we map $\langle\mathrm{sign}\rangle$ throughout the $(\theta/\pi,m/g)$ plane, identifying the regions most affected by the sign problem and tracking how those regions change with lattice volume.

\begin{figure}[htbp]
    \centering

    \makebox[\textwidth][c]{%
    \resizebox{1.12\textwidth}{!}{%
    \begin{minipage}{0.49\textwidth}
        \centering
        \includegraphics[width=\linewidth]{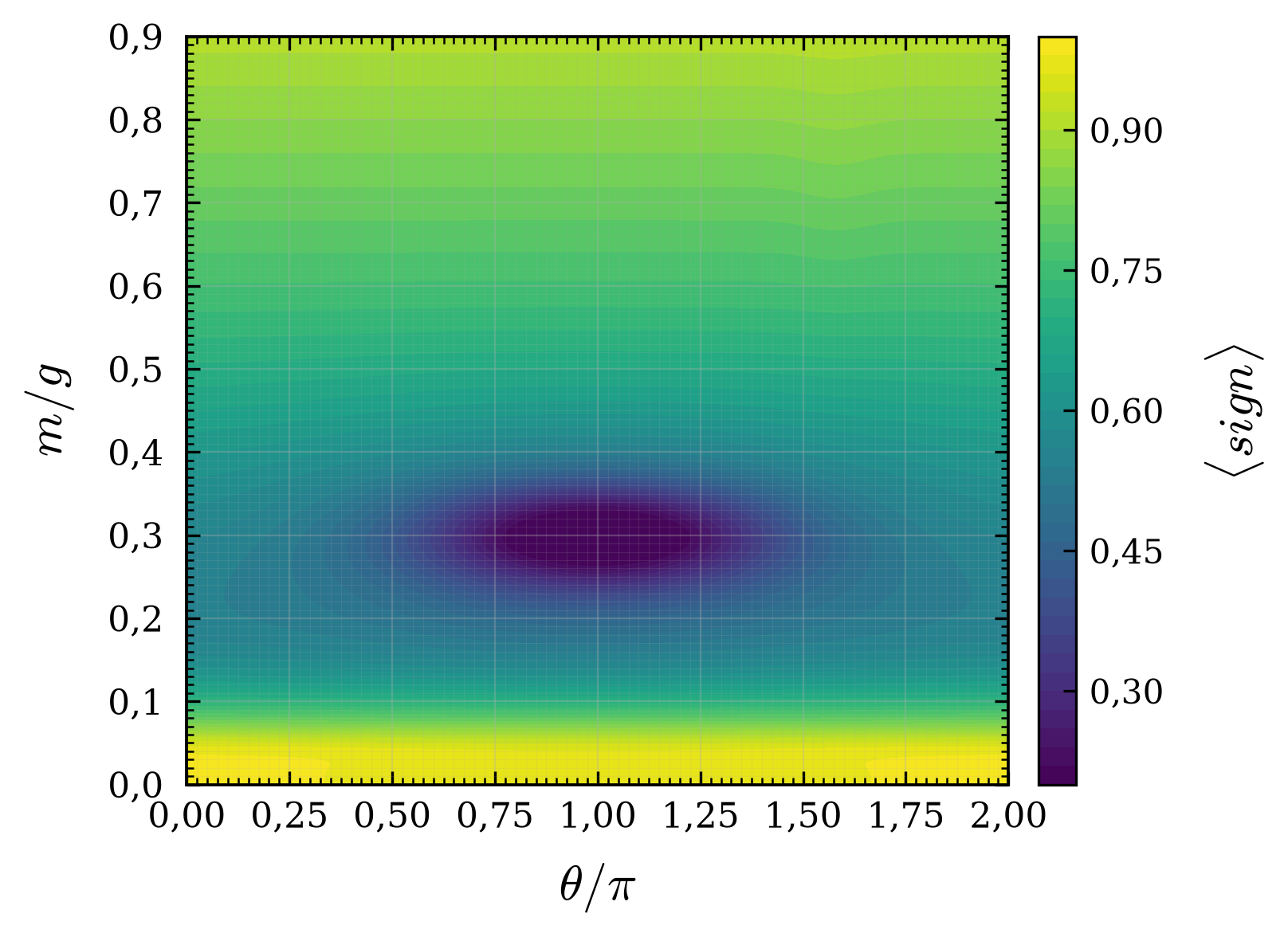}
        
        \vspace{0.1cm}
        {\small (a) $V=8\times8$, $\beta=6.4$.}
    \end{minipage}
    \hspace{-0.01\textwidth}
    \begin{minipage}{0.49\textwidth}
        \centering
        \includegraphics[width=\linewidth]{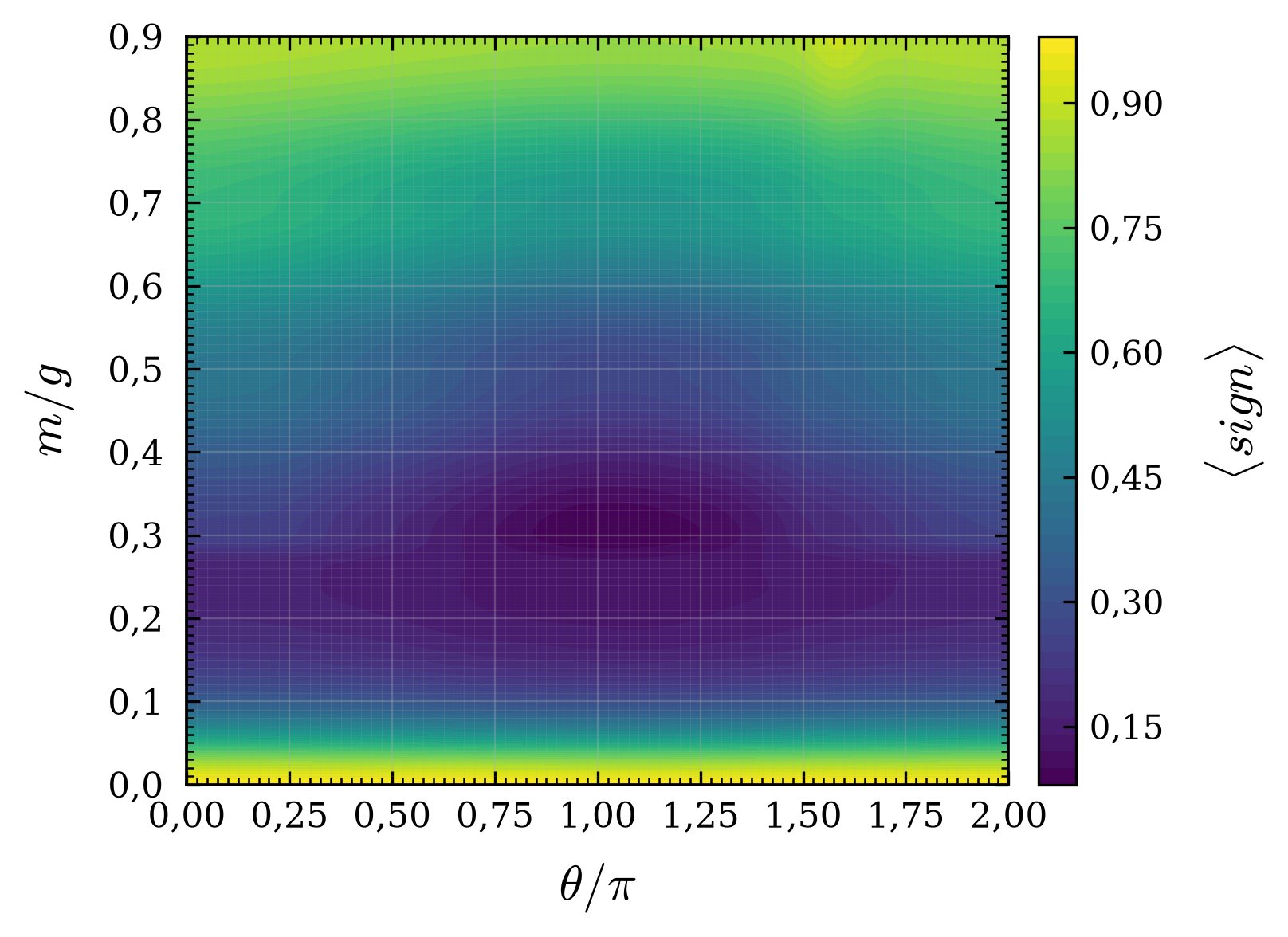}
        
        \vspace{0.1cm}
        {\small (b) $V=12\times12$, $\beta=14.4$.}
    \end{minipage}
    }%
    }

    \caption[Heat maps of the average sign at two simulation parameter sets]{Heat maps of the average sign $\langle\mathrm{sign}\rangle$ in the $(\theta/\pi,m/g)$ plane for two sets of lattice parameters. The region with smaller average sign lies near $\theta/\pi\simeq1$, where topological observables vary most strongly. Both $V$ and $\beta$ change between panels; the figure alone cannot isolate a volume effect.}
    \label{fig:heatmaps_signo_volumen}
\end{figure}

Comparing the heat maps in Figure~\ref{fig:heatmaps_signo_volumen} shows a larger low-sign region in the second simulation. Both the lattice dimensions and $\beta$ differ, so this comparison alone does not isolate a pure volume dependence of the sign problem. At the smaller lattice size the problematic region is more localised around $\theta/\pi\simeq1$, whereas it extends over a wider part of the $(\theta/\pi,m/g)$ plane in the second simulation. The observed loss of precision affects an extended region rather than one point; attributing its change exclusively to physical volume would require fixed-coupling comparisons.

\vspace{0.3 cm}

This is particularly relevant because the same region exhibits the strongest variations in the topological observables. The sign problem thus limits numerical access to the potentially critical part of the model.

\vspace{0.3 cm}

It is important to distinguish a physical signal from the ability of a numerical algorithm to resolve it. A small average sign does not by itself prove that the structure of topological observables is artificial; rather, it shows that these estimates depend on delicate cancellations between sectors of opposite sign. The growth of uncertainties should therefore be interpreted as a limitation of sampling and reweighting, not evidence that vacuum rearrangement disappears.
\vspace{0.3 cm}

The results are consistent with a qualitative rearrangement of the vacuum near $\theta=\pi$, associated with the physical mechanism of the Schwinger effect. However, the degradation of the average sign prevents reliable quantitative conclusions about the precise nature of a transition or its continuum extrapolation.

\bigskip

\section{Conclusions and Outlook}

Our results show that the dual representation provides a useful formulation for studying the Schwinger model numerically in the presence of a topological angle. In the massless theory, geometric sign cancellation allows simulations with positive weights and reproduces reference results, supporting the dual approach as a practical Monte Carlo tool. Extending the simulation to finite fermion mass gives access to richer behaviour associated with monomers and changes in the dominant configurations.

\vspace{0.3 cm}

In particular, maps in the $(\theta/\pi,m/g)$ plane reveal a special region around $\theta=\pi$: as the mass increases, the topological charge density changes more sharply and the topological susceptibility develops a more pronounced peak. This is consistent with a rearrangement of the vacuum near $\theta=\pi$ and can be interpreted qualitatively in terms of the Schwinger effect. The angle $\theta$ acts as a background electric field that can be screened through fermion-pair production. Screening is more effective for smaller masses, while at larger masses pair creation is suppressed and the change between electric-flux sectors becomes concentrated near $\theta=\pi$.

\vspace{0.3 cm}

However, this interpretation requires care. Although the observables suggest possible phase-transition behaviour, the present data do not establish a transition rigorously: a controlled analysis of both the thermodynamic and continuum limits would be required. The main limitation is the sign problem, since the region of interest coincides with a significant decrease in the average sign. Reweighting then becomes inefficient and uncertainties increase rapidly with lattice volume. Thus, while the dual representation improves some parameter regimes, it does not by itself remove the exponential difficulty of the sign problem for non-zero mass and $\theta\neq0$.

\vspace{0.3 cm}

Future work should improve statistical and algorithmic control, optimise update efficiency and investigate methods for mitigating the sign problem. Dual formulations and geometrical algorithms are promising in this respect, since in some models they reduce \textit{critical slowing down} \cite{Azcoiti2009GeometricAlgorithm}. Another possibility is to explore meron-based algorithms or similar constructions developed for particular sign problems \cite{Chandrasekharan1999MeronCluster}.

\vspace{0.3 cm}

In summary, the dual formulation gives access to non-trivial signatures of the massive theory with a topological term and suggests a possible rearrangement of the vacuum around $\theta=\pi$. However, the sign problem prevents these indications from becoming firm quantitative conclusions at present. Further progress will therefore rely on algorithmic improvements or alternative representations capable of controlling this region.

\clearpage
\phantomsection
\addcontentsline{toc}{section}{References}

{\huge\bfseries References\par}
\vspace{0.5cm}

\printbibliography[heading=none]

\appendix

\titleformat{\section}[display]
  {\normalfont\bfseries}
  {\Large Appendix \thesection}
  {0.1cm}
  {\huge}

\section{Bosonisation of the Schwinger Model}\label{app:bosonizacion}

An alternative way to study the Schwinger model is through bosonisation. This formulation rewrites the fermionic theory in terms of a real scalar field and provides an intuitive description of dynamical mass generation and the role of the topological angle $\theta$.

\vspace{0.3 cm}

The central idea of bosonisation is to express a fermionic theory in terms of a bosonic field. Rather than working directly with the fermion field $\psi$, certain operators, such as the current, are represented using a real scalar field $\phi$. This correspondence is particularly useful in $1+1$ dimensions: instead of describing $\psi$ and $\bar\psi$ directly, one can characterise spatial variations of fermionic charge density through the bosonic field $\phi$.

\vspace{0.3 cm}

To bosonise the Schwinger model, we first introduce the basic rules and their motivation. The next subsection therefore presents the correspondence between the fermionic current and the bosonic field, together with the rule used to rewrite the fermionic kinetic term.

\subsection{Bosonisation rules}

We begin by identifying how the fermionic current is represented in terms of the bosonic field. The starting point is the global $U(1)$ symmetry of the fermionic field,
\begin{equation}
\psi \longrightarrow e^{i\alpha}\psi,
\qquad
\bar{\psi} \longrightarrow \bar{\psi}e^{-i\alpha},
\end{equation}

where $\alpha$ is constant. By Noether's theorem, this symmetry corresponds to a conserved current. In the Schwinger model, it is
\begin{equation}
j^\mu = \bar{\psi}\gamma^\mu\psi .
\end{equation}

Current conservation is expressed by the continuity equation,
\begin{equation}
\partial_\mu j^\mu = 0 .
\end{equation}

In $1+1$ dimensions, this condition allows the current to be written in terms of a single scalar field. To see this, we write the continuity equation explicitly:
\begin{equation}
\partial_0 j^0 + \partial_1 j^1 = 0 .
\end{equation}

This equation is satisfied automatically when the two components of the current are expressed in terms of a scalar field $\phi$ such that
\begin{equation}
\label{current_boso}
j^0 =
-\frac{1}{\sqrt{\pi}}\partial_1 \phi,
\qquad
j^1 =
\frac{1}{\sqrt{\pi}}\partial_0 \phi ,
\end{equation}

where the factor $1/\sqrt\pi$ fixes the normalisation of the bosonic field. Indeed, because derivatives commute, we have
\begin{equation}
\partial_0 j^0 + \partial_1 j^1
=
-\frac{1}{\sqrt{\pi}}\partial_0\partial_1\phi
+
\frac{1}{\sqrt{\pi}}\partial_1\partial_0\phi
=
0 .
\end{equation}

In compact notation, Eqs.~(\ref{current_boso}) can be written as
\begin{equation}
\label{rule_boso_1}
j^\mu
=
-\frac{1}{\sqrt{\pi}}
\epsilon^{\mu\nu}\partial_\nu \phi .
\end{equation}

where $\epsilon^{\mu\nu}$ is the two-index Levi--Civita tensor. To preserve the sign convention used so far, we take $\epsilon^{01}=1$.

\vspace{0.3 cm}

The relation above expresses the fermionic current through $\phi$. To complete the bosonised formulation, we must also express the fermionic kinetic term using this field. With the current normalisation chosen here, bosonisation gives \cite{Coleman1976MassiveSchwinger}
\begin{equation}
\label{rule_boso_2}
\bar{\psi}i\gamma^\mu\partial_\mu\psi
\quad \longrightarrow \quad
\frac{1}{2}\partial_\mu\phi\,\partial^\mu\phi .
\end{equation}

That is, the fermionic kinetic term is replaced by the usual kinetic term of a real scalar field.

\subsection{Bosonisation of the massless Lagrangian}
\label{app:bosonizacion_sinmasa}
Having established the bosonisation rules, we apply them to the Schwinger model with a topological term. We start from the Lagrangian containing the topological angle,
\begin{equation}
\mathcal{L}
=
-\frac{1}{4}F_{\mu\nu}F^{\mu\nu}
+
\bar{\psi}
\left(
i\gamma^\mu D_\mu - m
\right)
\psi
+
\frac{g\theta}{2\pi}F_{01}.
\end{equation}

We derive the bosonised form of the massless-fermion Lagrangian, $m=0$. The fermionic term can be expanded by writing the covariant derivative (\ref{deriv_cov}) explicitly, giving
\begin{equation}
\bar{\psi}i\gamma^\mu D_\mu\psi
=
\bar{\psi}i\gamma^\mu\partial_\mu\psi
-
gA_\mu j^\mu .
\end{equation}

Applying the bosonisation rules (\ref{rule_boso_1}) and (\ref{rule_boso_2}), the massless bosonised Lagrangian becomes
\begin{equation}
\mathcal{L}
=
\frac{1}{2}\partial_\mu\phi\,\partial^\mu\phi
-
\frac{1}{4}F_{\mu\nu}F^{\mu\nu}
+
\frac{g}{\sqrt{\pi}}
A_\mu\epsilon^{\mu\nu}\partial_\nu\phi
+
\frac{g\theta}{2\pi}F_{01}.
\end{equation}

Within the action, the derivative acting on $\phi$ can be transferred to the gauge field by integration by parts, neglecting boundary terms. Thus,
\begin{equation}
A_\mu\epsilon^{\mu\nu}\partial_\nu\phi
\simeq
-
\phi\epsilon^{\mu\nu}\partial_\nu A_\mu .
\end{equation}

\vspace{0.3 cm}

With the sign convention chosen here, this combination is related to the field-strength tensor. Recalling its definition in (\ref{deriv_cov}), we find
\begin{equation}
-\epsilon^{\mu\nu}\partial_\nu A_\mu
=
F_{01}.
\end{equation}

The interaction term can therefore be written as
\begin{equation}
\frac{g}{\sqrt{\pi}}
A_\mu\epsilon^{\mu\nu}\partial_\nu\phi
\simeq
\frac{g}{\sqrt{\pi}}\phi F_{01}.
\end{equation}

Here $F_{01}$ is identified with the electric field $E$, the only independent component of the field-strength tensor in $1+1$ dimensions. 

The electromagnetic part of the Lagrangian can then be written as
\begin{equation}
-\frac{1}{4}F_{\mu\nu}F^{\mu\nu}
=
\frac{1}{2}E^2.
\end{equation}

The bosonised massless Schwinger-model Lagrangian thus takes the form
\begin{equation}
\mathcal{L}
=
\frac{1}{2}\partial_\mu\phi\,\partial^\mu\phi
+
\frac{1}{2}E^2
+
\frac{g}{\sqrt{\pi}}\phi E
+
\frac{g\theta}{2\pi}E .
\end{equation}

This expression illustrates the usefulness of bosonisation in the massless model. The electric field $E$ appears without derivatives and can be eliminated algebraically. The $E$-dependent part of the Lagrangian is
\begin{equation}
\mathcal{L}_E
=
\frac{1}{2}E^2
+
\frac{g}{\sqrt{\pi}}\phi E
+
\frac{g\theta}{2\pi}E .
\end{equation}

Collecting the terms linear in $E$,
\begin{equation}
\mathcal{L}_E
=
\frac{1}{2}E^2
+
g
\left(
\frac{\phi}{\sqrt{\pi}}
+
\frac{\theta}{2\pi}
\right)
E .
\end{equation}

Completing the square,
\begin{equation}
\mathcal{L}_E
=
\frac{1}{2}
\left[
E+
g
\left(
\frac{\phi}{\sqrt{\pi}}
+
\frac{\theta}{2\pi}
\right)
\right]^2
-
\frac{g^2}{2}
\left(
\frac{\phi}{\sqrt{\pi}}
+
\frac{\theta}{2\pi}
\right)^2 .
\end{equation}

In this bosonised description $E$ can be eliminated algebraically by imposing its stationary-field equation $\partial\mathcal{L}_E/\partial E=0$. Completing the square gives
\begin{equation}
\label{E_boso}
E
=
-
g
\left(
\frac{\phi}{\sqrt{\pi}}
+
\frac{\theta}{2\pi}
\right).
\end{equation}

This shows that the electric field responds to two contributions: one from the bosonic field $\phi$, interpreted in terms of the fermionic charge distribution, and one from the topological angle $\theta$, corresponding to a background electric field.

\vspace{0.3 cm}

Eliminating $E$ leaves a term depending only on the bosonic field, interpreted as an effective potential for $\phi$:
\begin{equation}
V(\phi)
=
\frac{g^2}{2}
\left(
\frac{\phi}{\sqrt{\pi}}
+
\frac{\theta}{2\pi}
\right)^2
=
\frac{g^2}{2\pi}
\left(
\phi+
\frac{\theta}{2\sqrt{\pi}}
\right)^2 .
\end{equation}

This potential measures the energy cost of moving $\phi$ away from its equilibrium value.

\vspace{0.3 cm}

For $\theta=0$, the potential reduces to
\begin{equation}
V(\phi)
=
\frac{g^2}{2\pi}\phi^2 .
\end{equation}

Comparing this with the potential of a massive scalar field,
\begin{equation}
V(\phi)
=
\frac{1}{2}m_\gamma^2\phi^2 ,
\end{equation}

we can identify the corresponding effective mass:
\begin{equation}
m_\gamma
=
\frac{g}{\sqrt{\pi}} .
\end{equation}

Thus, even for massless fermions, the gauge interaction dynamically generates a mass for the bosonic excitation.

\vspace{0.3 cm}

The topological angle $\theta$ also appears as a shift in the potential. In general, the minimum is attained at
\begin{equation}
\label{min_boso}
\phi_{\text{min}}
=
-
\frac{\theta}{2\sqrt{\pi}} .
\end{equation}

The value of $\theta$ therefore shifts the minimum of the bosonic field. Substituting $\phi_{\mathrm{min}}=-\theta/(2\sqrt{\pi})$ into Eq.~\eqref{E_boso} gives $E_{\mathrm{vac}}=0$ in this massless, one-flavour description: the background field is fully screened in the vacuum. The shifting scalar minimum should not be interpreted as a non-zero unscreened vacuum electric field.

\subsection{Effect of the mass term}
\label{app:termino_masa}
So far, we have considered massless fermions. In that case, bosonisation gives a quadratic potential for $\phi$, whose minimum shifts continuously with $\theta$. When $m\neq0$, the Minkowski Lagrangian in Eq.~\eqref{modelo_con_masa} contains the mass contribution $-m\bar{\psi}\psi$. We consider the corresponding fermion bilinear,
\begin{equation}
m\bar{\psi}\psi ,
\end{equation}

This term can also be bosonised, although its derivation is more involved than that of the current. In particular, fermionic chiral bilinears must be represented through the bosonic field. Using the standard result of Refs.~\cite{Coleman1976MassiveSchwinger}\cite{Jentsch2022MassiveSchwingerFRG},
\begin{equation}
m\bar{\psi}\psi
\quad \longrightarrow \quad
- m C \cos\left(2\sqrt{\pi}\phi\right),
\end{equation}

where $C$ is a constant depending on the prescription used to bosonise the mass term. Including the fermion mass then gives the effective potential
\begin{equation}
V(\phi)
=
\frac{g^2}{2\pi}
\left(
\phi+
\frac{\theta}{2\sqrt{\pi}}
\right)^2
-
m C \cos\left(2\sqrt{\pi}\phi\right).
\end{equation}

The massive model therefore contains two contributions. The quadratic term, arising from the gauge field, has a minimum shifted by $\theta$. The periodic term from the fermion mass introduces a structure of competing minima for $\phi$.

\vspace{0.3 cm}

The competition between these contributions makes the massive vacuum more sensitive than in the massless model. In the one-continuum-flavour bosonised theory, two nearly degenerate minimum-energy configurations can arise near $\theta=\pi$. This discussion motivates, but does not quantitatively determine, the phase structure of the unrooted staggered lattice theory investigated numerically in this thesis.

On the one hand, the mass term favours periodic values of the field,
\begin{equation}
\phi_{min} = n\sqrt{\pi},
\qquad n\in\mathbb{Z},
\end{equation}

whereas the quadratic term is minimised at (\ref{min_boso}). For $\theta=\pi$, it follows that
\begin{equation}
\phi_{min} =
-\frac{\sqrt{\pi}}{2},
\end{equation}

that is, precisely between two consecutive minima of the periodic term,
\begin{equation}
\phi_{min}^{n=0}=0
\qquad \text{and} \qquad
\phi_{min}^{n=-1} =-\sqrt{\pi}.
\end{equation}

These two values have equal contributions and lie at the same distance from the quadratic minimum. Thus, at $\theta=\pi$, two vacuum configurations can compete symmetrically. As the mass term becomes more important, the system may select one of them as the ground state. In the thermodynamic limit, this change can manifest as a phase transition.

\vspace{0.3 cm}

The mass term distinguishes two regimes. For small masses, the cosine contribution is weak compared with the quadratic term generated by the gauge field. The effective potential then has a single dominant minimum, shifting smoothly as $\theta$ varies. The vacuum therefore responds continuously to the background electric field.

\vspace{0.3 cm} 

For sufficiently large masses, the cosine term becomes important and may produce multiple competing minima. Near $\theta=\pi$, the competition can cause the ground state to change abruptly from one configuration to another, resulting in a possible phase transition.

\vspace{0.3 cm} 

This discussion connects with the Schwinger effect. The angle $\theta$ can be interpreted as a background electric field, so its influence on the vacuum probes the response of the theory to an external field. In this picture, vacuum screening can be understood in terms of particle--antiparticle pair production.

\vspace{0.3 cm}

At small masses, pair production is less suppressed and the vacuum can screen the field more smoothly. At larger masses, pair production costs more energy and screening is less favourable. The fermion mass therefore strongly affects the response of the vacuum to the background field, particularly near $\theta=\pi$.

\section{Discretisation of the Schwinger Model}
\label{app:discretizado}
This appendix derives the lattice discretisation of the three contributions to the Euclidean Schwinger model: the gauge, fermionic and topological terms. We start from (\ref{terminos_Schwinger}) and use the conventions introduced in Sections~\ref{Ferrmion_reticulo} and \ref{sec:schwingermono}.

\subsection{Fermionic term}

We first discretise the fermionic part of the Euclidean Lagrangian,

\begin{equation}
\mathcal{L}_F
=
\bar{\psi}
\left(
\gamma^\mu D_\mu + m
\right)
\psi .
\end{equation}

As discussed in Section~\ref{Ferrmion_reticulo}, staggered fermions are used, replacing the spinor field $\psi$ with the one-component Grassmann fields defined in (\ref{grassman_unacomp}).
\begin{equation}
m\bar{\psi}(x)\psi(x)
\quad \longrightarrow \quad
m\,\bar{\chi}(n)\chi(n).
\end{equation}

The kinetic term contains the covariant derivative and couples neighbouring lattice sites. We start from the symmetric gauge-covariant discretisation defined in (\ref{deriv_cov_discret}). Substituting this expression gives
\begin{equation}
\bar{\psi}(x)\gamma^\mu D_\mu\psi(x)
\quad \longrightarrow \quad
\frac{1}{2a}
\sum_{\mu=0}^{1}
\bar{\psi}(n)\gamma^\mu
\left[
U_\mu(n)\psi(n+\hat{\mu})
-
U_\mu^\dagger(n-\hat{\mu})\psi(n-\hat{\mu})
\right].
\end{equation}

In the staggered formulation, the spinor fields are replaced by one-component Grassmann fields $\chi(n)$ and $\bar\chi(n)$, and the matrix structure is encoded in the staggered phases $\eta^\mu(n)$. The expression therefore becomes
\begin{equation}
\bar{\psi}(n)\gamma^\mu U_\mu(n)\psi(n+\hat{\mu})
\quad \longrightarrow \quad
\eta^\mu(n)\bar{\chi}(n)U_\mu(n)\chi(n+\hat{\mu}),
\end{equation}
whereas
\begin{equation}
\bar{\psi}(n)\gamma^\mu U_\mu^\dagger(n-\hat{\mu})\psi(n-\hat{\mu})
\quad \longrightarrow \quad
\eta^\mu(n)\bar{\chi}(n)U_\mu^\dagger(n-\hat{\mu})\chi(n-\hat{\mu}).
\end{equation}

The kinetic term is then
\begin{equation}
\frac{1}{2a}
\sum_{\mu=0}^{1}
\eta^\mu(n)
\left[
\bar{\chi}(n)U_\mu(n)\chi(n+\hat{\mu})
-
\bar{\chi}(n)U_\mu^\dagger(n-\hat{\mu})\chi(n-\hat{\mu})
\right].
\end{equation}

To express both hopping terms through a link starting at the same site, shift the summation index $n\to n+\hat\mu$ in the backward-hopping contribution. This is allowed because the sum includes every lattice site. The staggered phase is unchanged under this shift because it does not depend on the coordinate $n_\mu$, as follows from Eq.~\eqref{staggered_phase}: 
\begin{equation}
\eta^\mu(n+\hat{\mu})=\eta^\mu(n).
\end{equation}
Therefore,
\begin{equation}
\sum_n
\eta^\mu(n)
\bar{\chi}(n)U_\mu^\dagger(n-\hat{\mu})\chi(n-\hat{\mu})
=
\sum_n
\eta^\mu(n)
\bar{\chi}(n+\hat{\mu})U_\mu^\dagger(n)\chi(n).
\end{equation}

Hence,
\begin{equation}
\bar{\psi}(x)\gamma^\mu D_\mu\psi(x)
\quad \longrightarrow \quad
\frac{1}{2a}
\sum_{\mu=0}^{1}
\eta^\mu(n)
\left[
\bar{\chi}(n)U_\mu(n)\chi(n+\hat{\mu})
-
\bar{\chi}(n+\hat{\mu})U_\mu^\dagger(n)\chi(n)
\right].
\end{equation}

Using the discretisation rule for the integral (\ref{int_to_sum}), the lattice fermionic action is finally
\begin{equation}
S_F
=
a^2\sum_n m\,\bar{\chi}(n)\chi(n)
+
\frac{a}{2}
\sum_{n,\mu}
\eta^\mu(n)
\left[
\bar{\chi}(n)U_\mu(n)\chi(n+\hat{\mu})
-
\bar{\chi}(n+\hat{\mu})U_\mu^\dagger(n)\chi(n)
\right].
\end{equation}

\subsection{Gauge term}
\label{T_gauge_anexo}

The continuum gauge term is
\begin{equation}
\mathcal{L}_G(x)
=
\frac{1}{4}F_{\mu\nu}(x)F^{\mu\nu}(x).
\end{equation}

In $1+1$ dimensions, $F_{\mu\nu}F^{\mu\nu}$ can be written using the only independent field-strength component. In Euclidean space, antisymmetry gives
\begin{equation}
F_{\mu\nu}F^{\mu\nu}
=
F_{01}F^{01}+F_{10}F^{10}
=
2F_{01}^2.
\end{equation}

The continuum gauge action is therefore
\begin{equation}
\label{cont_gauge}
S_G
=
\int d^2x\,\mathcal{L}_G(x)
=
\frac{1}{2}
\int d^2x\,F_{01}^2(x).
\end{equation}

On the lattice, the gauge contribution must be built from link variables while preserving local gauge invariance. The elementary gauge-invariant object is the plaquette defined in (\ref{plaqueta}). Thus, the lattice action is expressed through $U_p(n)$.

\vspace{0.3 cm}

To recover the continuum field strength explicitly, we start from the relation between link variables and the gauge field in (\ref{fase_en_continuo}). Using the plaquette definition (\ref{plaqueta}) and the Abelian nature of $U(1)$, we obtain
\begin{equation}
U_p(n)
=
\exp\left\{
iag
\left[
A_0(n)
+
A_1(n+\hat{0})
-
A_0(n+\hat{1})
-
A_1(n)
\right]
\right\}.
\end{equation}

In the continuum limit, fields vary smoothly between neighbouring sites, so
\begin{equation}
A_1(n+\hat{0})
=
A_1(n)
+
a\,\partial_0A_1(n)
+
\mathcal{O}(a^2),
\end{equation}
\begin{equation}
A_0(n+\hat{1})
=
A_0(n)
+
a\,\partial_1A_0(n)
+
\mathcal{O}(a^2).
\end{equation}

Substituting these expansions and using the field-strength definition (\ref{deriv_cov}) gives
\begin{equation}
U_p(n)
=
\exp\left\{
ia^2gF_{01}(n)
+
\mathcal{O}(a^3)
\right\}.
\end{equation}

Expanding consistently to the leading order required here gives
\begin{equation}
\label{desarrollo_plaqueta_continuo}
U_p(n)=1+ia^2gF_{01}(n)+\mathcal{O}(a^3).
\end{equation}
The plaquette phase is real. Taking the real part of the full exponential, rather than truncating its imaginary terms at different orders, therefore yields the gauge-action expansion below.

\vspace{0.3 cm}

The lattice gauge action is not unique, but it must be local, gauge invariant, real and approach the continuum action (\ref{cont_gauge}) as $a\to0$. One standard choice with these properties is
\begin{equation}
\label{accion_gauge_provisional}
S_G
=
\beta
\sum_n
\left[
1-\operatorname{Re}U_p(n)
\right].
\end{equation}

Using the phase of the plaquette from the exponential expression above, the constant term cancels the $1$ in the action and the leading real correction is quadratic in the phase. Hence,
\begin{equation}
1-\operatorname{Re}U_p(n)
=
\frac{a^4g^2}{2}F_{01}^2(n)
+
\mathcal{O}(a^5).
\end{equation}

Substitution into (\ref{accion_gauge_provisional}) yields
\begin{equation}
S_G
=
\beta
\sum_n
\left[
1-\operatorname{Re}U_p(n)
\right]
\simeq
\frac{\beta a^4g^2}{2}
\sum_n
F_{01}^2(n).
\end{equation}

On the other hand, the continuum action (\ref{cont_gauge}) is discretised by the rule (\ref{int_to_sum}) as
\begin{equation}
S_G
=
\frac{1}{2}
\int d^2x\,F_{01}^2(x)
\quad \longrightarrow \quad
\frac{a^2}{2}
\sum_n
F_{01}^2(n).
\end{equation}

For these expressions to agree in the continuum limit, we must have
\begin{equation}
\frac{\beta a^4g^2}{2}
\sum_n
F_{01}^2(n)
=
\frac{a^2}{2}
\sum_n
F_{01}^2(n).
\end{equation}
Simplifying gives
\begin{equation}
\beta a^4g^2=a^2,
\end{equation}
and therefore
\begin{equation}
\beta
=
\frac{1}{g^2a^2}.
\end{equation}

The discretised gauge term is thus
\begin{equation}
S_G
=
\beta
\sum_n
\left[
1-\operatorname{Re}U_p(n)
\right],
\qquad
\beta
=
\frac{1}{g^2a^2}.
\end{equation}

The parameter $\beta$ is called the lattice gauge coupling because it combines the continuum parameters $g$ and $a$ into a single quantity. We have therefore obtained a discrete form of the Schwinger-model gauge action.

\subsection{Topological term}
\label{app:s:theta}

The Euclidean topological term is
\begin{equation}
\mathcal{L}_\theta
=
i\frac{g\theta}{2\pi}F_{01}.
\end{equation}

Its contribution to the continuum action is therefore
\begin{equation}
S_\theta
=
i\frac{g\theta}{2\pi}
\int d^2x\,F_{01}(x).
\end{equation}

To discretise this contribution, we use the plaquette expansion in (\ref{desarrollo_plaqueta_continuo}). Since the plaquette phase obeys $\varphi_p(n)=a^2gF_{01}(n)+\mathcal{O}(a^3)$ for smooth fields, we have the exact identity in terms of the lattice phase
\begin{equation}
U_p(n)-U_p^{-1}(n)=2i\sin\varphi_p(n),
\end{equation}
and thus, at leading order in the lattice spacing, $U_p(n)-U_p^{-1}(n)\simeq 2i\sin\left(a^2gF_{01}(n)\right)$.

In the continuum limit,
\begin{equation}
\sin\left(a^2gF_{01}(n)\right)
\simeq
a^2gF_{01}(n),
\end{equation}
and therefore
\begin{equation}
U_p(n)-U_p^{-1}(n)
\simeq
2ia^2gF_{01}(n).
\end{equation}

This motivates defining the field-theoretical lattice topological charge as
\begin{equation}
Q_{\mathrm{lat}}[U]
=
\frac{1}{i4\pi}
\sum_n
\left[
U_p(n)-U_p^{-1}(n)
\right].
\end{equation}

Using the preceding expansion,
\begin{equation}
Q_{\mathrm{lat}}[U]
\simeq
\frac{1}{i4\pi}
\sum_n
2ia^2gF_{01}(n)
=
\frac{a^2g}{2\pi}
\sum_n
F_{01}(n).
\end{equation}

Finally, applying the discretisation rule (\ref{int_to_sum}) gives
\begin{equation}
Q_{\mathrm{lat}}[U]
\longrightarrow
\frac{g}{2\pi}
\int d^2x\,F_{01}(x).
\end{equation}

The topological contribution can therefore be written as
\begin{equation}
S_\theta = i\theta Q_{\mathrm{lat}}[U].
\end{equation}

Substituting the definition of $Q[U]$ yields
\begin{equation}
\label{S_theta2}
S_\theta
=
i\theta Q_{\mathrm{lat}}[U]
=
\frac{\theta}{4\pi}
\sum_n
\left[
U_p(n)-U_p^{-1}(n)
\right].
\end{equation}

Unlike the branch-dependent phase sum, the sine-based $Q_{\mathrm{lat}}$ is generally non-integer at finite spacing, so exact $2\pi$ periodicity in $\theta$ is not guaranteed for this lattice action. Both definitions agree at leading order for smooth continuum fields, as discussed in the discussion around Eq.~\eqref{carga_topo}. Although there is no explicit factor of $i$ in the final expression, the term remains complex because $U_p(n)-U_p^{-1}(n)$ is purely imaginary for $U_p(n)\in U(1)$. This motivates the dual formulation.

\section{Update Algorithm Pseudocode}
\label{pseudo}
This appendix provides pseudocode for the main algorithms used in the Monte Carlo simulations. An associated implementation is publicly available in the GitHub repository \cite{github}. The pseudocode provides an algorithmic summary; implementation-level details, including the full transition rules, are available in the linked source code.
\subsection{Worm algorithm}

First, choose uniformly a site \(x_0\) that does not belong to a loop. If no such site exists, reject the update. Then compute
\[
p_+ =
\begin{cases}
\min(1,2m), & m\neq 0,\\
0, & m=0,
\end{cases}
\qquad
p_- =
\begin{cases}
\min\left(1,\frac{1}{2m}\right), & m\neq 0,\\
0, & m=0.
\end{cases}
\]
If \(x_0\) contains a monomer, propose an open worm with probability \(p_-\). Otherwise, propose it with probability \(p_+\). If no open worm is proposed, perform a closed worm.
\subsubsection{Closed worm}
\begin{algorithm}[H]
\footnotesize

\KwData{\(x_0=(t_0,x_0)\): initial site; \(n_x\): monomer; \(b_\ell\): link}
\KwResult{Accepted or rejected configuration}
\vspace{1.1 mm}

Initialise head, tail, link types and iteration counter:
\[
x_h\leftarrow x_0,\quad x_t\leftarrow x_0,
\qquad
q_h\leftarrow 0,\quad q_t\leftarrow 0,
\qquad
k\leftarrow 0.
\]

\While{\TRUE}{
    \(k\leftarrow k+1\)\;

    construct \(\mathcal{D}_{q_h}(x_h)\) and \(\mathcal{D}_{q_t}(x_t)\)\;

    compute \(n_h=|\mathcal{D}_{q_h}(x_h)|\) and
    \(n_t=|\mathcal{D}_{q_t}(x_t)|\)\;

    \If{\(n_h=0\) and \(n_t=0\)}{
        \KwReject\;
    }

    \eIf{\(n_h>0\) and \(n_t>0\)}{
        choose \(e\in\{h,t\}\) at random\;
    }{
        choose the only available endpoint\;
    }

    choose \(a\in\mathcal{D}_{q_e}(x_e)\) at random\;

    \If{\(a\notin\{0,1,2,3\}\)}{
        \KwReject\;
    }

    identify \(\ell=(x_e,a)\) and update \(b_\ell\leftarrow 2-b_\ell\)\;

    move the selected endpoint: \(x_e\leftarrow x_e+a\)\;

    \eIf{\(n_{x_e}=1\)}{
        \(q_e\leftarrow 0\)\;
    }{
        \(q_e\leftarrow 1-q_e\)\;
    }

    \If{\(x_h=x_t\)}{
        \KwAccept\;
    }
}

\KwReject\;
\end{algorithm}

\clearpage

\subsubsection{Open worm}

\begin{algorithm}[H]
\footnotesize

\KwData{\(x_0=(t_0,x_0)\): initial site; \(n_x\): monomer; \(b_\ell\): link; \(p_+,p_-\): creation and annihilation probabilities}
\KwResult{Accepted or rejected configuration}
\vspace{1.1 mm}

Toggle the initial monomer:
\[
n_{x_0}\leftarrow 1-n_{x_0}.
\]

Initialise position, requested link type and iteration counter:
\[
x\leftarrow x_0,
\qquad
q\leftarrow (n_x=0),
\qquad
k\leftarrow 0.
\]

\While{\TRUE}{
    \(k\leftarrow k+1\)\;

    construct the set of admissible moves \(\mathcal{D}_q(x)\)\;

    \[
    \mathcal{D}_q(x)=
    \left\{
    \ell\sim x:
    \begin{array}{ll}
    b_\ell=0, & q=1,\\
    b_\ell=2, & q=0
    \end{array}
    \right\}.
    \]

    compute \(n_d=|\mathcal{D}_q(x)|\)\;

    check whether the worm can close at the current site:
    \[
    s=(q=0\ \mathrm{and}\ n_x=1)
    \quad \mathrm{or} \quad
    (q=1\ \mathrm{and}\ n_x=0).
    \]

    \If{\(n_d=0\) and \(s=0\)}{
        restore the local configuration from the backup\;
        \KwReject\;
    }

    choose an action at random \(a\)\;

    \If{\(s=1\)}{
        \(a\in\mathcal{D}_q(x)\cup\{\mathrm{close}\}\)\;
    }

    \If{\(s=0\)}{
        \(a\in\mathcal{D}_q(x)\)\;
    }

    \eIf{\(a=\mathrm{close}\)}{
        \eIf{\(n_x=0\)}{
            attempt \(n_x\leftarrow 1\) with probability \(p_+\)\;
        }{
            attempt \(n_x\leftarrow 0\) with probability \(p_-\)\;
        }

        \If{the change is accepted}{
            \KwAccept\;
        }
    }{
        \If{\(a\notin\{0,1,2,3\}\)}{
            restore the local configuration from the backup\;
            \KwReject\;
        }

        identify the traversed link \(\ell=(x,a)\)\;

        update the link:
        \[
        b_\ell\leftarrow 2-b_\ell.
        \]

        move the worm head:
        \[
        x\leftarrow x+a.
        \]

        update the requested link type\;

        \eIf{\(n_x=1\)}{
            \(q\leftarrow 0\)\;
        }{
            \(q\leftarrow 1-q\)\;
        }
    }
}
\KwReject\;

\end{algorithm}

\subsection{Local plaquette update}

\begin{algorithm}[H]
\footnotesize

\KwData{\(p_x\): plaquette field; \(n_x\): monomer; \(b_\ell\): link}
\KwResult{Accepted or rejected configuration}
\vspace{1.1 mm}

Choose a plaquette at random \(x_0=(t_0,x_0)\)\;

Read the four plaquette links:
\[
\ell=(b_0,b_1,b_2,b_3)
\]

\BlankLine

\textbf{1. Monomers, parallel dimers or loop \(1\times1\)}\;

Identify
\[
s_i =
\begin{cases}
M, & \text{four monomers},\\
D, & \text{two parallel dimers},\\
P, & \text{oriented loop }1\times1.
\end{cases}
\]

\If{a state exists \(s_i\)}{
    choose \(s_f\neq s_i\)\;

    compute the weights \(W_i\) and \(W_f\):
    \[
    W_M=16m^4\,w(p),\qquad
    W_D=2w(p),\qquad
    W_P=2w(p)
    \]

    \If{Metropolis\((W_i,W_f)\)}{
        update monomers, links and \(p_{x_0}\)\;
        \KwAccept\;
    }
}

\BlankLine

\textbf{2. Corner rotation}\;

\For{each corner \(i\)}{
    \If{there is a monomer at \(i\) and two links forming a corner}{
        propose the local corner rotation\;

        compute the weights \(W_i\) and \(W_f\)\;

        \If{Metropolis\((W_i,W_f)\)}{
            move the monomer to the opposite corner\;
            rotate the corner links\;
            update \(p_{x_0}\)\;
            \(\mathrm{sign}\leftarrow-\mathrm{sign}\)\;
            \KwAccept\;
        }
    }
}

\BlankLine

\textbf{3. Antiparallel links}\;

\If{two antiparallel links lie on opposite sides}{
    propose rotating them to the perpendicular direction\;

    compute the weights \(W_i\) and \(W_f\)\;

    \If{Metropolis\((W_i,W_f)\)}{
        update links and \(p_{x_0}\)\;
        \KwAccept\;
    }
}

\BlankLine

\textbf{4. Type-C configuration}\;

\For{each orientation \(i\)}{
    identify
    \[
    s_i\in\{M+\ell,\ D+\ell,\ C\}
    \]

    \If{a state exists \(s_i\)}{
        choose \(s_f\neq s_i\)\;
        compute the weights \(W_i\) and \(W_f\)\;

        \If{Metropolis\((W_i,W_f)\)}{
            update monomers, links and \(p_{x_0}\)\;
            \KwAccept\;
        }
    }
}

\KwReject\;

\end{algorithm}

\section{Auxiliary Figures}
\subsection{Thermalisation in the \texorpdfstring{$(m/g,\theta)$}{(m/g,theta)} plane}
\label{Termalizado_anexo}
\begin{figure}[H]
\centering
\includegraphics[width=0.97\textwidth]{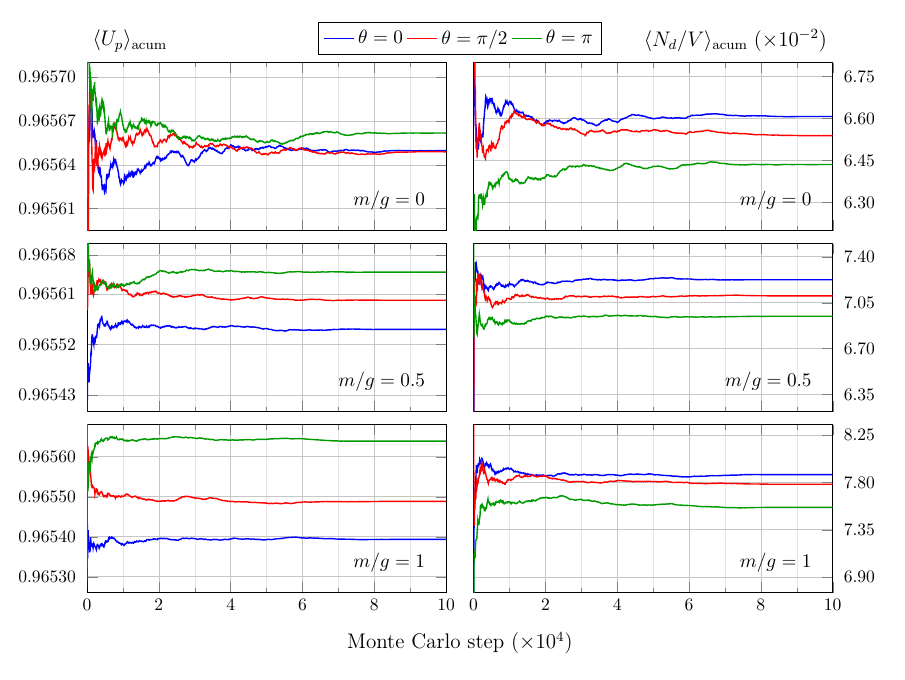}
\caption[Thermalisation across the $(m/g,\theta)$ plane]{Cumulative averages of the plaquette observable and dimer density versus Monte Carlo steps for several masses and topological angles. The simulations used $12\times12$ lattices, and the time axis is in units of $10^4$ Monte Carlo updates. Curves are reconstructed directly from the numerical tables included in the source package.}
\label{fig:termalizacion_masas_thetas_simple}
\end{figure}

\newpage

\subsection{Slices of the \texorpdfstring{$(m/g,\theta)$}{(m/g,theta)} plane}
\label{Cortes_plano_anexo}
\begin{figure}[htbp]
\centering

\begin{tikzpicture}
\begin{axis}[
    width=0.82\textwidth,
    height=0.50\textwidth,
    xlabel={$\theta/\pi$},
    ylabel={$\langle U_p\rangle$},
    xmin=0,
    xmax=2,
    grid=both,
    scaled y ticks=false,
    xticklabel={\pgfmathprintnumber[fixed,zerofill,precision=1]{\tick}},
    yticklabel={\pgfmathprintnumber[fixed,zerofill,precision=4]{\tick}},
    legend style={
        at={(0.51,0.30)},
        anchor=center,
        draw=none,
        fill=white,
        fill opacity=0.75,
        text opacity=1
    },
    legend cell align=left,
    tick align=inside,
]

\addplot+[
    blue,
    dashed,
    mark=*,
    mark size=1.6pt,
    line width=0.9pt,
    error bars/y dir=both,
    error bars/y explicit,
]
table[
    x index=0,
    y index=1,
    y error index=2
]
{cortes/plaquette_theta_scan_target_mg_0p00_smoothed.dat};
\addlegendentry{$m/g=0.00$}

\addplot+[
    orange,
    dashed,
    mark=*,
    mark size=1.6pt,
    line width=0.9pt,
    error bars/y dir=both,
    error bars/y explicit,
]
table[
    x index=0,
    y index=1,
    y error index=2
]
{cortes/plaquette_theta_scan_target_mg_0p10_smoothed.dat};
\addlegendentry{$m/g=0.10$}

\addplot+[
    green!60!black,
    dashed,
    mark=*,
    mark size=1.6pt,
    line width=0.9pt,
    error bars/y dir=both,
    error bars/y explicit,
]
table[
    x index=0,
    y index=1,
    y error index=2
]
{cortes/plaquette_theta_scan_target_mg_0p20_smoothed.dat};
\addlegendentry{$m/g=0.20$}

\addplot+[
    red,
    dashed,
    mark=*,
    mark size=1.6pt,
    line width=0.9pt,
    error bars/y dir=both,
    error bars/y explicit,
]
table[
    x index=0,
    y index=1,
    y error index=2
]
{cortes/plaquette_theta_scan_target_mg_0p25_smoothed.dat};
\addlegendentry{$m/g=0.25$}

\end{axis}
\end{tikzpicture}

\caption[Smoothed slices of the average-plaquette map]{Illustrative, Gaussian-smoothed slices of the average plaquette $\langle U_p\rangle$ versus $\theta/\pi$ for several masses, at $V=12\times12$ and $\beta=14.4$. Only the central values were smoothed; the displayed jackknife errors refer to the original unsmoothed estimates. These curves are not used for quantitative inference.}

\end{figure}
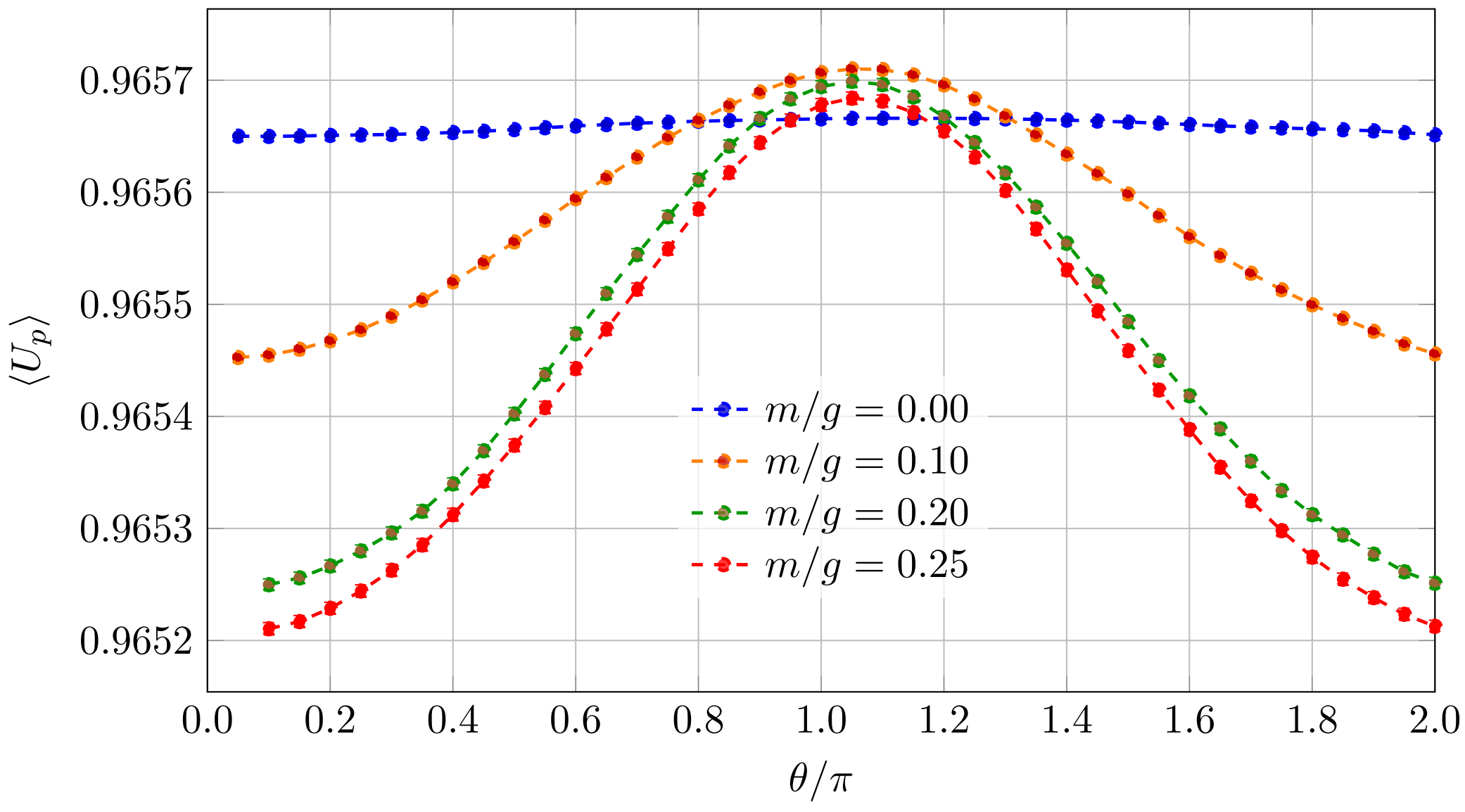
\vspace{0.5 cm}
\begin{figure}[htbp]
\centering

\begin{tikzpicture}
\begin{axis}[
    width=0.82\textwidth,
    height=0.50\textwidth,
    xlabel={$m/g$},
    ylabel={$\langle \bar{\chi}\chi\rangle$},
    xmin=-0.0,
    xmax=0.27,
    xtick={0,0.05,0.10,0.15,0.20,0.25,0.30},
    grid=both,
    scaled y ticks=false,
    xticklabel={\pgfmathprintnumber[fixed,zerofill,precision=2]{\tick}},
    yticklabel={\pgfmathprintnumber[fixed,zerofill,precision=1]{\tick}},
    legend style={
        at={(0.52,0.28)},
        anchor=center,
        draw=none,
        fill=white,
        fill opacity=0.75,
        text opacity=1
    },
    legend cell align=left,
    tick align=inside,
]

\addplot+[
    blue,
    dashed,
    mark=*,
    mark size=1.6pt,
    line width=0.9pt,
    error bars/y dir=both,
    error bars/y explicit,
]
table[
    header=false,
    x index=0,
    y index=1,
    y error index=2
]
{cortes/chiral_condensate_mass_scan_theta_0.dat};
\addlegendentry{$\theta/\pi=0.0$}

\addplot+[
    orange,
    dashed,
    mark=*,
    mark size=1.6pt,
    line width=0.9pt,
    error bars/y dir=both,
    error bars/y explicit,
]
table[
    header=false,
    x index=0,
    y index=1,
    y error index=2
]
{cortes/chiral_condensate_mass_scan_theta_pi_over_2.dat};
\addlegendentry{$\theta/\pi=0.5$}

\addplot+[
    green!60!black,
    dashed,
    mark=*,
    mark size=1.6pt,
    line width=0.9pt,
    error bars/y dir=both,
    error bars/y explicit,
]
table[
    header=false,
    x index=0,
    y index=1,
    y error index=2
]
{cortes/chiral_condensate_mass_scan_theta_pi.dat};
\addlegendentry{$\theta/\pi=1.0$}

\addplot+[
    red,
    dashed,
    mark=*,
    mark size=1.6pt,
    line width=0.9pt,
    error bars/y dir=both,
    error bars/y explicit,
]
table[
    header=false,
    x index=0,
    y index=1,
    y error index=2
]
{cortes/chiral_condensate_mass_scan_theta_3pi_over_2.dat};
\addlegendentry{$\theta/\pi=1.5$}

\addplot+[
    purple,
    dashed,
    mark=*,
    mark size=1.6pt,
    line width=0.9pt,
    error bars/y dir=both,
    error bars/y explicit,
]
table[
    header=false,
    x index=0,
    y index=1,
    y error index=2
]
{cortes/chiral_condensate_mass_scan_theta_2pi.dat};
\addlegendentry{$\theta/\pi=2.0$}

\end{axis}
\end{tikzpicture}

\caption[Vertical slices of the chiral-condensate map]{Chiral condensate $\langle\bar{\chi}\chi\rangle$ versus $m/g$ for different $\theta/\pi$, with $V=12\times12$ and $\beta=14.4$. The condensate increases monotonically with mass for all angles considered. Differences between slices are most visible at low and intermediate masses, particularly near $\theta/\pi\simeq1$, while the curves approach one another at larger masses.}

\newpage

\end{figure}
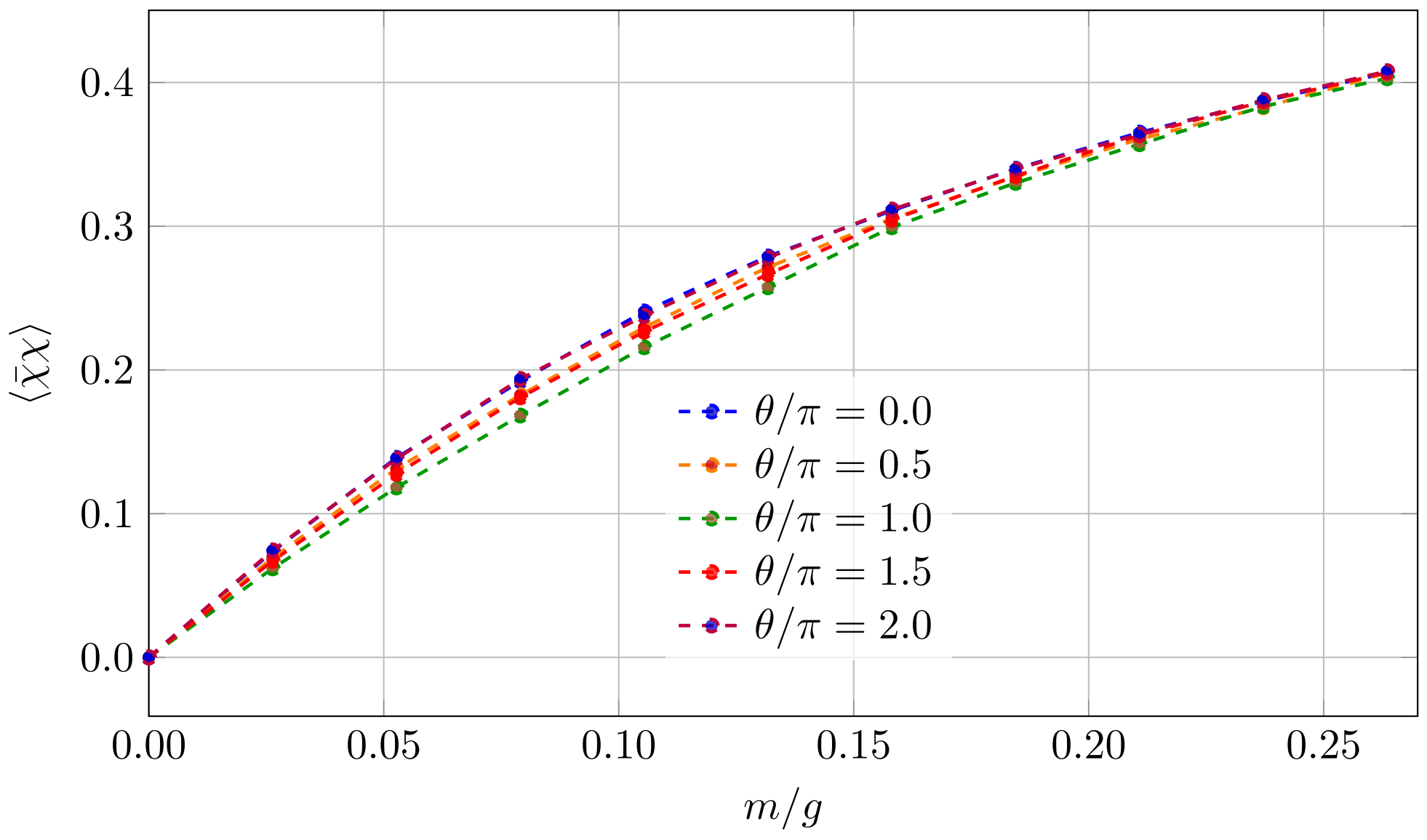
\end{document}